\documentclass[aps,prl,twocolumn,10pt, showkeys,showpacs,amsmath,amsfonts,floatfix,superscriptaddress,nofootinbib,longbibliography]{revtex4-2}

\usepackage{chemformula}

\usepackage{amssymb,amsmath,amstext}                
\usepackage{graphicx}                                               
\usepackage{epstopdf}                                               
\usepackage{color}       
\usepackage{bm}
\usepackage{blindtext}
\usepackage{appendix}                                              
\usepackage[utf8]{inputenc}
\usepackage{ulem}
\usepackage{latexsym}

\usepackage[colorlinks=true,citecolor=blue,linkcolor=magenta]{hyperref}
\def\be{\begin{equation}}
\def\ee{\end{equation}}
\def\bea{\begin{eqnarray}}
\def\eea{\end{eqnarray}}
\def\bi{\begin{itemize}}
\def\ei{\end{itemize}}

\newcommand{\ket}[1]{\mbox{$| #1 \rangle$}}

\makeatletter
\def\paragraph{%
  \@startsection
    {paragraph}%
    {4}%
    {\parindent}%
    {\z@}%
    {-0em}%
    {\normalfont\normalsize\itshape}%
}%
\makeatother

\begin{document}

\title{Superfluid dome in the spatially modulated two-dimensional XY model}

\author{Feng-Feng Song}
\email{song@issp.u-tokyo.ac.jp}
\affiliation{Institute for Solid State Physics, The University of Tokyo, Kashiwa, Chiba 277-8581, Japan}

\author{Aditya Chugh}
\affiliation{Max Planck Institute for the Physics of Complex Systems, N\"{o}thnitzer Strasse 38, Dresden 01187, Germany}

\author{Hanggai Nuomin}
\affiliation{Department of Chemistry, Duke University, Durham, North Carolina 27708, United States}

\author{Naoki Kawashima}
\email{kawashima@issp.u-tokyo.ac.jp}
\affiliation{Institute for Solid State Physics, The University of Tokyo, Kashiwa, Chiba 277-8581, Japan}
\affiliation{Trans-scale Quantum Science Institute, The University of Tokyo, Bunkyo, Tokyo 113-0033, Japan}

\author{Alexander Wietek}
\email{awietek@pks.mpg.de}
\affiliation{Max Planck Institute for the Physics of Complex Systems, N\"{o}thnitzer Strasse 38, Dresden 01187, Germany}

\date{\today}

\begin{abstract}
In strongly correlated electron systems, superconductivity and charge density waves often coexist in close proximity, suggesting a deeper relationship between these competing phases. Recent research indicates that these orders can intertwine, with the superconducting order parameter coupling to modulations in the electronic density. To elucidate this interplay, we study a two-dimensional XY model with a periodic modulation of the coupling strength in one spatial direction. Using a combination of tensor network methods and Monte Carlo simulations, we reveal a non-monotonic, dome-like dependence of $T_c$ on the modulation wavelength, with the peak $T_c$ shifting to longer wavelengths as the modulation strength grows. The origin of this phenomenon is traced back to an effective pinning of vortices in the valleys of the modulation, confirmed by a comparison to modulated $q$-state clock models. These findings shed new light on the phase behavior of intertwined superconducting and charge-ordered states, offering a deeper understanding of their complex interactions.
\end{abstract}

\maketitle
\paragraph{Introduction---} The interplay between charge orders and superconductivity (SC) leads to intriguing phenomena in a broad variety of strongly correlated electron materials. Coexistence of charge density waves (CDW) and SC was first observed in transition metal dichalcogenides such as $2H$-\ch{NbSe2} and $1T$-\ch{TiSe2}~\cite{Wilson1975,Moncton1975,Yokoya2001,Kusmartseva2009,Arguello2015,Lian2018}. Other material platforms with competing CDW and SC orders include the cuprate superconductors~\cite{Tranquada1995,Howald2003,Ghiringhelli2012,Comin2014,Tabis2014,Miao2021}, iron-based superconductors~\cite{Hu2025}, kagome metals~\cite{Ortiz2019,Ortiz2020,Uykur2021,Wenzel2022,Uykur2022}, \ch{SrPt2As2} and \ch{LaPt2Si2}~\cite{Kudo2010,Nagano2013,Kim2015}. From a theoretical perspective, it was proposed that CDW order can under particular circumstances enhance SC~\cite{Machida1987}, and form exotic intertwined orders~\cite{Vojta2009,Fradkin2015}, such as pair density waves~\cite{Agterberg2020}. 


Numerical techniques have meanwhile made tremendous progress towards building an understanding of unconventional superconductors by solving microscopic models of strongly interacting electrons. In particular, significant progress has been made in accurately simulating variants of the single-band Hubbard model relevant to cuprate high-temperature superconductors~\cite{LeBlanc2015,Zheng2017,Huang2017,Huang2018,Qin2020}. Several works have recently made the surprising discovery that CDW can coexist with SC, a state of matter sometimes referred to as a supersolid or superconducting stripes~\cite{Gong2021,Jiang2021,HCJiang2021,Jiang2023,Jiang2023b,Xu2024,Lu2024,Ponsioen2023,Wietek2022}. Detailed investigations using the Penrose-Onsager criterion of SC have revealed a remarkable feature of this state~\cite{Wietek2022}. In the presence of CDW, the SC condensate is fragmented, i.e., more than one Cooper-pair wave function is macroscopically occupied~\cite{Wietek2022,Baldelli2025}. 

Interestingly, the origin of the fragmentation has been understood as a consequence of coupling the superconducting order parameter to the charge density. More specifically, Ref.~\cite{Baldelli2025} considered an intertwined Ginzburg-Landau theory, where the free energy functional is of the form,   
\be
\label{eq:intertwinedgl}
\mathcal{F}[\psi] = \alpha(\bm{r})|\psi|^2 + \frac{\beta}{2} |\psi|^4 + \frac{1}{2m^*} \left| \left(-i \hbar \vec{\nabla} + 2e \bm{A}\right) \psi \right|^2
\ee
where $\alpha(\bm{r})$ and $\beta$ are phenomenological expansion coefficients, $m^*$ is the effective mass of Cooper condensates, and $\bm{A}$ is the vector potential. 

The crucial difference to standard Ginzburg-Landau theory is the position dependence of the coupling $\alpha(\bm{r})$. In the context of the Hubbard model a proper choice is $\alpha(\bm{r}) = \alpha n_h(\bm{r})$, where $n_h(\bm{r}) = 1- n(\bm{r})$ denotes the hole-density ($n(\bm{r})$ is the electron density). Eq. \eqref{eq:intertwinedgl} constitutes a minimalistic model describing the coupling between the charge degrees of freedom $n(\bm{r})$ and the superconducting order parameter $\psi(\bm{r})$. In this work, we aim to unravel the physical phenomena emerging when $\alpha(\bm{r})$ is periodically modulated due to the presence of a CDW.

\paragraph{Modulated XY Model---}
\begin{figure}[tbp]
    \centering
    \includegraphics[width=0.99\linewidth]{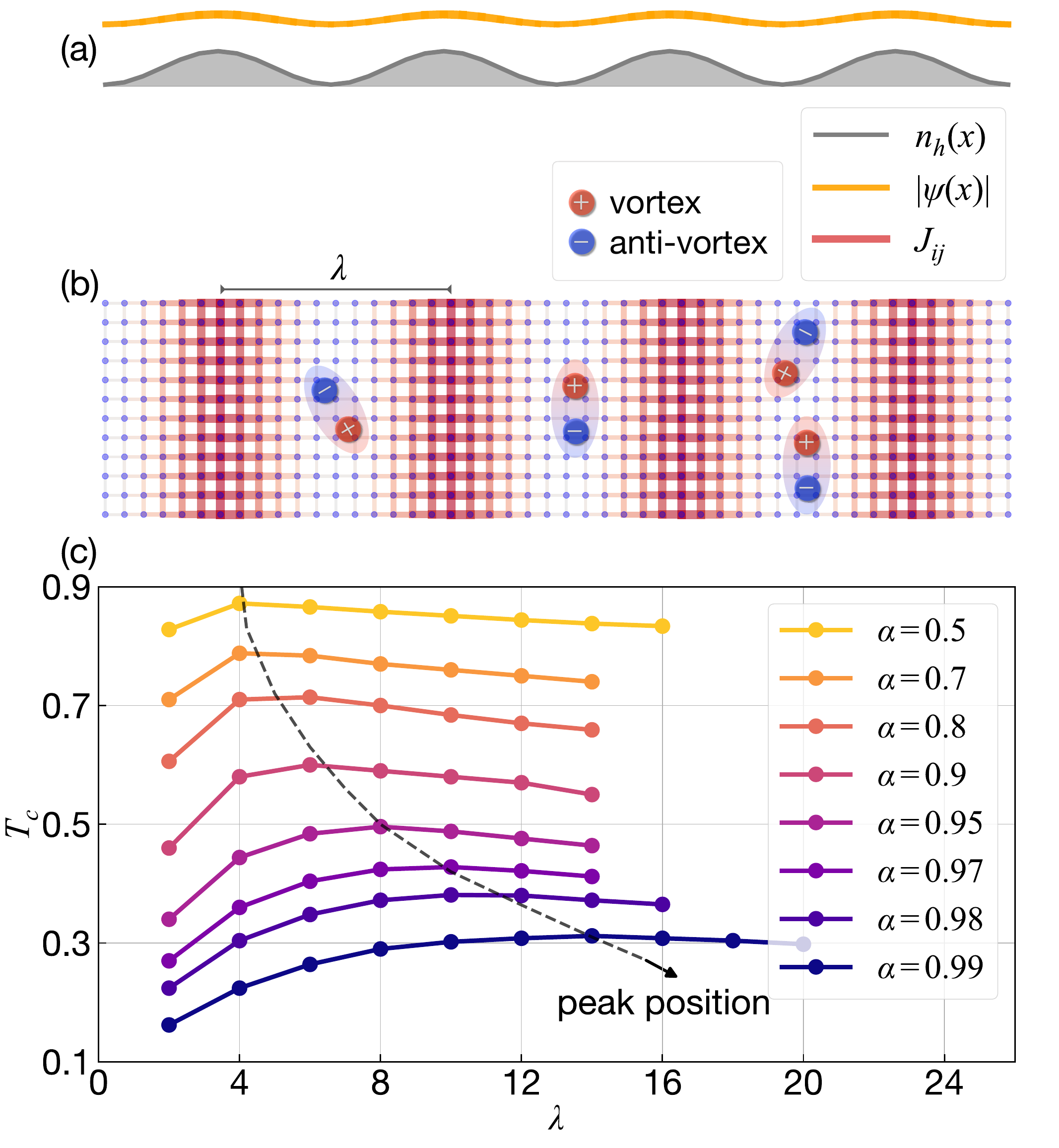}
    \caption{
    Illustration of the interplay between the CDW and the superconducting order. 
    (a) The hole density $n_h(x)$ modulated along $x$-direction gives rise to a spatial modulation in the amplitude of the superconducting order parameter $\psi(x)$. 
    (b) The modulated XY model with varying interaction strength $J_{ij}$ (red bonds) between NN spins (blue dots). The superconducting vortices are pinned between the CDW stripes and bound in pairs at low temperatures.
    (c) Dome structures in the dependence of $T_c$ on the modulation amplitudes $\alpha$ and wave lengths $\lambda$. The peak positions of $T_c$ shift to larger $\lambda$ with increasing $\alpha$. Dashed black line is a guide to eye.}
    \label{fig:model}
\end{figure}

The complex superconducting order parameter $\psi(\bm{r})$ can be expressed in terms of its amplitude and phase as $\psi(\bm{r}) = |\psi(\bm{r})| e^{i\theta(\bm{r})}$. Within the intertwined Ginzburg-Landau framework, the presence of charge order $n_h(x)$ induces a spatial modulation in the superconducting amplitude $|\psi(x)|$, as illustrated in Fig.~\ref{fig:model}(a). Near the superconducting transition temperature, amplitude fluctuations are typically suppressed compared to phase fluctuations. In this regime, the Ginzburg-Landau theory of Eq.~\eqref{eq:intertwinedgl} can be mapped onto a modulated XY model defined on a two-dimensional square lattice. The effective Hamiltonian, shown in Fig.~\ref{fig:model}(b), is given by
\be
H = -\sum_{\langle ij \rangle} J_{ij} \cos \left(\theta_{i} - \theta_{j}\right),
\label{eq:cdwxy}
\ee
where $J_{ij} = J_0 \left[1 + \alpha \sin(k x_i)\right]$ is the spatially modulated coupling along the $x$-direction with modulation amplitude $\alpha \in (0, 1)$ and period $\lambda = 2\pi/k$, with $x_i$ denoting the $x$-coordinate of the nearest-neighboring (NN) bonds $\langle ij\rangle$, and $\theta_i$'s are $U(1)$ spins associated to each lattice site. Since the influence of CDW is predominantly along the $x$-axis, we set $J_{ij}=J_0$ along the $y$-direction.

The spatial modulation of $J_{ij}$ generates stripe-like regions of alternating strong and weak coupling, as shown in Fig.~\ref{fig:model}(b). Vortices preferentially localize in the weakly coupled (valley) regions, where the energy cost for phase fluctuations is lower. In contrast, the strongly coupled regions promote quasi-long-range order (QLRO) by suppressing vortex proliferation.

For $\alpha = 0$, the model reduces to the standard 2D XY model, which undergoes a Berezinskii-Kosterlitz-Thouless (BKT) transition~\cite{Berezinsky1970,Kosterlitz1973}. In the opposite limit, $\alpha \to 1$, the system consists of decoupled one-dimensional stripes without finite-temperature phase transition. Thus, increasing $\alpha$ monotonically suppresses the critical temperature $T_c$ due to the enhanced anisotropy.

The dependence of $T_c$ on the modulation wave length $\lambda$ is more elaborate. One may expect that $T_c$ decreases with increasing $\lambda$, as the regions of weaker coupling expand. However, we find $T_c$ exhibits a non-monotonic dependence on $\lambda$. As shown in Fig.~\ref{fig:model}(c), $T_c$ increases with $\lambda$ up to an optimal value and then decreases as $\lambda$ becomes larger. The $T_c$-domes decrease monotonically and the peak positions shift right with increasing $\alpha$.

The modulated XY model thus captures important aspects observed in real materials, such as high-temperature cuprate superconductors. Since the modulation wavelength is closely related to the doping density~\cite{Baldelli2025}, the dome-like $T_c(\lambda)$ structure bears resemblance to the experimentally observed SC-CDW coexisting dome as a function of doping~\cite{Keimer2015,Xu2024}.

\paragraph{Critical properties---}
\begin{figure}[tbp]
    \centering
    \includegraphics[width=0.99\linewidth]{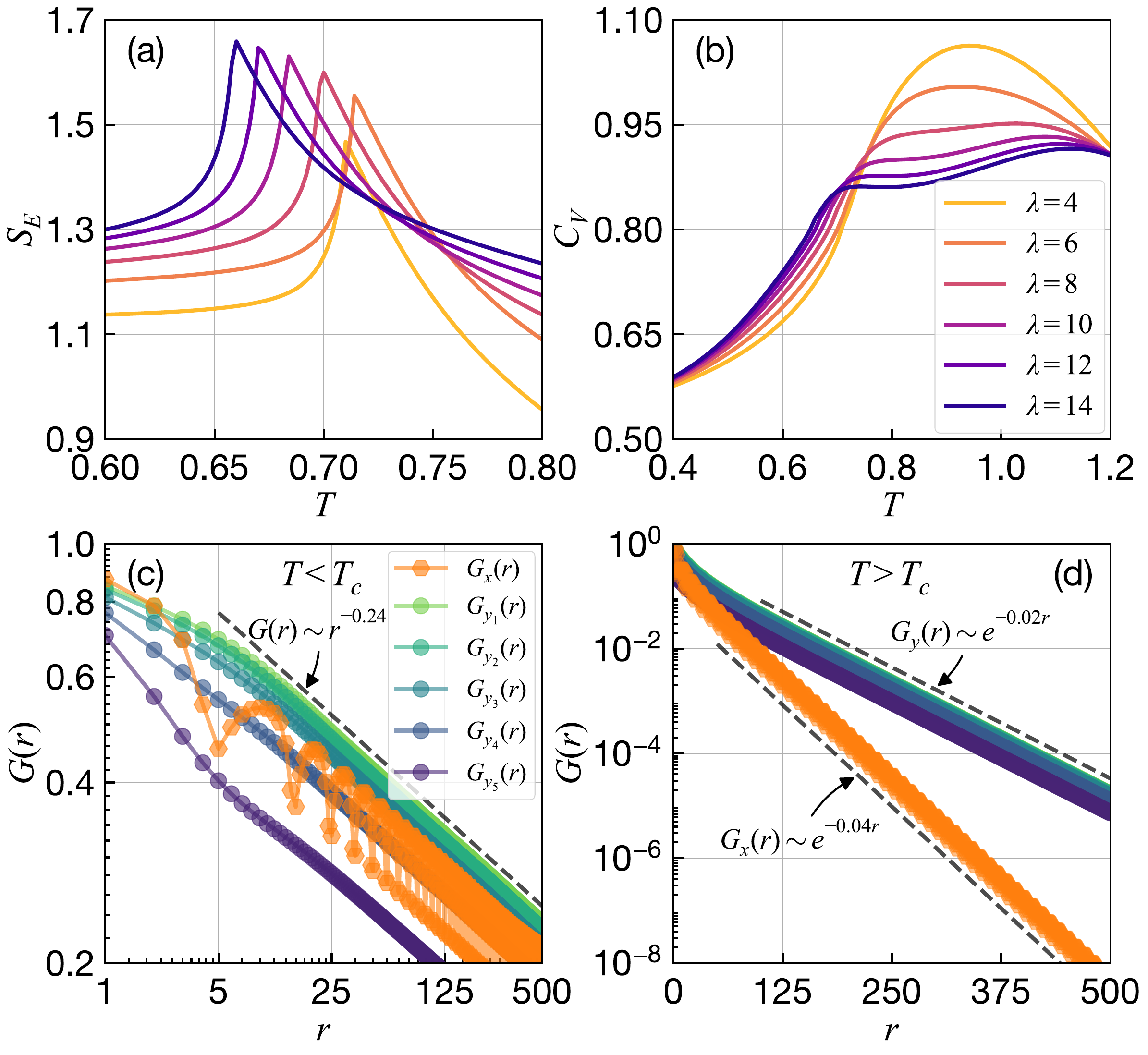}
    \caption{
    (a) The entanglement entropy for $\alpha=0.8$ as function of temperature under different modulation wave length $\lambda$.
    (b) The peaks of specific heat for $\alpha=0.8$ become broader with increasing $\lambda$.
    (c) The correlation functions show power-law decay $G(r)\sim r^{-\eta}$ at $T=0.65<T_c$ with $\alpha=0.8$, $\lambda=10$ and $T_c=0.684$, where the decay exponents $\eta$ are the same in both $x$ and $y$ directions.  
    (d) The correlation functions exhibit exponential decay $G(r)\sim e^{-r/\xi}$ at $T = 0.75 > T_c$ with the same modulation parameters, where correlations decay more rapidly along the $x$-direction than along $y$-direction. 
    $G_x(r)$ and $G_{y(i)}(r)$ denote the correlation functions along the horizontal direction and different vertical stripes, respectively.}
    \label{fig:eecv}
\end{figure}

We employ the tensor network methods to determine the critical temperature and study the critical properties in the thermodynamic limit. In the tensor network framework, the partition function is expressed as a contraction of local tensors on a square lattice, given by $Z=\mathrm{tTr}\prod_i O_{n_1,n_2}^{n_3,n_4}(i)$, with the local tensor $O$ obtained from a dual transformation~\cite{Yu2014,Vanderstraeten2019,Song2021}. For an infinite system, the partition function is determined by the dominant eigenvalues of the row-to-row transfer operator $\mathcal{T}$ consisting of an infinite sequence of $O$ tensors. The eigen-equation
$
\mathcal{T}|\Psi(A)\rangle=\Lambda_{\max}\ket{\Psi(A)}
$
can be accurately solved by the VUMPS algorithm~\cite{Haegeman2017,Zauner-Stauber2018,Vanderstraeten2019tan,Nietner2020}, where $\ket{\Psi(A)}$ is the leading eigenvector represented by matrix product states (MPS) made up of periodic local $A$ tensors with auxilary bond dimension $D$. Moreover, the expectation value of local observables and correlation functions can be evaluated using impurity tensors~\cite{Song2021,Song2022,Song2022ffxy,Song2023,Morita2019,Morita2025}. The details are given in the Supplemental Material~\cite{SM}.

Despite the lack of a local order parameter for the XY model, the entanglement entropy $S_E$ of the fixed-point MPS $\ket{\Psi(A)}$ provides an efficient measure to determine phase transitions~\cite{Vidal2003,Pollmann2009}. As shown in Fig.~\ref{fig:eecv}(a), $S_E$ displays pronounced peaks at the critical temperatures for various modulation wavelengths $\lambda$ with modulation amplitude $\alpha=0.8$. Notably, the peak positions of $S_E$ shift to lower temperatures for $\lambda > 6$, revealing the dome-like structure of $T_c$ in Fig.~\ref{fig:model}(c). The corresponding specific heat $C_V$ is displayed in Fig.~\ref{fig:eecv}(b), exhibiting rounded bumps. The peak of $C_V$ at higher temperatures than $T_c$ is a typical feature of BKT transition due to the progressive unbinding of vortex pairs as temperature increases~\cite{Solla1981,Kosterlitz2016}. The broadening of specific heat peaks with increasing $\lambda$ reflects an extended vortex unbinding process, as the enlarged valleys of weaker coupling allow vortex pairs to unbind over a broader temperature range.

Further insight into the BKT transition is provided by the correlation functions. Due to the modulation, we consider different correlation functions along the $y$-direction, $G_{y{(i)}}(r) = \langle \cos(\theta_{y(i)} - \theta_{y(i) + r}) \rangle$, with vertical stripe indices $i = 1, \ldots, \lambda$. Due to reflection symmetry along the $x$-direction, the correlation functions satisfy $G_{y(i)}(r) = G_{y({\lambda-i)}}(r)$. As shown in Fig.~\ref{fig:eecv}(c), at $T=0.65<T_c$ and $\alpha=0.8$, all correlation functions decay algebraically as $G(r)\sim r^{-\eta}$ with the same critical exponent $\eta\approx0.24$ close to the BKT exponent $1/4$, despite the modulation in $G_x(r)$ with period $\lambda=10$ along the $x$-direction. This is a key feature of the low-temperature critical phase of the XY model with QLRO. Within this phase, the central charge $c = 1$ is confirmed via logarithmic scaling of entanglement entropy with correlation length $S_E \propto (c/6)\ln\xi$~\cite{Tagliacozzo2008,Pollmann2009,Pirvu2012,SM}. In contrast, as the temperature increases above $T_c$, the correlation functions turn into exponential decay, $G(r)\sim e^{-r/\xi}$, as shown in Fig.~\ref{fig:eecv}(d) at $T=0.75$. This behavior indicates the loss of phase coherence, characteristic of the disordered phase. Notably, correlation functions $G_x(r)$ along the modulation direction decay more rapidly than $G_y(r)$, reflecting the anisotropic impact of the modulation.

\paragraph{Modulated clock model---}
\begin{figure}[tbp]
    \centering
    \includegraphics[width=0.99\linewidth]{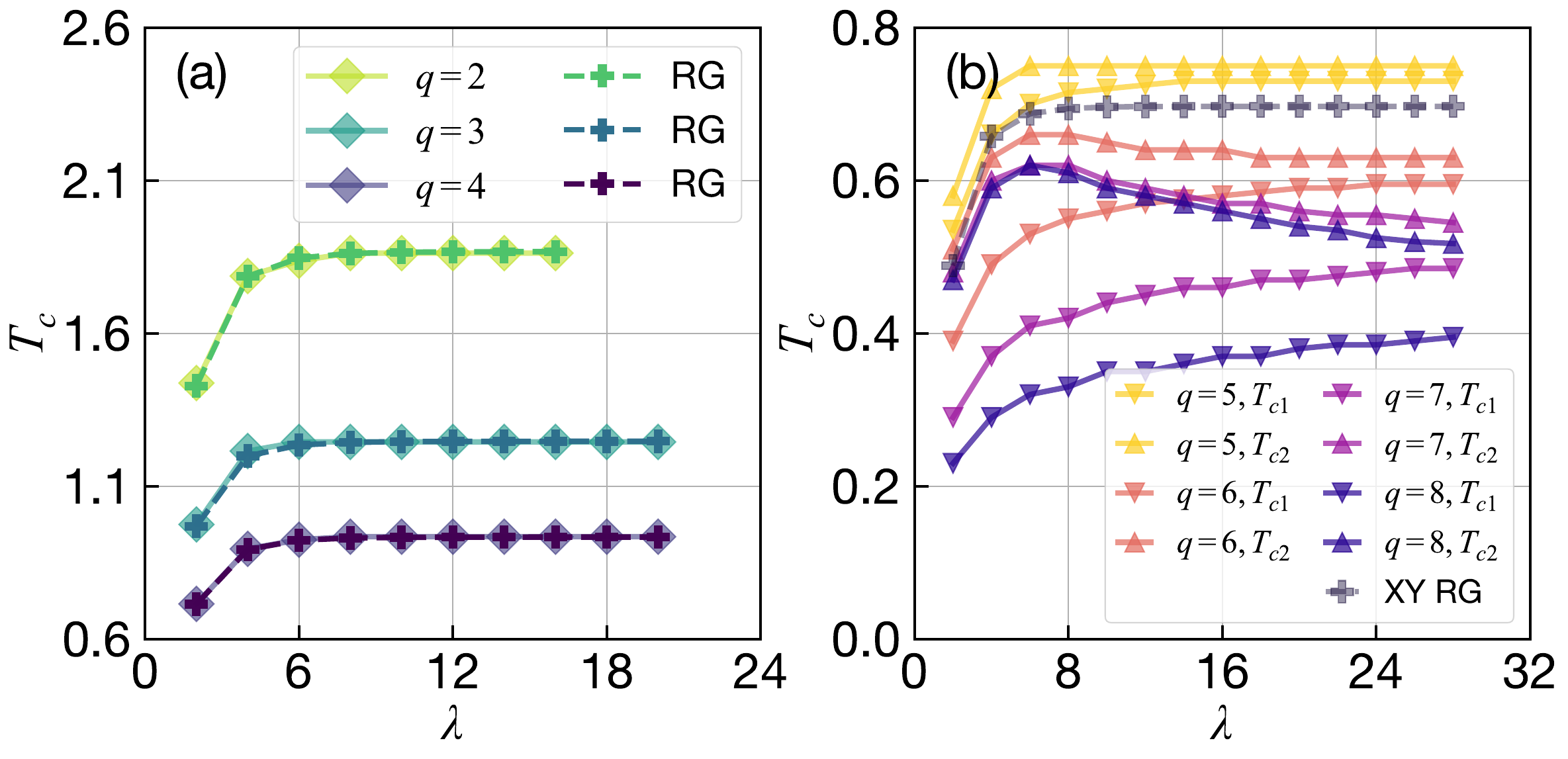}
    \caption{
    (a) The modulated clock model exhibits a single second-order phase transition for $q = 2, 3, 4$, with the critical temperature $T_c$ increasing monotonically with the modulation wavelength $\lambda$ under modulation amplitude $\alpha=0.9$. Dash lines with filled plus symbols are the $T_c$ evaluated from RG with duality equations. 
    (b) For $q \ge 5$, the system undergoes two BKT-like transitions. The lower critical temperature $T_{c1}$ increases monotonically while the higher critical temperature $T_{c2}$ exhibits a dome-shaped dependence on $\lambda$. Gray dashed line is the estimation for the XY limit.
    }
    \label{fig:clock}
\end{figure}

To elucidate the non-monotonic dependence of the critical temperature $T_c$ on modulation, we investigate the $q$-state clock model, where spin angles are discretized as $\theta_i = 2\pi l_i / q$ with $l_i = 1, \ldots, q$. For $q = 2, 3, 4$, the model exhibits a single second-order phase transition driven by domain wall excitations. For $q \ge 5$, two BKT-like transitions occur: a lower transition at $T_{c1}$ driven by domain wall excitations and a higher transition at $T_{c2}$ driven by vortex proliferation~\cite{Jose1977,Elitzur1979,Einhorn1980,Lapilli2006,Ortiz2012,Li2020}. These critical temperatures satisfy a duality relation $T_{c1} T_{c2} = T_{sd}^2$, where $T_{sd}$ is the self-dual temperature~\cite{Kramers1941,Li2020}. As $q$ increases, the macroscopic thermal averages become identical to those of the XY model ($q=\infty$) characterized by an emergent $U(1)$ symmetry~\cite{Lapilli2006}. Thus, the $q$-state clock model provides an ideal playground for understanding the nature of phase transitions driven by topological excitations.

By introducing the same spatial modulation $J_{ij}$ as in the modulated XY model, we obtain the modulated $q$-state clock model. The phase transitions are determined from the singularities of the entanglement entropy of the modulated model and its dual representation. As shown in Fig.~\ref{fig:clock}(a), for $q = 2, 3, 4$, there remains a single second-order transition, with the critical temperature $T_c$ increasing monotonically and saturating as the modulation period $\lambda$ increases. However, as displayed in Fig.~\ref{fig:clock}(b), for $q \ge 5$, the duality between $T_{c1}$ and $T_{c2}$ is broken: $T_{c1}$ increases with $\lambda$, while $T_{c2}$ exhibits a non-monotonic, dome-like dependence on $\lambda$. Both $T_{c1}$ and $T_{c2}$ decrease with larger modulation amplitude $\alpha$. For fixed $\alpha$ and $\lambda$, increasing $q$ leads to a reduction in both $T_{c1}$ and $T_{c2}$, similarly to the original $q$-state clock model. In the large-$q$ limit, $T_{c1}$ drops to zero while $T_{c2}$ gradually approaches the critical temperature $T_c$ of the modulated XY model. This indicates that vortex excitations are crucial in driving the dome-shaped structure of the critical temperatures.

\paragraph{Renormalization group analysis---}
\begin{figure}
    \centering
    \includegraphics[width=0.99\linewidth]{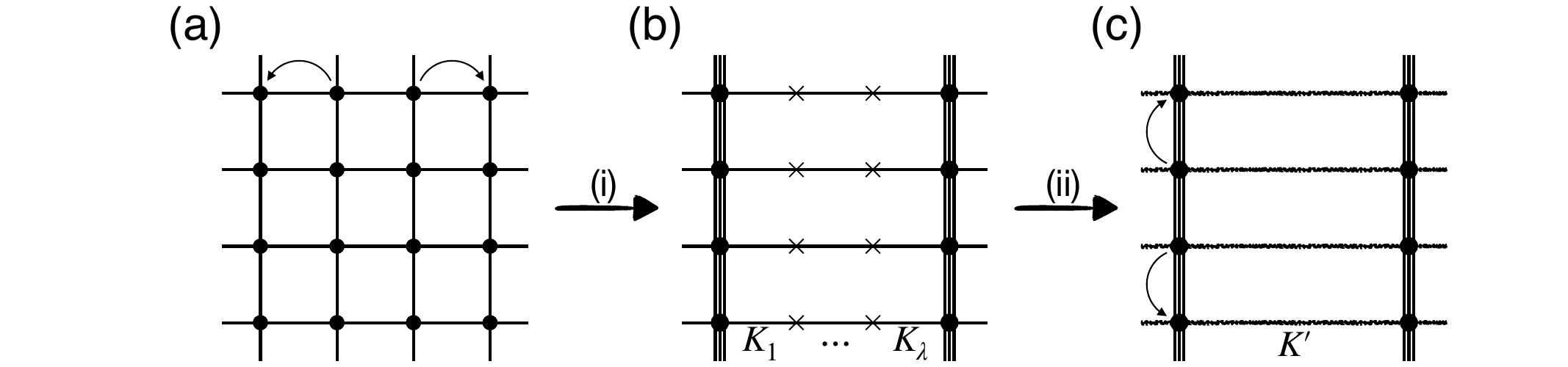}
    \caption{Half of the Migdal-Kadanoff transformation in 2D along horizontal direction via bond bunching followed by decimation. The spine indicated by ``$\times$'' are traced out by 1D decimation. The vertical transformation is obtained by a 90° rotation of the lattice and repeating the procedure.}
    \label{fig:rg}
\end{figure}

The dependence of $T_c$ on the modulation wavelength can be captured by the Migdal-Kadanoff (MK) renormalization group (RG) approach~\cite{Migdal1975,Kadanoff1976}. As illustrated in Fig.~\ref{fig:rg}, the RG transformation for a $d$-dimensional lattice is approximated by first combining $b$ parallel bonds (bond-bunching), then tracing out intermediate spins (decimation). These steps are applied alternately along horizontal and vertical directions to establish the recursion relations.

Despite its simplicity, the MK-RG approximation has notable limitations. In particular, it breaks certain lattice symmetries, and further improvement could be made by taking the limit $b \to 1$, which restores symmetry in the RG procedure~\cite{Jose1977}. However, for systems with spatial modulation, this limit is not straightforward to implement, since tracing over only a single bond does not easily yield a uniform lattice. 

To address the limitations of the MK-RG method, we adopt a hybrid approach that combines a single MK-RG step with duality equations~\cite{SM}. We first use MK-RG to map the modulated system onto an anisotropic model by tracing over $\lambda$ horizontal bonds $K_n$, where $K_n \equiv J_{ij}/T$, as shown in step (ii). For the clock model, this corresponds to forming a matrix product $W' = \prod_{n=1}^{\lambda} W_n$, with $W_n(\theta_i, \theta_j) = \exp(K_n \cos(\theta_i - \theta_j))$. The MK approach assumes the effective interaction retains the original form, so we seek an effective coupling $K'$ such that $W'(\theta_i, \theta_j) \propto \exp(K' \cos(\theta_i - \theta_j))$. With the resulting anisotropic model, we apply Kramers-Wannier duality~\cite{Kramers1941} to determine $T_c$. For $q=2$, the duality equation is $\sinh 2K_x \sinh 2K_y = 1$, which gives the exact transition temperature. For $q \ge 5$, there is no exact self-dual equation, so we simply estimate $T_c$ in the $q \to \infty$ (XY) limit using $T_c \approx T_{BKT}\sqrt{K_x/K_y}$, with $T_{BKT} = 0.893$ for the standard XY model~\cite{Vanderstraeten2019}. Although simpler than the self-consistent approach~\cite{Spivsak1993}, this estimate captures the same trend with anisotropy.

This hybrid approach greatly reduces uncontrolled RG approximations, as shown by the $T_c$ estimates in Fig.~\ref{fig:clock}. For $q = 2, 3, 4$, our results match tensor network calculations, but for larger $q$, $T_c$ is overestimated compared to $T_{c2}$. We  find the estimated $T_c$ decreases with increasing $\alpha$, reflecting the suppression of order by stronger modulation, while it increases monotonically with $\lambda$ due to the enhancement of the renormalized interaction. This behavior is expected for domain wall-driven transitions at $T_{c1}$, since the domain wall energy scale with the coupling strength ($E_{dw} \propto J$). However, vortex-related transitions are more subtle, as they depend not only on vortex energy but also on the modulation of vortex distributions. A more refined RG treatment that explicitly includes vortex effects will be needed to fully capture these transitions, which we leave for future work.

\paragraph{Modulated vortex distribution---}
\begin{figure}[tbp]
    \centering
    \includegraphics[width=0.99\linewidth]{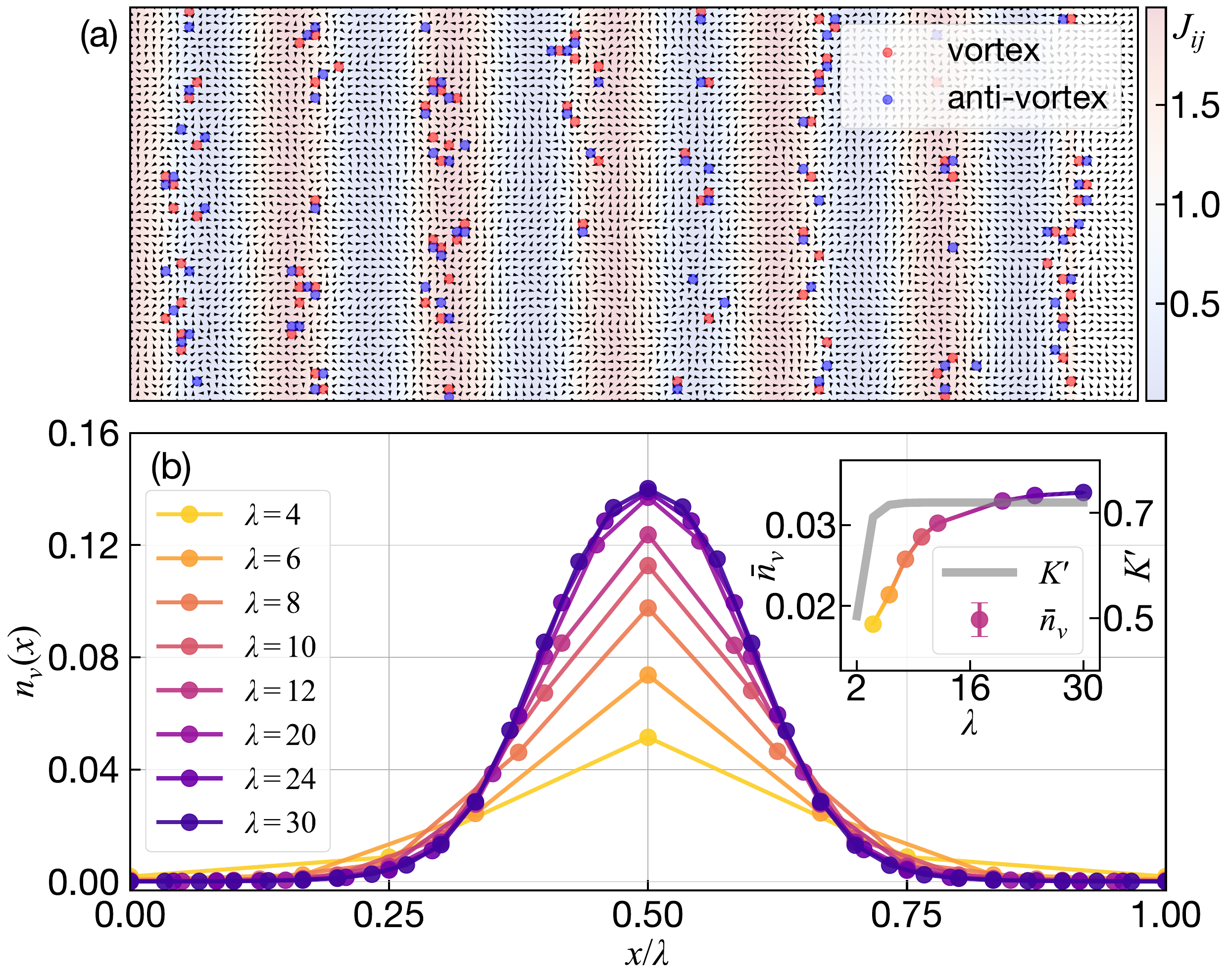}
    \caption{
    (a) Snapshot of the spin configuration and vortex distribution from MC simulations for system size $L=128$ at $T=0.8$, with modulation parameters $\alpha=0.8$ and $\lambda=16$. Background color indicates the spatial modulation of interaction strength $J_{ij}$. Vortices are predominantly localized in the valleys (blue regions).
    (b) Averaged vortex density along the modulation direction with $L=120$, folded onto a single period, shown for different modulation wavelengths at $T=0.8$ and $\alpha=0.8$. Inset: Total vortex density as a function of wavelength. The statistical error is smaller than the symbol size. The gray line indicates the renormalized interaction strength $K'$.
    }
    \label{fig:mc}
\end{figure}

To investigate vortex excitations in the modulated XY model, we perform Monte Carlo (MC) simulations using both the Metropolis and Wolff algorithms~\cite{Wolff1989,Gubernatis2016} under periodic boundary conditions. System sizes up to $N = L \times L$ with $L = 128$ are studied. Each simulation is equilibrated for $10^5$ MC steps, followed by $10^6$ steps for measuring equilibrium properties.

Typical vortex configurations from MC simulations at $T = 0.8$, slightly above $T_c$, are shown in Fig.~\ref{fig:mc}(a). A vortex (antivortex) with vorticity $v_p = \pm 1$ in a plaquette center is identified by a change of total angle of $\pm 2\pi$ when traversing the surrounding spins anticlockwise. The spatial distribution of the NN interaction strength $J_{ij}$ is indicated by the background color, with modulation amplitude $\alpha=0.8$ and period $\lambda=16$. It is evident that vortices tend to localize in regions where $J_{ij} < J_0$, i.e., the valleys of the modulation. These modulated vortex distributions are consistent with the ground-state solutions of the intertwined Ginzburg-Landau theory, where vortices are pinned between charge density stripes~\cite{Baldelli2025}.

Fig.~\ref{fig:mc}(b) shows how the average vortex density $n_v(x) = \frac{1}{L_y} \sum_y n_v(x, y)$ in each vertical stripe depends on the spatial modulation. The modulation periods $\lambda$ are chosen as factors of the system size $L = 120$, and the density profile is rescaled as $x/\lambda$ within one modulation period. As $\lambda$ increases, vortices accumulate more in the valleys until saturation. This enhanced anisotropy explains the directional dependence of the correlation functions above $T_c$. Vortex pairs unbind more readily along the $x$-direction than along the $y$-direction. This anisotropic unbinding process also gives rise to a broadened specific heat peak.

Based on these observations, we can understand the dome-shaped dependence of $T_c$ from a thermodynamic perspective. The BKT transition results from a competition between vortex energy and entropy, with the transition temperature set by $T_c = E/S$, where $F = E - TS$ is the free energy. For a fixed modulation amplitude $\alpha$, consider gradually increasing the modulation period $\lambda$. When $\lambda$ is small, the vortex energy $E$ rises rapidly due to the stronger effective interaction $K'$, as shown in the inset of Fig.~\ref{fig:mc}(b), leading to an initial increase in $T_c$. As $\lambda$ increases further, vortex excitations become easier, enhancing the entropy $S$, as reflected by the total vortex density $\bar{n}_v$ in the inset, which then dominates and causes $T_c$ to decrease. For large $\lambda$, both energy and entropy contributions saturate, resulting in a stable $T_c$. Notably, increasing $\alpha$ shifts the saturation of the effective interaction $K'$ to larger $\lambda$, which in turn moves the $T_c$ peak to the right as shown in Fig~\ref{fig:model}(c).

\paragraph{Conclusion and outlook---}
Motivated by the interplay between charge density waves and superconductivity, we investigated a two-dimensional XY model with spatially modulated coupling strength, capturing essential features of these intertwined orders in strongly correlated systems~\cite{Baldelli2025}. Using tensor network methods and Monte Carlo simulations, we uncovered an emergent, non-monotonic ``superfluid dome'' in the critical temperature $T_c$ as a function of the modulation wavelength $\lambda$, with peak positions shifting to larger $\lambda$ as the modulation amplitude $\alpha$ increases. This dome emerges from the effective pinning of vortices in the weakly coupled regions of the modulation, which is further demonstrated by comparisons with modulated $q$-state clock models, where vortex-driven transitions at $T_{c2}$ exhibit similar dome-shaped structures for $q \ge 5$. Our renormalization group analysis effectively describes the domain wall-driven phase transitions but overestimates $T_c$ for XY-like systems, highlighting the critical role of modulated vortex distributions in shaping the dome structure, which could be understood through the interplay of vortex energy and entropy in the Berezinskii-Kosterlitz-Thouless (BKT) transition.

Our work suggests studying vortex pinning to charge density waves experimentally, e.g., using scanning tunnelling microscopy~\cite{Suderow2014,Ge2016}, which could shed further light on the nature of these intertwined orders in actual materials. From a theoretical perspective, amplitude fluctuations $|\psi|$ of the order parameter in Eq.~\eqref{eq:intertwinedgl} introduce further non-trivial effects such as the stabilization of soliton solutions~\cite{Baldelli2025}. A future study of their precise role could reveal further physical phenomena. While in Eq.~\eqref{eq:intertwinedgl}, we considered a static charge distribution $n_h(\bm{r})$, an interesting future direction is the study of additional dynamical charge degrees of freedom allowing to understand the role of superconducting pairing in relation to charge orders, such as fluctuating stripes~\cite{Tranquada1995,Zheng2017,Huang2017,Xu2024} or phase-separated regimes~\cite{Pan2001,McElroy2005,Campi2015,Sinha2024}.


\paragraph{Acknowledgments---}
We would like to thank Anders Sandvik, Jun Takahashi and Misawa Takahiro for helpful discussions.
This work is supported by JSPS KAKENHI (Grant No.\ 23K25789). 
FF.S. acknowledges support by  JSPS Grant-in-Aid for Early-Career Scientists (Grant No.\ 25K17311).
A.W.\ acknowledges support by the DFG through the Emmy Noether program (Grant No.\ 509755282). 
Part of the computation in this work has been done using the facilities of the Supercomputer Center, the Institute for Solid State Physics, the University of Tokyo.



\bibliographystyle{apsrev4-2}
\bibliography{ref}

\begin{thebibliography}{80}%
\makeatletter
\providecommand \@ifxundefined [1]{%
 \@ifx{#1\undefined}
}%
\providecommand \@ifnum [1]{%
 \ifnum #1\expandafter \@firstoftwo
 \else \expandafter \@secondoftwo
 \fi
}%
\providecommand \@ifx [1]{%
 \ifx #1\expandafter \@firstoftwo
 \else \expandafter \@secondoftwo
 \fi
}%
\providecommand \natexlab [1]{#1}%
\providecommand \enquote  [1]{``#1''}%
\providecommand \bibnamefont  [1]{#1}%
\providecommand \bibfnamefont [1]{#1}%
\providecommand \citenamefont [1]{#1}%
\providecommand \href@noop [0]{\@secondoftwo}%
\providecommand \href [0]{\begingroup \@sanitize@url \@href}%
\providecommand \@href[1]{\@@startlink{#1}\@@href}%
\providecommand \@@href[1]{\endgroup#1\@@endlink}%
\providecommand \@sanitize@url [0]{\catcode `\\12\catcode `\$12\catcode
  `\&12\catcode `\#12\catcode `\^12\catcode `\_12\catcode `\%12\relax}%
\providecommand \@@startlink[1]{}%
\providecommand \@@endlink[0]{}%
\providecommand \url  [0]{\begingroup\@sanitize@url \@url }%
\providecommand \@url [1]{\endgroup\@href {#1}{\urlprefix }}%
\providecommand \urlprefix  [0]{URL }%
\providecommand \Eprint [0]{\href }%
\providecommand \doibase [0]{https://doi.org/}%
\providecommand \selectlanguage [0]{\@gobble}%
\providecommand \bibinfo  [0]{\@secondoftwo}%
\providecommand \bibfield  [0]{\@secondoftwo}%
\providecommand \translation [1]{[#1]}%
\providecommand \BibitemOpen [0]{}%
\providecommand \bibitemStop [0]{}%
\providecommand \bibitemNoStop [0]{.\EOS\space}%
\providecommand \EOS [0]{\spacefactor3000\relax}%
\providecommand \BibitemShut  [1]{\csname bibitem#1\endcsname}%
\let\auto@bib@innerbib\@empty
\bibitem [{\citenamefont {Wilson}\ \emph {et~al.}(1975)\citenamefont {Wilson},
  \citenamefont {Di~Salvo},\ and\ \citenamefont {Mahajan}}]{Wilson1975}%
  \BibitemOpen
  \bibfield  {author} {\bibinfo {author} {\bibfnamefont {J.}~\bibnamefont
  {Wilson}}, \bibinfo {author} {\bibfnamefont {F.}~\bibnamefont {Di~Salvo}},\
  and\ \bibinfo {author} {\bibfnamefont {S.}~\bibnamefont {Mahajan}},\ }\href
  {https://doi.org/10.1080/00018737500101391} {\bibfield  {journal} {\bibinfo
  {journal} {Adv. Phys.}\ }\textbf {\bibinfo {volume} {24}},\ \bibinfo {pages}
  {117–201} (\bibinfo {year} {1975})}\BibitemShut {NoStop}%
\bibitem [{\citenamefont {Moncton}\ \emph {et~al.}(1975)\citenamefont
  {Moncton}, \citenamefont {Axe},\ and\ \citenamefont {DiSalvo}}]{Moncton1975}%
  \BibitemOpen
  \bibfield  {author} {\bibinfo {author} {\bibfnamefont {D.~E.}\ \bibnamefont
  {Moncton}}, \bibinfo {author} {\bibfnamefont {J.~D.}\ \bibnamefont {Axe}},\
  and\ \bibinfo {author} {\bibfnamefont {F.~J.}\ \bibnamefont {DiSalvo}},\
  }\href {https://doi.org/10.1103/PhysRevLett.34.734} {\bibfield  {journal}
  {\bibinfo  {journal} {Phys. Rev. Lett.}\ }\textbf {\bibinfo {volume} {34}},\
  \bibinfo {pages} {734} (\bibinfo {year} {1975})}\BibitemShut {NoStop}%
\bibitem [{\citenamefont {Yokoya}\ \emph {et~al.}(2001)\citenamefont {Yokoya},
  \citenamefont {Kiss}, \citenamefont {Chainani}, \citenamefont {Shin},
  \citenamefont {Nohara},\ and\ \citenamefont {Takagi}}]{Yokoya2001}%
  \BibitemOpen
  \bibfield  {author} {\bibinfo {author} {\bibfnamefont {T.}~\bibnamefont
  {Yokoya}}, \bibinfo {author} {\bibfnamefont {T.}~\bibnamefont {Kiss}},
  \bibinfo {author} {\bibfnamefont {A.}~\bibnamefont {Chainani}}, \bibinfo
  {author} {\bibfnamefont {S.}~\bibnamefont {Shin}}, \bibinfo {author}
  {\bibfnamefont {M.}~\bibnamefont {Nohara}},\ and\ \bibinfo {author}
  {\bibfnamefont {H.}~\bibnamefont {Takagi}},\ }\href
  {https://doi.org/10.1126/science.1065068} {\bibfield  {journal} {\bibinfo
  {journal} {Science}\ }\textbf {\bibinfo {volume} {294}},\ \bibinfo {pages}
  {2518–2520} (\bibinfo {year} {2001})}\BibitemShut {NoStop}%
\bibitem [{\citenamefont {Kusmartseva}\ \emph {et~al.}(2009)\citenamefont
  {Kusmartseva}, \citenamefont {Sipos}, \citenamefont {Berger}, \citenamefont
  {Forr\'o},\ and\ \citenamefont {Tuti\ifmmode~\check{s}\else
  \v{s}\fi{}}}]{Kusmartseva2009}%
  \BibitemOpen
  \bibfield  {author} {\bibinfo {author} {\bibfnamefont {A.~F.}\ \bibnamefont
  {Kusmartseva}}, \bibinfo {author} {\bibfnamefont {B.}~\bibnamefont {Sipos}},
  \bibinfo {author} {\bibfnamefont {H.}~\bibnamefont {Berger}}, \bibinfo
  {author} {\bibfnamefont {L.}~\bibnamefont {Forr\'o}},\ and\ \bibinfo {author}
  {\bibfnamefont {E.}~\bibnamefont {Tuti\ifmmode~\check{s}\else \v{s}\fi{}}},\
  }\href {https://doi.org/10.1103/PhysRevLett.103.236401} {\bibfield  {journal}
  {\bibinfo  {journal} {Phys. Rev. Lett.}\ }\textbf {\bibinfo {volume} {103}},\
  \bibinfo {pages} {236401} (\bibinfo {year} {2009})}\BibitemShut {NoStop}%
\bibitem [{\citenamefont {Arguello}\ \emph {et~al.}(2015)\citenamefont
  {Arguello}, \citenamefont {Rosenthal}, \citenamefont {Andrade}, \citenamefont
  {Jin}, \citenamefont {Yeh}, \citenamefont {Zaki}, \citenamefont {Jia},
  \citenamefont {Cava}, \citenamefont {Fernandes}, \citenamefont {Millis},
  \citenamefont {Valla}, \citenamefont {Osgood},\ and\ \citenamefont
  {Pasupathy}}]{Arguello2015}%
  \BibitemOpen
  \bibfield  {author} {\bibinfo {author} {\bibfnamefont {C.~J.}\ \bibnamefont
  {Arguello}}, \bibinfo {author} {\bibfnamefont {E.~P.}\ \bibnamefont
  {Rosenthal}}, \bibinfo {author} {\bibfnamefont {E.~F.}\ \bibnamefont
  {Andrade}}, \bibinfo {author} {\bibfnamefont {W.}~\bibnamefont {Jin}},
  \bibinfo {author} {\bibfnamefont {P.~C.}\ \bibnamefont {Yeh}}, \bibinfo
  {author} {\bibfnamefont {N.}~\bibnamefont {Zaki}}, \bibinfo {author}
  {\bibfnamefont {S.}~\bibnamefont {Jia}}, \bibinfo {author} {\bibfnamefont
  {R.~J.}\ \bibnamefont {Cava}}, \bibinfo {author} {\bibfnamefont {R.~M.}\
  \bibnamefont {Fernandes}}, \bibinfo {author} {\bibfnamefont {A.~J.}\
  \bibnamefont {Millis}}, \bibinfo {author} {\bibfnamefont {T.}~\bibnamefont
  {Valla}}, \bibinfo {author} {\bibfnamefont {R.~M.}\ \bibnamefont {Osgood}},\
  and\ \bibinfo {author} {\bibfnamefont {A.~N.}\ \bibnamefont {Pasupathy}},\
  }\href {https://doi.org/10.1103/PhysRevLett.114.037001} {\bibfield  {journal}
  {\bibinfo  {journal} {Phys. Rev. Lett.}\ }\textbf {\bibinfo {volume} {114}},\
  \bibinfo {pages} {037001} (\bibinfo {year} {2015})}\BibitemShut {NoStop}%
\bibitem [{\citenamefont {Lian}\ \emph {et~al.}(2018)\citenamefont {Lian},
  \citenamefont {Si},\ and\ \citenamefont {Duan}}]{Lian2018}%
  \BibitemOpen
  \bibfield  {author} {\bibinfo {author} {\bibfnamefont {C.-S.}\ \bibnamefont
  {Lian}}, \bibinfo {author} {\bibfnamefont {C.}~\bibnamefont {Si}},\ and\
  \bibinfo {author} {\bibfnamefont {W.}~\bibnamefont {Duan}},\ }\href
  {https://doi.org/10.1021/acs.nanolett.8b00237} {\bibfield  {journal}
  {\bibinfo  {journal} {Nano Lett.}\ }\textbf {\bibinfo {volume} {18}},\
  \bibinfo {pages} {2924–2929} (\bibinfo {year} {2018})}\BibitemShut
  {NoStop}%
\bibitem [{\citenamefont {Tranquada}\ \emph {et~al.}(1995)\citenamefont
  {Tranquada}, \citenamefont {Sternlieb}, \citenamefont {Axe}, \citenamefont
  {Nakamura},\ and\ \citenamefont {Uchida}}]{Tranquada1995}%
  \BibitemOpen
  \bibfield  {author} {\bibinfo {author} {\bibfnamefont {J.~M.}\ \bibnamefont
  {Tranquada}}, \bibinfo {author} {\bibfnamefont {B.~J.}\ \bibnamefont
  {Sternlieb}}, \bibinfo {author} {\bibfnamefont {J.~D.}\ \bibnamefont {Axe}},
  \bibinfo {author} {\bibfnamefont {Y.}~\bibnamefont {Nakamura}},\ and\
  \bibinfo {author} {\bibfnamefont {S.}~\bibnamefont {Uchida}},\ }\href
  {https://doi.org/10.1038/375561a0} {\bibfield  {journal} {\bibinfo  {journal}
  {Nature}\ }\textbf {\bibinfo {volume} {375}},\ \bibinfo {pages} {561–563}
  (\bibinfo {year} {1995})}\BibitemShut {NoStop}%
\bibitem [{\citenamefont {Howald}\ \emph {et~al.}(2003)\citenamefont {Howald},
  \citenamefont {Eisaki}, \citenamefont {Kaneko},\ and\ \citenamefont
  {Kapitulnik}}]{Howald2003}%
  \BibitemOpen
  \bibfield  {author} {\bibinfo {author} {\bibfnamefont {C.}~\bibnamefont
  {Howald}}, \bibinfo {author} {\bibfnamefont {H.}~\bibnamefont {Eisaki}},
  \bibinfo {author} {\bibfnamefont {N.}~\bibnamefont {Kaneko}},\ and\ \bibinfo
  {author} {\bibfnamefont {A.}~\bibnamefont {Kapitulnik}},\ }\href
  {https://doi.org/10.1073/pnas.1233768100} {\bibfield  {journal} {\bibinfo
  {journal} {Proc. Natl. Acad. Sci. U.S.A.}\ }\textbf {\bibinfo {volume}
  {100}},\ \bibinfo {pages} {9705–9709} (\bibinfo {year} {2003})}\BibitemShut
  {NoStop}%
\bibitem [{\citenamefont {Ghiringhelli}\ \emph {et~al.}(2012)\citenamefont
  {Ghiringhelli}, \citenamefont {Le~Tacon}, \citenamefont {Minola},
  \citenamefont {Blanco-Canosa}, \citenamefont {Mazzoli}, \citenamefont
  {Brookes}, \citenamefont {De~Luca}, \citenamefont {Frano}, \citenamefont
  {Hawthorn}, \citenamefont {He}, \citenamefont {Loew}, \citenamefont {Sala},
  \citenamefont {Peets}, \citenamefont {Salluzzo}, \citenamefont {Schierle},
  \citenamefont {Sutarto}, \citenamefont {Sawatzky}, \citenamefont {Weschke},
  \citenamefont {Keimer},\ and\ \citenamefont {Braicovich}}]{Ghiringhelli2012}%
  \BibitemOpen
  \bibfield  {author} {\bibinfo {author} {\bibfnamefont {G.}~\bibnamefont
  {Ghiringhelli}}, \bibinfo {author} {\bibfnamefont {M.}~\bibnamefont
  {Le~Tacon}}, \bibinfo {author} {\bibfnamefont {M.}~\bibnamefont {Minola}},
  \bibinfo {author} {\bibfnamefont {S.}~\bibnamefont {Blanco-Canosa}}, \bibinfo
  {author} {\bibfnamefont {C.}~\bibnamefont {Mazzoli}}, \bibinfo {author}
  {\bibfnamefont {N.~B.}\ \bibnamefont {Brookes}}, \bibinfo {author}
  {\bibfnamefont {G.~M.}\ \bibnamefont {De~Luca}}, \bibinfo {author}
  {\bibfnamefont {A.}~\bibnamefont {Frano}}, \bibinfo {author} {\bibfnamefont
  {D.~G.}\ \bibnamefont {Hawthorn}}, \bibinfo {author} {\bibfnamefont
  {F.}~\bibnamefont {He}}, \bibinfo {author} {\bibfnamefont {T.}~\bibnamefont
  {Loew}}, \bibinfo {author} {\bibfnamefont {M.~M.}\ \bibnamefont {Sala}},
  \bibinfo {author} {\bibfnamefont {D.~C.}\ \bibnamefont {Peets}}, \bibinfo
  {author} {\bibfnamefont {M.}~\bibnamefont {Salluzzo}}, \bibinfo {author}
  {\bibfnamefont {E.}~\bibnamefont {Schierle}}, \bibinfo {author}
  {\bibfnamefont {R.}~\bibnamefont {Sutarto}}, \bibinfo {author} {\bibfnamefont
  {G.~A.}\ \bibnamefont {Sawatzky}}, \bibinfo {author} {\bibfnamefont
  {E.}~\bibnamefont {Weschke}}, \bibinfo {author} {\bibfnamefont
  {B.}~\bibnamefont {Keimer}},\ and\ \bibinfo {author} {\bibfnamefont
  {L.}~\bibnamefont {Braicovich}},\ }\href
  {https://doi.org/10.1126/science.1223532} {\bibfield  {journal} {\bibinfo
  {journal} {Science}\ }\textbf {\bibinfo {volume} {337}},\ \bibinfo {pages}
  {821–825} (\bibinfo {year} {2012})}\BibitemShut {NoStop}%
\bibitem [{\citenamefont {Comin}\ \emph {et~al.}(2014)\citenamefont {Comin},
  \citenamefont {Frano}, \citenamefont {Yee}, \citenamefont {Yoshida},
  \citenamefont {Eisaki}, \citenamefont {Schierle}, \citenamefont {Weschke},
  \citenamefont {Sutarto}, \citenamefont {He}, \citenamefont {Soumyanarayanan},
  \citenamefont {He}, \citenamefont {Le~Tacon}, \citenamefont {Elfimov},
  \citenamefont {Hoffman}, \citenamefont {Sawatzky}, \citenamefont {Keimer},\
  and\ \citenamefont {Damascelli}}]{Comin2014}%
  \BibitemOpen
  \bibfield  {author} {\bibinfo {author} {\bibfnamefont {R.}~\bibnamefont
  {Comin}}, \bibinfo {author} {\bibfnamefont {A.}~\bibnamefont {Frano}},
  \bibinfo {author} {\bibfnamefont {M.~M.}\ \bibnamefont {Yee}}, \bibinfo
  {author} {\bibfnamefont {Y.}~\bibnamefont {Yoshida}}, \bibinfo {author}
  {\bibfnamefont {H.}~\bibnamefont {Eisaki}}, \bibinfo {author} {\bibfnamefont
  {E.}~\bibnamefont {Schierle}}, \bibinfo {author} {\bibfnamefont
  {E.}~\bibnamefont {Weschke}}, \bibinfo {author} {\bibfnamefont
  {R.}~\bibnamefont {Sutarto}}, \bibinfo {author} {\bibfnamefont
  {F.}~\bibnamefont {He}}, \bibinfo {author} {\bibfnamefont {A.}~\bibnamefont
  {Soumyanarayanan}}, \bibinfo {author} {\bibfnamefont {Y.}~\bibnamefont {He}},
  \bibinfo {author} {\bibfnamefont {M.}~\bibnamefont {Le~Tacon}}, \bibinfo
  {author} {\bibfnamefont {I.~S.}\ \bibnamefont {Elfimov}}, \bibinfo {author}
  {\bibfnamefont {J.~E.}\ \bibnamefont {Hoffman}}, \bibinfo {author}
  {\bibfnamefont {G.~A.}\ \bibnamefont {Sawatzky}}, \bibinfo {author}
  {\bibfnamefont {B.}~\bibnamefont {Keimer}},\ and\ \bibinfo {author}
  {\bibfnamefont {A.}~\bibnamefont {Damascelli}},\ }\href
  {https://doi.org/10.1126/science.1242996} {\bibfield  {journal} {\bibinfo
  {journal} {Science}\ }\textbf {\bibinfo {volume} {343}},\ \bibinfo {pages}
  {390–392} (\bibinfo {year} {2014})}\BibitemShut {NoStop}%
\bibitem [{\citenamefont {Tabis}\ \emph {et~al.}(2014)\citenamefont {Tabis},
  \citenamefont {Li}, \citenamefont {Tacon}, \citenamefont {Braicovich},
  \citenamefont {Kreyssig}, \citenamefont {Minola}, \citenamefont {Dellea},
  \citenamefont {Weschke}, \citenamefont {Veit}, \citenamefont {Ramazanoglu},
  \citenamefont {Goldman}, \citenamefont {Schmitt}, \citenamefont
  {Ghiringhelli}, \citenamefont {Bari{\v s}i{\'c}}, \citenamefont {Chan},
  \citenamefont {Dorow}, \citenamefont {Yu}, \citenamefont {Zhao},
  \citenamefont {Keimer},\ and\ \citenamefont {Greven}}]{Tabis2014}%
  \BibitemOpen
  \bibfield  {author} {\bibinfo {author} {\bibfnamefont {W.}~\bibnamefont
  {Tabis}}, \bibinfo {author} {\bibfnamefont {Y.}~\bibnamefont {Li}}, \bibinfo
  {author} {\bibfnamefont {M.~L.}\ \bibnamefont {Tacon}}, \bibinfo {author}
  {\bibfnamefont {L.}~\bibnamefont {Braicovich}}, \bibinfo {author}
  {\bibfnamefont {A.}~\bibnamefont {Kreyssig}}, \bibinfo {author}
  {\bibfnamefont {M.}~\bibnamefont {Minola}}, \bibinfo {author} {\bibfnamefont
  {G.}~\bibnamefont {Dellea}}, \bibinfo {author} {\bibfnamefont
  {E.}~\bibnamefont {Weschke}}, \bibinfo {author} {\bibfnamefont {M.~J.}\
  \bibnamefont {Veit}}, \bibinfo {author} {\bibfnamefont {M.}~\bibnamefont
  {Ramazanoglu}}, \bibinfo {author} {\bibfnamefont {A.~I.}\ \bibnamefont
  {Goldman}}, \bibinfo {author} {\bibfnamefont {T.}~\bibnamefont {Schmitt}},
  \bibinfo {author} {\bibfnamefont {G.}~\bibnamefont {Ghiringhelli}}, \bibinfo
  {author} {\bibfnamefont {N.}~\bibnamefont {Bari{\v s}i{\'c}}}, \bibinfo
  {author} {\bibfnamefont {M.~K.}\ \bibnamefont {Chan}}, \bibinfo {author}
  {\bibfnamefont {C.~J.}\ \bibnamefont {Dorow}}, \bibinfo {author}
  {\bibfnamefont {G.}~\bibnamefont {Yu}}, \bibinfo {author} {\bibfnamefont
  {X.}~\bibnamefont {Zhao}}, \bibinfo {author} {\bibfnamefont {B.}~\bibnamefont
  {Keimer}},\ and\ \bibinfo {author} {\bibfnamefont {M.}~\bibnamefont
  {Greven}},\ }\href {https://doi.org/10.1038/ncomms6875} {\bibfield  {journal}
  {\bibinfo  {journal} {Nat. Commun.}\ }\textbf {\bibinfo {volume} {5}},\
  \bibinfo {pages} {5875} (\bibinfo {year} {2014})}\BibitemShut {NoStop}%
\bibitem [{\citenamefont {Miao}\ \emph {et~al.}(2021)\citenamefont {Miao},
  \citenamefont {Fabbris}, \citenamefont {Koch}, \citenamefont {Mazzone},
  \citenamefont {Nelson}, \citenamefont {Acevedo-Esteves}, \citenamefont {Gu},
  \citenamefont {Li}, \citenamefont {Yilimaz}, \citenamefont {Kaznatcheev},
  \citenamefont {Vescovo}, \citenamefont {Oda}, \citenamefont {Kurosawa},
  \citenamefont {Momono}, \citenamefont {Assefa}, \citenamefont {Robinson},
  \citenamefont {Bozin}, \citenamefont {Tranquada}, \citenamefont {Johnson},\
  and\ \citenamefont {Dean}}]{Miao2021}%
  \BibitemOpen
  \bibfield  {author} {\bibinfo {author} {\bibfnamefont {H.}~\bibnamefont
  {Miao}}, \bibinfo {author} {\bibfnamefont {G.}~\bibnamefont {Fabbris}},
  \bibinfo {author} {\bibfnamefont {R.~J.}\ \bibnamefont {Koch}}, \bibinfo
  {author} {\bibfnamefont {D.~G.}\ \bibnamefont {Mazzone}}, \bibinfo {author}
  {\bibfnamefont {C.~S.}\ \bibnamefont {Nelson}}, \bibinfo {author}
  {\bibfnamefont {R.}~\bibnamefont {Acevedo-Esteves}}, \bibinfo {author}
  {\bibfnamefont {G.~D.}\ \bibnamefont {Gu}}, \bibinfo {author} {\bibfnamefont
  {Y.}~\bibnamefont {Li}}, \bibinfo {author} {\bibfnamefont {T.}~\bibnamefont
  {Yilimaz}}, \bibinfo {author} {\bibfnamefont {K.}~\bibnamefont
  {Kaznatcheev}}, \bibinfo {author} {\bibfnamefont {E.}~\bibnamefont
  {Vescovo}}, \bibinfo {author} {\bibfnamefont {M.}~\bibnamefont {Oda}},
  \bibinfo {author} {\bibfnamefont {T.}~\bibnamefont {Kurosawa}}, \bibinfo
  {author} {\bibfnamefont {N.}~\bibnamefont {Momono}}, \bibinfo {author}
  {\bibfnamefont {T.}~\bibnamefont {Assefa}}, \bibinfo {author} {\bibfnamefont
  {I.~K.}\ \bibnamefont {Robinson}}, \bibinfo {author} {\bibfnamefont {E.~S.}\
  \bibnamefont {Bozin}}, \bibinfo {author} {\bibfnamefont {J.~M.}\ \bibnamefont
  {Tranquada}}, \bibinfo {author} {\bibfnamefont {P.~D.}\ \bibnamefont
  {Johnson}},\ and\ \bibinfo {author} {\bibfnamefont {M.~P.~M.}\ \bibnamefont
  {Dean}},\ }\href {https://doi.org/10.1038/s41535-021-00327-4} {\bibfield
  {journal} {\bibinfo  {journal} {npj Quantum Mater.}\ }\textbf {\bibinfo
  {volume} {6}},\ \bibinfo {pages} {31} (\bibinfo {year} {2021})}\BibitemShut
  {NoStop}%
\bibitem [{\citenamefont {Hu}\ \emph {et~al.}(2025)\citenamefont {Hu},
  \citenamefont {Zheng}, \citenamefont {Xu}, \citenamefont {Deng},
  \citenamefont {Liang}, \citenamefont {Yang}, \citenamefont {Wang},
  \citenamefont {Grinenko}, \citenamefont {Lv}, \citenamefont {Ding},\ and\
  \citenamefont {Yim}}]{Hu2025}%
  \BibitemOpen
  \bibfield  {author} {\bibinfo {author} {\bibfnamefont {Q.}~\bibnamefont
  {Hu}}, \bibinfo {author} {\bibfnamefont {Y.}~\bibnamefont {Zheng}}, \bibinfo
  {author} {\bibfnamefont {H.}~\bibnamefont {Xu}}, \bibinfo {author}
  {\bibfnamefont {J.}~\bibnamefont {Deng}}, \bibinfo {author} {\bibfnamefont
  {C.}~\bibnamefont {Liang}}, \bibinfo {author} {\bibfnamefont
  {F.}~\bibnamefont {Yang}}, \bibinfo {author} {\bibfnamefont {Z.}~\bibnamefont
  {Wang}}, \bibinfo {author} {\bibfnamefont {V.}~\bibnamefont {Grinenko}},
  \bibinfo {author} {\bibfnamefont {B.}~\bibnamefont {Lv}}, \bibinfo {author}
  {\bibfnamefont {H.}~\bibnamefont {Ding}},\ and\ \bibinfo {author}
  {\bibfnamefont {C.~M.}\ \bibnamefont {Yim}},\ }\href
  {https://doi.org/10.1038/s41467-024-55368-7} {\bibfield  {journal} {\bibinfo
  {journal} {Nat. Commun.}\ }\textbf {\bibinfo {volume} {16}},\ \bibinfo
  {pages} {253} (\bibinfo {year} {2025})}\BibitemShut {NoStop}%
\bibitem [{\citenamefont {Ortiz}\ \emph {et~al.}(2019)\citenamefont {Ortiz},
  \citenamefont {Gomes}, \citenamefont {Morey}, \citenamefont {Winiarski},
  \citenamefont {Bordelon}, \citenamefont {Mangum}, \citenamefont {Oswald},
  \citenamefont {Rodriguez-Rivera}, \citenamefont {Neilson}, \citenamefont
  {Wilson}, \citenamefont {Ertekin}, \citenamefont {McQueen},\ and\
  \citenamefont {Toberer}}]{Ortiz2019}%
  \BibitemOpen
  \bibfield  {author} {\bibinfo {author} {\bibfnamefont {B.~R.}\ \bibnamefont
  {Ortiz}}, \bibinfo {author} {\bibfnamefont {L.~C.}\ \bibnamefont {Gomes}},
  \bibinfo {author} {\bibfnamefont {J.~R.}\ \bibnamefont {Morey}}, \bibinfo
  {author} {\bibfnamefont {M.}~\bibnamefont {Winiarski}}, \bibinfo {author}
  {\bibfnamefont {M.}~\bibnamefont {Bordelon}}, \bibinfo {author}
  {\bibfnamefont {J.~S.}\ \bibnamefont {Mangum}}, \bibinfo {author}
  {\bibfnamefont {I.~W.~H.}\ \bibnamefont {Oswald}}, \bibinfo {author}
  {\bibfnamefont {J.~A.}\ \bibnamefont {Rodriguez-Rivera}}, \bibinfo {author}
  {\bibfnamefont {J.~R.}\ \bibnamefont {Neilson}}, \bibinfo {author}
  {\bibfnamefont {S.~D.}\ \bibnamefont {Wilson}}, \bibinfo {author}
  {\bibfnamefont {E.}~\bibnamefont {Ertekin}}, \bibinfo {author} {\bibfnamefont
  {T.~M.}\ \bibnamefont {McQueen}},\ and\ \bibinfo {author} {\bibfnamefont
  {E.~S.}\ \bibnamefont {Toberer}},\ }\href
  {https://doi.org/10.1103/PhysRevMaterials.3.094407} {\bibfield  {journal}
  {\bibinfo  {journal} {Phys. Rev. Mater.}\ }\textbf {\bibinfo {volume} {3}},\
  \bibinfo {pages} {094407} (\bibinfo {year} {2019})}\BibitemShut {NoStop}%
\bibitem [{\citenamefont {Ortiz}\ \emph {et~al.}(2020)\citenamefont {Ortiz},
  \citenamefont {Teicher}, \citenamefont {Hu}, \citenamefont {Zuo},
  \citenamefont {Sarte}, \citenamefont {Schueller}, \citenamefont {Abeykoon},
  \citenamefont {Krogstad}, \citenamefont {Rosenkranz}, \citenamefont {Osborn},
  \citenamefont {Seshadri}, \citenamefont {Balents}, \citenamefont {He},\ and\
  \citenamefont {Wilson}}]{Ortiz2020}%
  \BibitemOpen
  \bibfield  {author} {\bibinfo {author} {\bibfnamefont {B.~R.}\ \bibnamefont
  {Ortiz}}, \bibinfo {author} {\bibfnamefont {S.~M.~L.}\ \bibnamefont
  {Teicher}}, \bibinfo {author} {\bibfnamefont {Y.}~\bibnamefont {Hu}},
  \bibinfo {author} {\bibfnamefont {J.~L.}\ \bibnamefont {Zuo}}, \bibinfo
  {author} {\bibfnamefont {P.~M.}\ \bibnamefont {Sarte}}, \bibinfo {author}
  {\bibfnamefont {E.~C.}\ \bibnamefont {Schueller}}, \bibinfo {author}
  {\bibfnamefont {A.~M.~M.}\ \bibnamefont {Abeykoon}}, \bibinfo {author}
  {\bibfnamefont {M.~J.}\ \bibnamefont {Krogstad}}, \bibinfo {author}
  {\bibfnamefont {S.}~\bibnamefont {Rosenkranz}}, \bibinfo {author}
  {\bibfnamefont {R.}~\bibnamefont {Osborn}}, \bibinfo {author} {\bibfnamefont
  {R.}~\bibnamefont {Seshadri}}, \bibinfo {author} {\bibfnamefont
  {L.}~\bibnamefont {Balents}}, \bibinfo {author} {\bibfnamefont
  {J.}~\bibnamefont {He}},\ and\ \bibinfo {author} {\bibfnamefont {S.~D.}\
  \bibnamefont {Wilson}},\ }\href
  {https://doi.org/10.1103/PhysRevLett.125.247002} {\bibfield  {journal}
  {\bibinfo  {journal} {Phys. Rev. Lett.}\ }\textbf {\bibinfo {volume} {125}},\
  \bibinfo {pages} {247002} (\bibinfo {year} {2020})}\BibitemShut {NoStop}%
\bibitem [{\citenamefont {Uykur}\ \emph {et~al.}(2021)\citenamefont {Uykur},
  \citenamefont {Ortiz}, \citenamefont {Iakutkina}, \citenamefont {Wenzel},
  \citenamefont {Wilson}, \citenamefont {Dressel},\ and\ \citenamefont
  {Tsirlin}}]{Uykur2021}%
  \BibitemOpen
  \bibfield  {author} {\bibinfo {author} {\bibfnamefont {E.}~\bibnamefont
  {Uykur}}, \bibinfo {author} {\bibfnamefont {B.~R.}\ \bibnamefont {Ortiz}},
  \bibinfo {author} {\bibfnamefont {O.}~\bibnamefont {Iakutkina}}, \bibinfo
  {author} {\bibfnamefont {M.}~\bibnamefont {Wenzel}}, \bibinfo {author}
  {\bibfnamefont {S.~D.}\ \bibnamefont {Wilson}}, \bibinfo {author}
  {\bibfnamefont {M.}~\bibnamefont {Dressel}},\ and\ \bibinfo {author}
  {\bibfnamefont {A.~A.}\ \bibnamefont {Tsirlin}},\ }\href
  {https://doi.org/10.1103/PhysRevB.104.045130} {\bibfield  {journal} {\bibinfo
   {journal} {Phys. Rev. B}\ }\textbf {\bibinfo {volume} {104}},\ \bibinfo
  {pages} {045130} (\bibinfo {year} {2021})}\BibitemShut {NoStop}%
\bibitem [{\citenamefont {Wenzel}\ \emph {et~al.}(2022)\citenamefont {Wenzel},
  \citenamefont {Ortiz}, \citenamefont {Wilson}, \citenamefont {Dressel},
  \citenamefont {Tsirlin},\ and\ \citenamefont {Uykur}}]{Wenzel2022}%
  \BibitemOpen
  \bibfield  {author} {\bibinfo {author} {\bibfnamefont {M.}~\bibnamefont
  {Wenzel}}, \bibinfo {author} {\bibfnamefont {B.~R.}\ \bibnamefont {Ortiz}},
  \bibinfo {author} {\bibfnamefont {S.~D.}\ \bibnamefont {Wilson}}, \bibinfo
  {author} {\bibfnamefont {M.}~\bibnamefont {Dressel}}, \bibinfo {author}
  {\bibfnamefont {A.~A.}\ \bibnamefont {Tsirlin}},\ and\ \bibinfo {author}
  {\bibfnamefont {E.}~\bibnamefont {Uykur}},\ }\href
  {https://doi.org/10.1103/PhysRevB.105.245123} {\bibfield  {journal} {\bibinfo
   {journal} {Phys. Rev. B}\ }\textbf {\bibinfo {volume} {105}},\ \bibinfo
  {pages} {245123} (\bibinfo {year} {2022})}\BibitemShut {NoStop}%
\bibitem [{\citenamefont {Uykur}\ \emph {et~al.}(2022)\citenamefont {Uykur},
  \citenamefont {Ortiz}, \citenamefont {Wilson}, \citenamefont {Dressel},\ and\
  \citenamefont {Tsirlin}}]{Uykur2022}%
  \BibitemOpen
  \bibfield  {author} {\bibinfo {author} {\bibfnamefont {E.}~\bibnamefont
  {Uykur}}, \bibinfo {author} {\bibfnamefont {B.~R.}\ \bibnamefont {Ortiz}},
  \bibinfo {author} {\bibfnamefont {S.~D.}\ \bibnamefont {Wilson}}, \bibinfo
  {author} {\bibfnamefont {M.}~\bibnamefont {Dressel}},\ and\ \bibinfo {author}
  {\bibfnamefont {A.~A.}\ \bibnamefont {Tsirlin}},\ }\href
  {https://doi.org/10.1038/s41535-021-00420-8} {\bibfield  {journal} {\bibinfo
  {journal} {npj Quantum Mater.}\ }\textbf {\bibinfo {volume} {7}},\ \bibinfo
  {pages} {16} (\bibinfo {year} {2022})}\BibitemShut {NoStop}%
\bibitem [{\citenamefont {Kudo}\ \emph {et~al.}(2010)\citenamefont {Kudo},
  \citenamefont {Nishikubo},\ and\ \citenamefont {Nohara}}]{Kudo2010}%
  \BibitemOpen
  \bibfield  {author} {\bibinfo {author} {\bibfnamefont {K.}~\bibnamefont
  {Kudo}}, \bibinfo {author} {\bibfnamefont {Y.}~\bibnamefont {Nishikubo}},\
  and\ \bibinfo {author} {\bibfnamefont {M.}~\bibnamefont {Nohara}},\ }\href
  {https://doi.org/10.1143/JPSJ.79.123710} {\bibfield  {journal} {\bibinfo
  {journal} {J. Phys. Soc. Jpn.}\ }\textbf {\bibinfo {volume} {79}},\ \bibinfo
  {pages} {123710} (\bibinfo {year} {2010})}\BibitemShut {NoStop}%
\bibitem [{\citenamefont {Nagano}\ \emph {et~al.}(2013)\citenamefont {Nagano},
  \citenamefont {Araoka}, \citenamefont {Mitsuda}, \citenamefont {Yayama},
  \citenamefont {Wada}, \citenamefont {Ichihara}, \citenamefont {Isobe},\ and\
  \citenamefont {Ueda}}]{Nagano2013}%
  \BibitemOpen
  \bibfield  {author} {\bibinfo {author} {\bibfnamefont {Y.}~\bibnamefont
  {Nagano}}, \bibinfo {author} {\bibfnamefont {N.}~\bibnamefont {Araoka}},
  \bibinfo {author} {\bibfnamefont {A.}~\bibnamefont {Mitsuda}}, \bibinfo
  {author} {\bibfnamefont {H.}~\bibnamefont {Yayama}}, \bibinfo {author}
  {\bibfnamefont {H.}~\bibnamefont {Wada}}, \bibinfo {author} {\bibfnamefont
  {M.}~\bibnamefont {Ichihara}}, \bibinfo {author} {\bibfnamefont
  {M.}~\bibnamefont {Isobe}},\ and\ \bibinfo {author} {\bibfnamefont
  {Y.}~\bibnamefont {Ueda}},\ }\href {https://doi.org/10.7566/JPSJ.82.064715}
  {\bibfield  {journal} {\bibinfo  {journal} {J. Phys. Soc. Jpn.}\ }\textbf
  {\bibinfo {volume} {82}},\ \bibinfo {pages} {064715} (\bibinfo {year}
  {2013})}\BibitemShut {NoStop}%
\bibitem [{\citenamefont {Kim}\ \emph {et~al.}(2015)\citenamefont {Kim},
  \citenamefont {Kim},\ and\ \citenamefont {Min}}]{Kim2015}%
  \BibitemOpen
  \bibfield  {author} {\bibinfo {author} {\bibfnamefont {S.}~\bibnamefont
  {Kim}}, \bibinfo {author} {\bibfnamefont {K.}~\bibnamefont {Kim}},\ and\
  \bibinfo {author} {\bibfnamefont {B.~I.}\ \bibnamefont {Min}},\ }\href
  {https://doi.org/10.1038/srep15052} {\bibfield  {journal} {\bibinfo
  {journal} {Sci. Rep.}\ }\textbf {\bibinfo {volume} {5}},\ \bibinfo {pages}
  {15052} (\bibinfo {year} {2015})}\BibitemShut {NoStop}%
\bibitem [{\citenamefont {Machida}\ and\ \citenamefont
  {Kato}(1987)}]{Machida1987}%
  \BibitemOpen
  \bibfield  {author} {\bibinfo {author} {\bibfnamefont {K.}~\bibnamefont
  {Machida}}\ and\ \bibinfo {author} {\bibfnamefont {M.}~\bibnamefont {Kato}},\
  }\href {https://doi.org/10.1103/PhysRevB.36.854} {\bibfield  {journal}
  {\bibinfo  {journal} {Phys. Rev. B}\ }\textbf {\bibinfo {volume} {36}},\
  \bibinfo {pages} {854} (\bibinfo {year} {1987})}\BibitemShut {NoStop}%
\bibitem [{\citenamefont {Vojta}(2009)}]{Vojta2009}%
  \BibitemOpen
  \bibfield  {author} {\bibinfo {author} {\bibfnamefont {M.}~\bibnamefont
  {Vojta}},\ }\href {https://doi.org/10.1080/00018730903122242} {\bibfield
  {journal} {\bibinfo  {journal} {Adv. Phys.}\ }\textbf {\bibinfo {volume}
  {58}},\ \bibinfo {pages} {699–820} (\bibinfo {year} {2009})}\BibitemShut
  {NoStop}%
\bibitem [{\citenamefont {Fradkin}\ \emph {et~al.}(2015)\citenamefont
  {Fradkin}, \citenamefont {Kivelson},\ and\ \citenamefont
  {Tranquada}}]{Fradkin2015}%
  \BibitemOpen
  \bibfield  {author} {\bibinfo {author} {\bibfnamefont {E.}~\bibnamefont
  {Fradkin}}, \bibinfo {author} {\bibfnamefont {S.~A.}\ \bibnamefont
  {Kivelson}},\ and\ \bibinfo {author} {\bibfnamefont {J.~M.}\ \bibnamefont
  {Tranquada}},\ }\href {https://doi.org/10.1103/RevModPhys.87.457} {\bibfield
  {journal} {\bibinfo  {journal} {Rev. Mod. Phys.}\ }\textbf {\bibinfo {volume}
  {87}},\ \bibinfo {pages} {457} (\bibinfo {year} {2015})}\BibitemShut
  {NoStop}%
\bibitem [{\citenamefont {Agterberg}\ \emph {et~al.}(2020)\citenamefont
  {Agterberg}, \citenamefont {Davis}, \citenamefont {Edkins}, \citenamefont
  {Fradkin}, \citenamefont {Van~Harlingen}, \citenamefont {Kivelson},
  \citenamefont {Lee}, \citenamefont {Radzihovsky}, \citenamefont {Tranquada},\
  and\ \citenamefont {Wang}}]{Agterberg2020}%
  \BibitemOpen
  \bibfield  {author} {\bibinfo {author} {\bibfnamefont {D.~F.}\ \bibnamefont
  {Agterberg}}, \bibinfo {author} {\bibfnamefont {J.~S.}\ \bibnamefont
  {Davis}}, \bibinfo {author} {\bibfnamefont {S.~D.}\ \bibnamefont {Edkins}},
  \bibinfo {author} {\bibfnamefont {E.}~\bibnamefont {Fradkin}}, \bibinfo
  {author} {\bibfnamefont {D.~J.}\ \bibnamefont {Van~Harlingen}}, \bibinfo
  {author} {\bibfnamefont {S.~A.}\ \bibnamefont {Kivelson}}, \bibinfo {author}
  {\bibfnamefont {P.~A.}\ \bibnamefont {Lee}}, \bibinfo {author} {\bibfnamefont
  {L.}~\bibnamefont {Radzihovsky}}, \bibinfo {author} {\bibfnamefont {J.~M.}\
  \bibnamefont {Tranquada}},\ and\ \bibinfo {author} {\bibfnamefont
  {Y.}~\bibnamefont {Wang}},\ }\href
  {https://doi.org/10.1146/annurev-conmatphys-031119-050711} {\bibfield
  {journal} {\bibinfo  {journal} {Annu. Rev. Condens. Matter Phys.}\ }\textbf
  {\bibinfo {volume} {11}},\ \bibinfo {pages} {231–270} (\bibinfo {year}
  {2020})}\BibitemShut {NoStop}%
\bibitem [{\citenamefont {LeBlanc}\ \emph {et~al.}(2015)\citenamefont
  {LeBlanc}, \citenamefont {Antipov}, \citenamefont {Becca}, \citenamefont
  {Bulik}, \citenamefont {Chan}, \citenamefont {Chung}, \citenamefont {Deng},
  \citenamefont {Ferrero}, \citenamefont {Henderson}, \citenamefont
  {Jim\'enez-Hoyos}, \citenamefont {Kozik}, \citenamefont {Liu}, \citenamefont
  {Millis}, \citenamefont {Prokof'ev}, \citenamefont {Qin}, \citenamefont
  {Scuseria}, \citenamefont {Shi}, \citenamefont {Svistunov}, \citenamefont
  {Tocchio}, \citenamefont {Tupitsyn}, \citenamefont {White}, \citenamefont
  {Zhang}, \citenamefont {Zheng}, \citenamefont {Zhu},\ and\ \citenamefont
  {Gull}}]{LeBlanc2015}%
  \BibitemOpen
  \bibfield  {author} {\bibinfo {author} {\bibfnamefont {J.~P.~F.}\
  \bibnamefont {LeBlanc}}, \bibinfo {author} {\bibfnamefont {A.~E.}\
  \bibnamefont {Antipov}}, \bibinfo {author} {\bibfnamefont {F.}~\bibnamefont
  {Becca}}, \bibinfo {author} {\bibfnamefont {I.~W.}\ \bibnamefont {Bulik}},
  \bibinfo {author} {\bibfnamefont {G.~K.-L.}\ \bibnamefont {Chan}}, \bibinfo
  {author} {\bibfnamefont {C.-M.}\ \bibnamefont {Chung}}, \bibinfo {author}
  {\bibfnamefont {Y.}~\bibnamefont {Deng}}, \bibinfo {author} {\bibfnamefont
  {M.}~\bibnamefont {Ferrero}}, \bibinfo {author} {\bibfnamefont {T.~M.}\
  \bibnamefont {Henderson}}, \bibinfo {author} {\bibfnamefont {C.~A.}\
  \bibnamefont {Jim\'enez-Hoyos}}, \bibinfo {author} {\bibfnamefont
  {E.}~\bibnamefont {Kozik}}, \bibinfo {author} {\bibfnamefont {X.-W.}\
  \bibnamefont {Liu}}, \bibinfo {author} {\bibfnamefont {A.~J.}\ \bibnamefont
  {Millis}}, \bibinfo {author} {\bibfnamefont {N.~V.}\ \bibnamefont
  {Prokof'ev}}, \bibinfo {author} {\bibfnamefont {M.}~\bibnamefont {Qin}},
  \bibinfo {author} {\bibfnamefont {G.~E.}\ \bibnamefont {Scuseria}}, \bibinfo
  {author} {\bibfnamefont {H.}~\bibnamefont {Shi}}, \bibinfo {author}
  {\bibfnamefont {B.~V.}\ \bibnamefont {Svistunov}}, \bibinfo {author}
  {\bibfnamefont {L.~F.}\ \bibnamefont {Tocchio}}, \bibinfo {author}
  {\bibfnamefont {I.~S.}\ \bibnamefont {Tupitsyn}}, \bibinfo {author}
  {\bibfnamefont {S.~R.}\ \bibnamefont {White}}, \bibinfo {author}
  {\bibfnamefont {S.}~\bibnamefont {Zhang}}, \bibinfo {author} {\bibfnamefont
  {B.-X.}\ \bibnamefont {Zheng}}, \bibinfo {author} {\bibfnamefont
  {Z.}~\bibnamefont {Zhu}},\ and\ \bibinfo {author} {\bibfnamefont
  {E.}~\bibnamefont {Gull}} (\bibinfo {collaboration} {Simons Collaboration on
  the Many-Electron Problem}),\ }\href
  {https://doi.org/10.1103/PhysRevX.5.041041} {\bibfield  {journal} {\bibinfo
  {journal} {Phys. Rev. X}\ }\textbf {\bibinfo {volume} {5}},\ \bibinfo {pages}
  {041041} (\bibinfo {year} {2015})}\BibitemShut {NoStop}%
\bibitem [{\citenamefont {Zheng}\ \emph {et~al.}(2017)\citenamefont {Zheng},
  \citenamefont {Chung}, \citenamefont {Corboz}, \citenamefont {Ehlers},
  \citenamefont {Qin}, \citenamefont {Noack}, \citenamefont {Shi},
  \citenamefont {White}, \citenamefont {Zhang},\ and\ \citenamefont
  {Chan}}]{Zheng2017}%
  \BibitemOpen
  \bibfield  {author} {\bibinfo {author} {\bibfnamefont {B.-X.}\ \bibnamefont
  {Zheng}}, \bibinfo {author} {\bibfnamefont {C.-M.}\ \bibnamefont {Chung}},
  \bibinfo {author} {\bibfnamefont {P.}~\bibnamefont {Corboz}}, \bibinfo
  {author} {\bibfnamefont {G.}~\bibnamefont {Ehlers}}, \bibinfo {author}
  {\bibfnamefont {M.-P.}\ \bibnamefont {Qin}}, \bibinfo {author} {\bibfnamefont
  {R.~M.}\ \bibnamefont {Noack}}, \bibinfo {author} {\bibfnamefont
  {H.}~\bibnamefont {Shi}}, \bibinfo {author} {\bibfnamefont {S.~R.}\
  \bibnamefont {White}}, \bibinfo {author} {\bibfnamefont {S.}~\bibnamefont
  {Zhang}},\ and\ \bibinfo {author} {\bibfnamefont {G.~K.-L.}\ \bibnamefont
  {Chan}},\ }\href {https://doi.org/10.1126/science.aam7127} {\bibfield
  {journal} {\bibinfo  {journal} {Science}\ }\textbf {\bibinfo {volume}
  {358}},\ \bibinfo {pages} {1155} (\bibinfo {year} {2017})}\BibitemShut
  {NoStop}%
\bibitem [{\citenamefont {Huang}\ \emph {et~al.}(2017)\citenamefont {Huang},
  \citenamefont {Mendl}, \citenamefont {Liu}, \citenamefont {Johnston},
  \citenamefont {Jiang}, \citenamefont {Moritz},\ and\ \citenamefont
  {Devereaux}}]{Huang2017}%
  \BibitemOpen
  \bibfield  {author} {\bibinfo {author} {\bibfnamefont {E.~W.}\ \bibnamefont
  {Huang}}, \bibinfo {author} {\bibfnamefont {C.~B.}\ \bibnamefont {Mendl}},
  \bibinfo {author} {\bibfnamefont {S.}~\bibnamefont {Liu}}, \bibinfo {author}
  {\bibfnamefont {S.}~\bibnamefont {Johnston}}, \bibinfo {author}
  {\bibfnamefont {H.-C.}\ \bibnamefont {Jiang}}, \bibinfo {author}
  {\bibfnamefont {B.}~\bibnamefont {Moritz}},\ and\ \bibinfo {author}
  {\bibfnamefont {T.~P.}\ \bibnamefont {Devereaux}},\ }\href
  {https://doi.org/10.1126/science.aak9546} {\bibfield  {journal} {\bibinfo
  {journal} {Science}\ }\textbf {\bibinfo {volume} {358}},\ \bibinfo {pages}
  {1161} (\bibinfo {year} {2017})}\BibitemShut {NoStop}%
\bibitem [{\citenamefont {Huang}\ \emph {et~al.}(2018)\citenamefont {Huang},
  \citenamefont {Mendl}, \citenamefont {Jiang}, \citenamefont {Moritz},\ and\
  \citenamefont {Devereaux}}]{Huang2018}%
  \BibitemOpen
  \bibfield  {author} {\bibinfo {author} {\bibfnamefont {E.~W.}\ \bibnamefont
  {Huang}}, \bibinfo {author} {\bibfnamefont {C.~B.}\ \bibnamefont {Mendl}},
  \bibinfo {author} {\bibfnamefont {H.-C.}\ \bibnamefont {Jiang}}, \bibinfo
  {author} {\bibfnamefont {B.}~\bibnamefont {Moritz}},\ and\ \bibinfo {author}
  {\bibfnamefont {T.~P.}\ \bibnamefont {Devereaux}},\ }\href
  {https://doi.org/10.1038/s41535-018-0097-0} {\bibfield  {journal} {\bibinfo
  {journal} {npj Quantum Mater.}\ }\textbf {\bibinfo {volume} {3}},\ \bibinfo
  {pages} {22} (\bibinfo {year} {2018})}\BibitemShut {NoStop}%
\bibitem [{\citenamefont {Qin}\ \emph {et~al.}(2020)\citenamefont {Qin},
  \citenamefont {Chung}, \citenamefont {Shi}, \citenamefont {Vitali},
  \citenamefont {Hubig}, \citenamefont {Schollw\"ock}, \citenamefont {White},\
  and\ \citenamefont {Zhang}}]{Qin2020}%
  \BibitemOpen
  \bibfield  {author} {\bibinfo {author} {\bibfnamefont {M.}~\bibnamefont
  {Qin}}, \bibinfo {author} {\bibfnamefont {C.-M.}\ \bibnamefont {Chung}},
  \bibinfo {author} {\bibfnamefont {H.}~\bibnamefont {Shi}}, \bibinfo {author}
  {\bibfnamefont {E.}~\bibnamefont {Vitali}}, \bibinfo {author} {\bibfnamefont
  {C.}~\bibnamefont {Hubig}}, \bibinfo {author} {\bibfnamefont
  {U.}~\bibnamefont {Schollw\"ock}}, \bibinfo {author} {\bibfnamefont {S.~R.}\
  \bibnamefont {White}},\ and\ \bibinfo {author} {\bibfnamefont
  {S.}~\bibnamefont {Zhang}} (\bibinfo {collaboration} {Simons Collaboration on
  the Many-Electron Problem}),\ }\href
  {https://doi.org/10.1103/PhysRevX.10.031016} {\bibfield  {journal} {\bibinfo
  {journal} {Phys. Rev. X}\ }\textbf {\bibinfo {volume} {10}},\ \bibinfo
  {pages} {031016} (\bibinfo {year} {2020})}\BibitemShut {NoStop}%
\bibitem [{\citenamefont {Gong}\ \emph {et~al.}(2021)\citenamefont {Gong},
  \citenamefont {Zhu},\ and\ \citenamefont {Sheng}}]{Gong2021}%
  \BibitemOpen
  \bibfield  {author} {\bibinfo {author} {\bibfnamefont {S.}~\bibnamefont
  {Gong}}, \bibinfo {author} {\bibfnamefont {W.}~\bibnamefont {Zhu}},\ and\
  \bibinfo {author} {\bibfnamefont {D.~N.}\ \bibnamefont {Sheng}},\ }\href
  {https://doi.org/10.1103/PhysRevLett.127.097003} {\bibfield  {journal}
  {\bibinfo  {journal} {Phys. Rev. Lett.}\ }\textbf {\bibinfo {volume} {127}},\
  \bibinfo {pages} {097003} (\bibinfo {year} {2021})}\BibitemShut {NoStop}%
\bibitem [{\citenamefont {Jiang}\ \emph {et~al.}(2021)\citenamefont {Jiang},
  \citenamefont {Scalapino},\ and\ \citenamefont {White}}]{Jiang2021}%
  \BibitemOpen
  \bibfield  {author} {\bibinfo {author} {\bibfnamefont {S.}~\bibnamefont
  {Jiang}}, \bibinfo {author} {\bibfnamefont {D.~J.}\ \bibnamefont
  {Scalapino}},\ and\ \bibinfo {author} {\bibfnamefont {S.~R.}\ \bibnamefont
  {White}},\ }\href {https://www.pnas.org/content/118/44/e2109978118}
  {\bibfield  {journal} {\bibinfo  {journal} {Proc. Natl. Acad. Sci. U.S.A.}\
  }\textbf {\bibinfo {volume} {118}},\ \bibinfo {pages} {e2109978118} (\bibinfo
  {year} {2021})}\BibitemShut {NoStop}%
\bibitem [{\citenamefont {Jiang}\ and\ \citenamefont
  {Kivelson}(2021)}]{HCJiang2021}%
  \BibitemOpen
  \bibfield  {author} {\bibinfo {author} {\bibfnamefont {H.-C.}\ \bibnamefont
  {Jiang}}\ and\ \bibinfo {author} {\bibfnamefont {S.~A.}\ \bibnamefont
  {Kivelson}},\ }\href {https://doi.org/10.1103/PhysRevLett.127.097002}
  {\bibfield  {journal} {\bibinfo  {journal} {Phys. Rev. Lett.}\ }\textbf
  {\bibinfo {volume} {127}},\ \bibinfo {pages} {097002} (\bibinfo {year}
  {2021})}\BibitemShut {NoStop}%
\bibitem [{\citenamefont {Jiang}\ \emph {et~al.}(2023)\citenamefont {Jiang},
  \citenamefont {Devereaux},\ and\ \citenamefont {Jiang}}]{Jiang2023}%
  \BibitemOpen
  \bibfield  {author} {\bibinfo {author} {\bibfnamefont {Y.-F.}\ \bibnamefont
  {Jiang}}, \bibinfo {author} {\bibfnamefont {T.~P.}\ \bibnamefont
  {Devereaux}},\ and\ \bibinfo {author} {\bibfnamefont {H.-C.}\ \bibnamefont
  {Jiang}},\ }\href@noop {} {\bibinfo {title} {Ground state phase diagram and
  superconductivity of the doped hubbard model on six-leg square cylinders}}
  (\bibinfo {year} {2023}),\ \Eprint {https://arxiv.org/abs/arXiv:2303.15541}
  {arXiv:2303.15541} \BibitemShut {NoStop}%
\bibitem [{\citenamefont {Jiang}(2023)}]{Jiang2023b}%
  \BibitemOpen
  \bibfield  {author} {\bibinfo {author} {\bibfnamefont {H.-C.}\ \bibnamefont
  {Jiang}},\ }\href {https://doi.org/10.1103/PhysRevB.107.214504} {\bibfield
  {journal} {\bibinfo  {journal} {Phys. Rev. B}\ }\textbf {\bibinfo {volume}
  {107}},\ \bibinfo {pages} {214504} (\bibinfo {year} {2023})}\BibitemShut
  {NoStop}%
\bibitem [{\citenamefont {Xu}\ \emph {et~al.}(2024)\citenamefont {Xu},
  \citenamefont {Chung}, \citenamefont {Qin}, \citenamefont {Schollwöck},
  \citenamefont {White},\ and\ \citenamefont {Zhang}}]{Xu2024}%
  \BibitemOpen
  \bibfield  {author} {\bibinfo {author} {\bibfnamefont {H.}~\bibnamefont
  {Xu}}, \bibinfo {author} {\bibfnamefont {C.-M.}\ \bibnamefont {Chung}},
  \bibinfo {author} {\bibfnamefont {M.}~\bibnamefont {Qin}}, \bibinfo {author}
  {\bibfnamefont {U.}~\bibnamefont {Schollwöck}}, \bibinfo {author}
  {\bibfnamefont {S.~R.}\ \bibnamefont {White}},\ and\ \bibinfo {author}
  {\bibfnamefont {S.}~\bibnamefont {Zhang}},\ }\href
  {https://doi.org/10.1126/science.adh7691} {\bibfield  {journal} {\bibinfo
  {journal} {Science}\ }\textbf {\bibinfo {volume} {384}},\ \bibinfo {pages}
  {eadh7691} (\bibinfo {year} {2024})}\BibitemShut {NoStop}%
\bibitem [{\citenamefont {Lu}\ \emph {et~al.}(2024)\citenamefont {Lu},
  \citenamefont {Chen}, \citenamefont {Zhu}, \citenamefont {Sheng},\ and\
  \citenamefont {Gong}}]{Lu2024}%
  \BibitemOpen
  \bibfield  {author} {\bibinfo {author} {\bibfnamefont {X.}~\bibnamefont
  {Lu}}, \bibinfo {author} {\bibfnamefont {F.}~\bibnamefont {Chen}}, \bibinfo
  {author} {\bibfnamefont {W.}~\bibnamefont {Zhu}}, \bibinfo {author}
  {\bibfnamefont {D.~N.}\ \bibnamefont {Sheng}},\ and\ \bibinfo {author}
  {\bibfnamefont {S.-S.}\ \bibnamefont {Gong}},\ }\href
  {https://doi.org/10.1103/PhysRevLett.132.066002} {\bibfield  {journal}
  {\bibinfo  {journal} {Phys. Rev. Lett.}\ }\textbf {\bibinfo {volume} {132}},\
  \bibinfo {pages} {066002} (\bibinfo {year} {2024})}\BibitemShut {NoStop}%
\bibitem [{\citenamefont {Ponsioen}\ \emph {et~al.}(2023)\citenamefont
  {Ponsioen}, \citenamefont {Chung},\ and\ \citenamefont
  {Corboz}}]{Ponsioen2023}%
  \BibitemOpen
  \bibfield  {author} {\bibinfo {author} {\bibfnamefont {B.}~\bibnamefont
  {Ponsioen}}, \bibinfo {author} {\bibfnamefont {S.~S.}\ \bibnamefont
  {Chung}},\ and\ \bibinfo {author} {\bibfnamefont {P.}~\bibnamefont
  {Corboz}},\ }\href {https://doi.org/10.1103/PhysRevB.108.205154} {\bibfield
  {journal} {\bibinfo  {journal} {Phys. Rev. B}\ }\textbf {\bibinfo {volume}
  {108}},\ \bibinfo {pages} {205154} (\bibinfo {year} {2023})}\BibitemShut
  {NoStop}%
\bibitem [{\citenamefont {Wietek}(2022)}]{Wietek2022}%
  \BibitemOpen
  \bibfield  {author} {\bibinfo {author} {\bibfnamefont {A.}~\bibnamefont
  {Wietek}},\ }\href {https://doi.org/10.1103/PhysRevLett.129.177001}
  {\bibfield  {journal} {\bibinfo  {journal} {Phys. Rev. Lett.}\ }\textbf
  {\bibinfo {volume} {129}},\ \bibinfo {pages} {177001} (\bibinfo {year}
  {2022})}\BibitemShut {NoStop}%
\bibitem [{\citenamefont {Baldelli}\ \emph {et~al.}(2025)\citenamefont
  {Baldelli}, \citenamefont {Karlsson}, \citenamefont {Kloss}, \citenamefont
  {Fishman},\ and\ \citenamefont {Wietek}}]{Baldelli2025}%
  \BibitemOpen
  \bibfield  {author} {\bibinfo {author} {\bibfnamefont {N.}~\bibnamefont
  {Baldelli}}, \bibinfo {author} {\bibfnamefont {H.}~\bibnamefont {Karlsson}},
  \bibinfo {author} {\bibfnamefont {B.}~\bibnamefont {Kloss}}, \bibinfo
  {author} {\bibfnamefont {M.}~\bibnamefont {Fishman}},\ and\ \bibinfo {author}
  {\bibfnamefont {A.}~\bibnamefont {Wietek}},\ }\href
  {https://doi.org/10.1038/s41535-024-00718-3} {\bibfield  {journal} {\bibinfo
  {journal} {npj Quantum Mater.}\ }\textbf {\bibinfo {volume} {10}},\ \bibinfo
  {pages} {22} (\bibinfo {year} {2025})}\BibitemShut {NoStop}%
\bibitem [{\citenamefont {Berezinsky}(1971)}]{Berezinsky1970}%
  \BibitemOpen
  \bibfield  {author} {\bibinfo {author} {\bibfnamefont {V.}~\bibnamefont
  {Berezinsky}},\ }\href
  {https://s3.cern.ch/inspire-prod-files-0/0f7b50c47ec26bed99a50ff199960259}
  {\bibfield  {journal} {\bibinfo  {journal} {Sov. Phys. JETP}\ }\textbf
  {\bibinfo {volume} {32}},\ \bibinfo {pages} {493} (\bibinfo {year}
  {1971})}\BibitemShut {NoStop}%
\bibitem [{\citenamefont {Kosterlitz}\ and\ \citenamefont
  {Thouless}(1973)}]{Kosterlitz1973}%
  \BibitemOpen
  \bibfield  {author} {\bibinfo {author} {\bibfnamefont {J.~M.}\ \bibnamefont
  {Kosterlitz}}\ and\ \bibinfo {author} {\bibfnamefont {D.~J.}\ \bibnamefont
  {Thouless}},\ }\href {https://doi.org/10.1088/0022-3719/6/7/010} {\bibfield
  {journal} {\bibinfo  {journal} {J. Phys. C: Solid State Phys.}\ }\textbf
  {\bibinfo {volume} {6}},\ \bibinfo {pages} {1181} (\bibinfo {year}
  {1973})}\BibitemShut {NoStop}%
\bibitem [{\citenamefont {Keimer}\ \emph {et~al.}(2015)\citenamefont {Keimer},
  \citenamefont {Kivelson}, \citenamefont {Norman}, \citenamefont {Uchida},\
  and\ \citenamefont {Zaanen}}]{Keimer2015}%
  \BibitemOpen
  \bibfield  {author} {\bibinfo {author} {\bibfnamefont {B.}~\bibnamefont
  {Keimer}}, \bibinfo {author} {\bibfnamefont {S.~A.}\ \bibnamefont
  {Kivelson}}, \bibinfo {author} {\bibfnamefont {M.~R.}\ \bibnamefont
  {Norman}}, \bibinfo {author} {\bibfnamefont {S.}~\bibnamefont {Uchida}},\
  and\ \bibinfo {author} {\bibfnamefont {J.}~\bibnamefont {Zaanen}},\ }\href
  {https://doi.org/10.1038/nature14165} {\bibfield  {journal} {\bibinfo
  {journal} {Nature}\ }\textbf {\bibinfo {volume} {518}},\ \bibinfo {pages}
  {179} (\bibinfo {year} {2015})}\BibitemShut {NoStop}%
\bibitem [{\citenamefont {Yu}\ \emph {et~al.}(2014)\citenamefont {Yu},
  \citenamefont {Xie}, \citenamefont {Meurice}, \citenamefont {Liu},
  \citenamefont {Denbleyker}, \citenamefont {Zou}, \citenamefont {Qin},
  \citenamefont {Chen},\ and\ \citenamefont {Xiang}}]{Yu2014}%
  \BibitemOpen
  \bibfield  {author} {\bibinfo {author} {\bibfnamefont {J.~F.}\ \bibnamefont
  {Yu}}, \bibinfo {author} {\bibfnamefont {Z.~Y.}\ \bibnamefont {Xie}},
  \bibinfo {author} {\bibfnamefont {Y.}~\bibnamefont {Meurice}}, \bibinfo
  {author} {\bibfnamefont {Y.}~\bibnamefont {Liu}}, \bibinfo {author}
  {\bibfnamefont {A.}~\bibnamefont {Denbleyker}}, \bibinfo {author}
  {\bibfnamefont {H.}~\bibnamefont {Zou}}, \bibinfo {author} {\bibfnamefont
  {M.~P.}\ \bibnamefont {Qin}}, \bibinfo {author} {\bibfnamefont
  {J.}~\bibnamefont {Chen}},\ and\ \bibinfo {author} {\bibfnamefont
  {T.}~\bibnamefont {Xiang}},\ }\href
  {https://doi.org/10.1103/PhysRevE.89.013308} {\bibfield  {journal} {\bibinfo
  {journal} {Phys. Rev. E}\ }\textbf {\bibinfo {volume} {89}},\ \bibinfo
  {pages} {013308} (\bibinfo {year} {2014})}\BibitemShut {NoStop}%
\bibitem [{\citenamefont {Vanderstraeten}\ \emph
  {et~al.}(2019{\natexlab{a}})\citenamefont {Vanderstraeten}, \citenamefont
  {Vanhecke}, \citenamefont {L\"auchli},\ and\ \citenamefont
  {Verstraete}}]{Vanderstraeten2019}%
  \BibitemOpen
  \bibfield  {author} {\bibinfo {author} {\bibfnamefont {L.}~\bibnamefont
  {Vanderstraeten}}, \bibinfo {author} {\bibfnamefont {B.}~\bibnamefont
  {Vanhecke}}, \bibinfo {author} {\bibfnamefont {A.~M.}\ \bibnamefont
  {L\"auchli}},\ and\ \bibinfo {author} {\bibfnamefont {F.}~\bibnamefont
  {Verstraete}},\ }\href {https://doi.org/10.1103/PhysRevE.100.062136}
  {\bibfield  {journal} {\bibinfo  {journal} {Phys. Rev. E}\ }\textbf {\bibinfo
  {volume} {100}},\ \bibinfo {pages} {062136} (\bibinfo {year}
  {2019}{\natexlab{a}})}\BibitemShut {NoStop}%
\bibitem [{\citenamefont {Song}\ and\ \citenamefont {Zhang}(2021)}]{Song2021}%
  \BibitemOpen
  \bibfield  {author} {\bibinfo {author} {\bibfnamefont {F.-F.}\ \bibnamefont
  {Song}}\ and\ \bibinfo {author} {\bibfnamefont {G.-M.}\ \bibnamefont
  {Zhang}},\ }\href {https://doi.org/10.1103/PhysRevB.103.024518} {\bibfield
  {journal} {\bibinfo  {journal} {Phys. Rev. B}\ }\textbf {\bibinfo {volume}
  {103}},\ \bibinfo {pages} {024518} (\bibinfo {year} {2021})}\BibitemShut
  {NoStop}%
\bibitem [{\citenamefont {Haegeman}\ and\ \citenamefont
  {Verstraete}(2017)}]{Haegeman2017}%
  \BibitemOpen
  \bibfield  {author} {\bibinfo {author} {\bibfnamefont {J.}~\bibnamefont
  {Haegeman}}\ and\ \bibinfo {author} {\bibfnamefont {F.}~\bibnamefont
  {Verstraete}},\ }\href
  {https://doi.org/https://doi.org/10.1146/annurev-conmatphys-031016-025507}
  {\bibfield  {journal} {\bibinfo  {journal} {Annu. Rev. Condens. Matter
  Phys.}\ }\textbf {\bibinfo {volume} {8}},\ \bibinfo {pages} {355} (\bibinfo
  {year} {2017})}\BibitemShut {NoStop}%
\bibitem [{\citenamefont {Zauner-Stauber}\ \emph {et~al.}(2018)\citenamefont
  {Zauner-Stauber}, \citenamefont {Vanderstraeten}, \citenamefont {Fishman},
  \citenamefont {Verstraete},\ and\ \citenamefont
  {Haegeman}}]{Zauner-Stauber2018}%
  \BibitemOpen
  \bibfield  {author} {\bibinfo {author} {\bibfnamefont {V.}~\bibnamefont
  {Zauner-Stauber}}, \bibinfo {author} {\bibfnamefont {L.}~\bibnamefont
  {Vanderstraeten}}, \bibinfo {author} {\bibfnamefont {M.~T.}\ \bibnamefont
  {Fishman}}, \bibinfo {author} {\bibfnamefont {F.}~\bibnamefont
  {Verstraete}},\ and\ \bibinfo {author} {\bibfnamefont {J.}~\bibnamefont
  {Haegeman}},\ }\href {https://doi.org/10.1103/PhysRevB.97.045145} {\bibfield
  {journal} {\bibinfo  {journal} {Phys. Rev. B}\ }\textbf {\bibinfo {volume}
  {97}},\ \bibinfo {pages} {045145} (\bibinfo {year} {2018})}\BibitemShut
  {NoStop}%
\bibitem [{\citenamefont {Vanderstraeten}\ \emph
  {et~al.}(2019{\natexlab{b}})\citenamefont {Vanderstraeten}, \citenamefont
  {Haegeman},\ and\ \citenamefont {Verstraete}}]{Vanderstraeten2019tan}%
  \BibitemOpen
  \bibfield  {author} {\bibinfo {author} {\bibfnamefont {L.}~\bibnamefont
  {Vanderstraeten}}, \bibinfo {author} {\bibfnamefont {J.}~\bibnamefont
  {Haegeman}},\ and\ \bibinfo {author} {\bibfnamefont {F.}~\bibnamefont
  {Verstraete}},\ }\href {https://doi.org/10.21468/SciPostPhysLectNotes.7}
  {\bibfield  {journal} {\bibinfo  {journal} {SciPost Phys. Lect. Notes}\ ,\
  \bibinfo {pages} {7}} (\bibinfo {year} {2019}{\natexlab{b}})}\BibitemShut
  {NoStop}%
\bibitem [{\citenamefont {Nietner}\ \emph {et~al.}(2020)\citenamefont
  {Nietner}, \citenamefont {Vanhecke}, \citenamefont {Verstraete},
  \citenamefont {Eisert},\ and\ \citenamefont {Vanderstraeten}}]{Nietner2020}%
  \BibitemOpen
  \bibfield  {author} {\bibinfo {author} {\bibfnamefont {A.}~\bibnamefont
  {Nietner}}, \bibinfo {author} {\bibfnamefont {B.}~\bibnamefont {Vanhecke}},
  \bibinfo {author} {\bibfnamefont {F.}~\bibnamefont {Verstraete}}, \bibinfo
  {author} {\bibfnamefont {J.}~\bibnamefont {Eisert}},\ and\ \bibinfo {author}
  {\bibfnamefont {L.}~\bibnamefont {Vanderstraeten}},\ }\href
  {https://doi.org/10.22331/q-2020-09-21-328} {\bibfield  {journal} {\bibinfo
  {journal} {{Quantum}}\ }\textbf {\bibinfo {volume} {4}},\ \bibinfo {pages}
  {328} (\bibinfo {year} {2020})}\BibitemShut {NoStop}%
\bibitem [{\citenamefont {Song}\ and\ \citenamefont
  {Zhang}(2022{\natexlab{a}})}]{Song2022}%
  \BibitemOpen
  \bibfield  {author} {\bibinfo {author} {\bibfnamefont {F.-F.}\ \bibnamefont
  {Song}}\ and\ \bibinfo {author} {\bibfnamefont {G.-M.}\ \bibnamefont
  {Zhang}},\ }\href {https://doi.org/10.1103/PhysRevLett.128.195301} {\bibfield
   {journal} {\bibinfo  {journal} {Phys. Rev. Lett.}\ }\textbf {\bibinfo
  {volume} {128}},\ \bibinfo {pages} {195301} (\bibinfo {year}
  {2022}{\natexlab{a}})}\BibitemShut {NoStop}%
\bibitem [{\citenamefont {Song}\ and\ \citenamefont
  {Zhang}(2022{\natexlab{b}})}]{Song2022ffxy}%
  \BibitemOpen
  \bibfield  {author} {\bibinfo {author} {\bibfnamefont {F.-F.}\ \bibnamefont
  {Song}}\ and\ \bibinfo {author} {\bibfnamefont {G.-M.}\ \bibnamefont
  {Zhang}},\ }\href {https://doi.org/10.1103/PhysRevB.105.134516} {\bibfield
  {journal} {\bibinfo  {journal} {Phys. Rev. B}\ }\textbf {\bibinfo {volume}
  {105}},\ \bibinfo {pages} {134516} (\bibinfo {year}
  {2022}{\natexlab{b}})}\BibitemShut {NoStop}%
\bibitem [{\citenamefont {Song}\ and\ \citenamefont {Zhang}(2023)}]{Song2023}%
  \BibitemOpen
  \bibfield  {author} {\bibinfo {author} {\bibfnamefont {F.-F.}\ \bibnamefont
  {Song}}\ and\ \bibinfo {author} {\bibfnamefont {G.-M.}\ \bibnamefont
  {Zhang}},\ }\href {https://doi.org/10.1103/PhysRevB.108.014424} {\bibfield
  {journal} {\bibinfo  {journal} {Phys. Rev. B}\ }\textbf {\bibinfo {volume}
  {108}},\ \bibinfo {pages} {014424} (\bibinfo {year} {2023})}\BibitemShut
  {NoStop}%
\bibitem [{\citenamefont {Morita}\ and\ \citenamefont
  {Kawashima}(2019)}]{Morita2019}%
  \BibitemOpen
  \bibfield  {author} {\bibinfo {author} {\bibfnamefont {S.}~\bibnamefont
  {Morita}}\ and\ \bibinfo {author} {\bibfnamefont {N.}~\bibnamefont
  {Kawashima}},\ }\href
  {https://doi.org/https://doi.org/10.1016/j.cpc.2018.10.014} {\bibfield
  {journal} {\bibinfo  {journal} {Comput. Phys. Commun.}\ }\textbf {\bibinfo
  {volume} {236}},\ \bibinfo {pages} {65} (\bibinfo {year} {2019})}\BibitemShut
  {NoStop}%
\bibitem [{\citenamefont {Morita}\ and\ \citenamefont
  {Kawashima}(2025)}]{Morita2025}%
  \BibitemOpen
  \bibfield  {author} {\bibinfo {author} {\bibfnamefont {S.}~\bibnamefont
  {Morita}}\ and\ \bibinfo {author} {\bibfnamefont {N.}~\bibnamefont
  {Kawashima}},\ }\href {https://doi.org/10.1103/PhysRevB.111.054433}
  {\bibfield  {journal} {\bibinfo  {journal} {Phys. Rev. B}\ }\textbf {\bibinfo
  {volume} {111}},\ \bibinfo {pages} {054433} (\bibinfo {year}
  {2025})}\BibitemShut {NoStop}%
\bibitem [{SM()}]{SM}%
  \BibitemOpen
  \href@noop {} {}\bibinfo {note} {See the Supplemental Material for details of
  the tensor methods, the modulated clock model, the duality equations, the
  Migdal-Kadanoff renormaization group analysis and more Monte Carlo
  results.}\BibitemShut {Stop}%
\bibitem [{\citenamefont {Vidal}\ \emph {et~al.}(2003)\citenamefont {Vidal},
  \citenamefont {Latorre}, \citenamefont {Rico},\ and\ \citenamefont
  {Kitaev}}]{Vidal2003}%
  \BibitemOpen
  \bibfield  {author} {\bibinfo {author} {\bibfnamefont {G.}~\bibnamefont
  {Vidal}}, \bibinfo {author} {\bibfnamefont {J.~I.}\ \bibnamefont {Latorre}},
  \bibinfo {author} {\bibfnamefont {E.}~\bibnamefont {Rico}},\ and\ \bibinfo
  {author} {\bibfnamefont {A.}~\bibnamefont {Kitaev}},\ }\href
  {https://doi.org/10.1103/PhysRevLett.90.227902} {\bibfield  {journal}
  {\bibinfo  {journal} {Phys. Rev. Lett.}\ }\textbf {\bibinfo {volume} {90}},\
  \bibinfo {pages} {227902} (\bibinfo {year} {2003})}\BibitemShut {NoStop}%
\bibitem [{\citenamefont {Pollmann}\ \emph {et~al.}(2009)\citenamefont
  {Pollmann}, \citenamefont {Mukerjee}, \citenamefont {Turner},\ and\
  \citenamefont {Moore}}]{Pollmann2009}%
  \BibitemOpen
  \bibfield  {author} {\bibinfo {author} {\bibfnamefont {F.}~\bibnamefont
  {Pollmann}}, \bibinfo {author} {\bibfnamefont {S.}~\bibnamefont {Mukerjee}},
  \bibinfo {author} {\bibfnamefont {A.~M.}\ \bibnamefont {Turner}},\ and\
  \bibinfo {author} {\bibfnamefont {J.~E.}\ \bibnamefont {Moore}},\ }\href
  {https://doi.org/10.1103/PhysRevLett.102.255701} {\bibfield  {journal}
  {\bibinfo  {journal} {Phys. Rev. Lett.}\ }\textbf {\bibinfo {volume} {102}},\
  \bibinfo {pages} {255701} (\bibinfo {year} {2009})}\BibitemShut {NoStop}%
\bibitem [{\citenamefont {Solla}\ and\ \citenamefont
  {Riedel}(1981)}]{Solla1981}%
  \BibitemOpen
  \bibfield  {author} {\bibinfo {author} {\bibfnamefont {S.~A.}\ \bibnamefont
  {Solla}}\ and\ \bibinfo {author} {\bibfnamefont {E.~K.}\ \bibnamefont
  {Riedel}},\ }\href {https://doi.org/10.1103/PhysRevB.23.6008} {\bibfield
  {journal} {\bibinfo  {journal} {Phys. Rev. B}\ }\textbf {\bibinfo {volume}
  {23}},\ \bibinfo {pages} {6008} (\bibinfo {year} {1981})}\BibitemShut
  {NoStop}%
\bibitem [{\citenamefont {Kosterlitz}(2016)}]{Kosterlitz2016}%
  \BibitemOpen
  \bibfield  {author} {\bibinfo {author} {\bibfnamefont {J.~M.}\ \bibnamefont
  {Kosterlitz}},\ }\href {https://doi.org/10.1088/0034-4885/79/2/026001}
  {\bibfield  {journal} {\bibinfo  {journal} {Rep. Prog. Phys.}\ }\textbf
  {\bibinfo {volume} {79}},\ \bibinfo {pages} {026001} (\bibinfo {year}
  {2016})}\BibitemShut {NoStop}%
\bibitem [{\citenamefont {Tagliacozzo}\ \emph {et~al.}(2008)\citenamefont
  {Tagliacozzo}, \citenamefont {de~Oliveira}, \citenamefont {Iblisdir},\ and\
  \citenamefont {Latorre}}]{Tagliacozzo2008}%
  \BibitemOpen
  \bibfield  {author} {\bibinfo {author} {\bibfnamefont {L.}~\bibnamefont
  {Tagliacozzo}}, \bibinfo {author} {\bibfnamefont {T.~R.}\ \bibnamefont
  {de~Oliveira}}, \bibinfo {author} {\bibfnamefont {S.}~\bibnamefont
  {Iblisdir}},\ and\ \bibinfo {author} {\bibfnamefont {J.~I.}\ \bibnamefont
  {Latorre}},\ }\href {https://doi.org/10.1103/PhysRevB.78.024410} {\bibfield
  {journal} {\bibinfo  {journal} {Phys. Rev. B}\ }\textbf {\bibinfo {volume}
  {78}},\ \bibinfo {pages} {024410} (\bibinfo {year} {2008})}\BibitemShut
  {NoStop}%
\bibitem [{\citenamefont {Pirvu}\ \emph {et~al.}(2012)\citenamefont {Pirvu},
  \citenamefont {Vidal}, \citenamefont {Verstraete},\ and\ \citenamefont
  {Tagliacozzo}}]{Pirvu2012}%
  \BibitemOpen
  \bibfield  {author} {\bibinfo {author} {\bibfnamefont {B.}~\bibnamefont
  {Pirvu}}, \bibinfo {author} {\bibfnamefont {G.}~\bibnamefont {Vidal}},
  \bibinfo {author} {\bibfnamefont {F.}~\bibnamefont {Verstraete}},\ and\
  \bibinfo {author} {\bibfnamefont {L.}~\bibnamefont {Tagliacozzo}},\ }\href
  {https://doi.org/10.1103/PhysRevB.86.075117} {\bibfield  {journal} {\bibinfo
  {journal} {Phys. Rev. B}\ }\textbf {\bibinfo {volume} {86}},\ \bibinfo
  {pages} {075117} (\bibinfo {year} {2012})}\BibitemShut {NoStop}%
\bibitem [{\citenamefont {Jos\'e}\ \emph {et~al.}(1977)\citenamefont {Jos\'e},
  \citenamefont {Kadanoff}, \citenamefont {Kirkpatrick},\ and\ \citenamefont
  {Nelson}}]{Jose1977}%
  \BibitemOpen
  \bibfield  {author} {\bibinfo {author} {\bibfnamefont {J.~V.}\ \bibnamefont
  {Jos\'e}}, \bibinfo {author} {\bibfnamefont {L.~P.}\ \bibnamefont
  {Kadanoff}}, \bibinfo {author} {\bibfnamefont {S.}~\bibnamefont
  {Kirkpatrick}},\ and\ \bibinfo {author} {\bibfnamefont {D.~R.}\ \bibnamefont
  {Nelson}},\ }\href {https://doi.org/10.1103/PhysRevB.16.1217} {\bibfield
  {journal} {\bibinfo  {journal} {Phys. Rev. B}\ }\textbf {\bibinfo {volume}
  {16}},\ \bibinfo {pages} {1217} (\bibinfo {year} {1977})}\BibitemShut
  {NoStop}%
\bibitem [{\citenamefont {Elitzur}\ \emph {et~al.}(1979)\citenamefont
  {Elitzur}, \citenamefont {Pearson},\ and\ \citenamefont
  {Shigemitsu}}]{Elitzur1979}%
  \BibitemOpen
  \bibfield  {author} {\bibinfo {author} {\bibfnamefont {S.}~\bibnamefont
  {Elitzur}}, \bibinfo {author} {\bibfnamefont {R.~B.}\ \bibnamefont
  {Pearson}},\ and\ \bibinfo {author} {\bibfnamefont {J.}~\bibnamefont
  {Shigemitsu}},\ }\href {https://doi.org/10.1103/PhysRevD.19.3698} {\bibfield
  {journal} {\bibinfo  {journal} {Phys. Rev. D}\ }\textbf {\bibinfo {volume}
  {19}},\ \bibinfo {pages} {3698} (\bibinfo {year} {1979})}\BibitemShut
  {NoStop}%
\bibitem [{\citenamefont {Einhorn}\ \emph {et~al.}(1980)\citenamefont
  {Einhorn}, \citenamefont {Savit},\ and\ \citenamefont
  {Rabinovici}}]{Einhorn1980}%
  \BibitemOpen
  \bibfield  {author} {\bibinfo {author} {\bibfnamefont {M.~B.}\ \bibnamefont
  {Einhorn}}, \bibinfo {author} {\bibfnamefont {R.}~\bibnamefont {Savit}},\
  and\ \bibinfo {author} {\bibfnamefont {E.}~\bibnamefont {Rabinovici}},\
  }\href {https://doi.org/https://doi.org/10.1016/0550-3213(80)90473-3}
  {\bibfield  {journal} {\bibinfo  {journal} {Nucl. Phys. B}\ }\textbf
  {\bibinfo {volume} {170}},\ \bibinfo {pages} {16} (\bibinfo {year}
  {1980})}\BibitemShut {NoStop}%
\bibitem [{\citenamefont {Lapilli}\ \emph {et~al.}(2006)\citenamefont
  {Lapilli}, \citenamefont {Pfeifer},\ and\ \citenamefont
  {Wexler}}]{Lapilli2006}%
  \BibitemOpen
  \bibfield  {author} {\bibinfo {author} {\bibfnamefont {C.~M.}\ \bibnamefont
  {Lapilli}}, \bibinfo {author} {\bibfnamefont {P.}~\bibnamefont {Pfeifer}},\
  and\ \bibinfo {author} {\bibfnamefont {C.}~\bibnamefont {Wexler}},\ }\href
  {https://doi.org/10.1103/PhysRevLett.96.140603} {\bibfield  {journal}
  {\bibinfo  {journal} {Phys. Rev. Lett.}\ }\textbf {\bibinfo {volume} {96}},\
  \bibinfo {pages} {140603} (\bibinfo {year} {2006})}\BibitemShut {NoStop}%
\bibitem [{\citenamefont {Ortiz}\ \emph {et~al.}(2012)\citenamefont {Ortiz},
  \citenamefont {Cobanera},\ and\ \citenamefont {Nussinov}}]{Ortiz2012}%
  \BibitemOpen
  \bibfield  {author} {\bibinfo {author} {\bibfnamefont {G.}~\bibnamefont
  {Ortiz}}, \bibinfo {author} {\bibfnamefont {E.}~\bibnamefont {Cobanera}},\
  and\ \bibinfo {author} {\bibfnamefont {Z.}~\bibnamefont {Nussinov}},\ }\href
  {https://doi.org/https://doi.org/10.1016/j.nuclphysb.2011.09.012} {\bibfield
  {journal} {\bibinfo  {journal} {Nucl. Phys. B}\ }\textbf {\bibinfo {volume}
  {854}},\ \bibinfo {pages} {780} (\bibinfo {year} {2012})}\BibitemShut
  {NoStop}%
\bibitem [{\citenamefont {Li}\ \emph {et~al.}(2020)\citenamefont {Li},
  \citenamefont {Yang}, \citenamefont {Xie}, \citenamefont {Tu}, \citenamefont
  {Liao},\ and\ \citenamefont {Xiang}}]{Li2020}%
  \BibitemOpen
  \bibfield  {author} {\bibinfo {author} {\bibfnamefont {Z.-Q.}\ \bibnamefont
  {Li}}, \bibinfo {author} {\bibfnamefont {L.-P.}\ \bibnamefont {Yang}},
  \bibinfo {author} {\bibfnamefont {Z.~Y.}\ \bibnamefont {Xie}}, \bibinfo
  {author} {\bibfnamefont {H.-H.}\ \bibnamefont {Tu}}, \bibinfo {author}
  {\bibfnamefont {H.-J.}\ \bibnamefont {Liao}},\ and\ \bibinfo {author}
  {\bibfnamefont {T.}~\bibnamefont {Xiang}},\ }\href
  {https://doi.org/10.1103/PhysRevE.101.060105} {\bibfield  {journal} {\bibinfo
   {journal} {Phys. Rev. E}\ }\textbf {\bibinfo {volume} {101}},\ \bibinfo
  {pages} {060105} (\bibinfo {year} {2020})}\BibitemShut {NoStop}%
\bibitem [{\citenamefont {Kramers}\ and\ \citenamefont
  {Wannier}(1941)}]{Kramers1941}%
  \BibitemOpen
  \bibfield  {author} {\bibinfo {author} {\bibfnamefont {H.~A.}\ \bibnamefont
  {Kramers}}\ and\ \bibinfo {author} {\bibfnamefont {G.~H.}\ \bibnamefont
  {Wannier}},\ }\href {https://doi.org/10.1103/PhysRev.60.252} {\bibfield
  {journal} {\bibinfo  {journal} {Phys. Rev.}\ }\textbf {\bibinfo {volume}
  {60}},\ \bibinfo {pages} {252} (\bibinfo {year} {1941})}\BibitemShut
  {NoStop}%
\bibitem [{\citenamefont {Migdal}(1975)}]{Migdal1975}%
  \BibitemOpen
  \bibfield  {author} {\bibinfo {author} {\bibfnamefont {A.}~\bibnamefont
  {Migdal}},\ }\href {http://jetp.ras.ru/cgi-bin/dn/e_042_04_0743.pdf}
  {\bibfield  {journal} {\bibinfo  {journal} {Zh. Eksp. Teor. Fiz}\ }\textbf
  {\bibinfo {volume} {69}},\ \bibinfo {pages} {1457} (\bibinfo {year}
  {1975})}\BibitemShut {NoStop}%
\bibitem [{\citenamefont {Kadanoff}(1976)}]{Kadanoff1976}%
  \BibitemOpen
  \bibfield  {author} {\bibinfo {author} {\bibfnamefont {L.~P.}\ \bibnamefont
  {Kadanoff}},\ }\href
  {https://www.sciencedirect.com/science/article/pii/000349167690066X}
  {\bibfield  {journal} {\bibinfo  {journal} {Ann. Phys.}\ }\textbf {\bibinfo
  {volume} {100}},\ \bibinfo {pages} {359} (\bibinfo {year}
  {1976})}\BibitemShut {NoStop}%
\bibitem [{\citenamefont {Spi{\v{s}}{\'a}k}(1993)}]{Spivsak1993}%
  \BibitemOpen
  \bibfield  {author} {\bibinfo {author} {\bibfnamefont {D.}~\bibnamefont
  {Spi{\v{s}}{\'a}k}},\ }\href
  {https://www.sciencedirect.com/science/article/pii/092145269390201G}
  {\bibfield  {journal} {\bibinfo  {journal} {Physica B}\ }\textbf {\bibinfo
  {volume} {190}},\ \bibinfo {pages} {407} (\bibinfo {year}
  {1993})}\BibitemShut {NoStop}%
\bibitem [{\citenamefont {Wolff}(1989)}]{Wolff1989}%
  \BibitemOpen
  \bibfield  {author} {\bibinfo {author} {\bibfnamefont {U.}~\bibnamefont
  {Wolff}},\ }\href {https://doi.org/10.1103/PhysRevLett.62.361} {\bibfield
  {journal} {\bibinfo  {journal} {Phys. Rev. Lett.}\ }\textbf {\bibinfo
  {volume} {62}},\ \bibinfo {pages} {361} (\bibinfo {year} {1989})}\BibitemShut
  {NoStop}%
\bibitem [{\citenamefont {Gubernatis}\ \emph {et~al.}(2016)\citenamefont
  {Gubernatis}, \citenamefont {Kawashima},\ and\ \citenamefont
  {Werner}}]{Gubernatis2016}%
  \BibitemOpen
  \bibfield  {author} {\bibinfo {author} {\bibfnamefont {J.}~\bibnamefont
  {Gubernatis}}, \bibinfo {author} {\bibfnamefont {N.}~\bibnamefont
  {Kawashima}},\ and\ \bibinfo {author} {\bibfnamefont {P.}~\bibnamefont
  {Werner}},\ }\href {https://doi.org/10.1017/cbo9780511902581} {\emph
  {\bibinfo {title} {Quantum Monte Carlo Methods}}}\ (\bibinfo  {publisher}
  {Cambridge University Press},\ \bibinfo {year} {2016})\BibitemShut {NoStop}%
\bibitem [{\citenamefont {Suderow}\ \emph {et~al.}(2014)\citenamefont
  {Suderow}, \citenamefont {Guillamón}, \citenamefont {Rodrigo},\ and\
  \citenamefont {Vieira}}]{Suderow2014}%
  \BibitemOpen
  \bibfield  {author} {\bibinfo {author} {\bibfnamefont {H.}~\bibnamefont
  {Suderow}}, \bibinfo {author} {\bibfnamefont {I.}~\bibnamefont {Guillamón}},
  \bibinfo {author} {\bibfnamefont {J.~G.}\ \bibnamefont {Rodrigo}},\ and\
  \bibinfo {author} {\bibfnamefont {S.}~\bibnamefont {Vieira}},\ }\href
  {https://doi.org/10.1088/0953-2048/27/6/063001} {\bibfield  {journal}
  {\bibinfo  {journal} {Supercond. Sci. Technol.}\ }\textbf {\bibinfo {volume}
  {27}},\ \bibinfo {pages} {063001} (\bibinfo {year} {2014})}\BibitemShut
  {NoStop}%
\bibitem [{\citenamefont {Ge}\ \emph {et~al.}(2016)\citenamefont {Ge},
  \citenamefont {Gladilin}, \citenamefont {Tempere}, \citenamefont {Xue},
  \citenamefont {Devreese}, \citenamefont {Van~de Vondel}, \citenamefont
  {Zhou},\ and\ \citenamefont {Moshchalkov}}]{Ge2016}%
  \BibitemOpen
  \bibfield  {author} {\bibinfo {author} {\bibfnamefont {J.-Y.}\ \bibnamefont
  {Ge}}, \bibinfo {author} {\bibfnamefont {V.~N.}\ \bibnamefont {Gladilin}},
  \bibinfo {author} {\bibfnamefont {J.}~\bibnamefont {Tempere}}, \bibinfo
  {author} {\bibfnamefont {C.}~\bibnamefont {Xue}}, \bibinfo {author}
  {\bibfnamefont {J.~T.}\ \bibnamefont {Devreese}}, \bibinfo {author}
  {\bibfnamefont {J.}~\bibnamefont {Van~de Vondel}}, \bibinfo {author}
  {\bibfnamefont {Y.}~\bibnamefont {Zhou}},\ and\ \bibinfo {author}
  {\bibfnamefont {V.~V.}\ \bibnamefont {Moshchalkov}},\ }\href
  {https://doi.org/10.1038/ncomms13880} {\bibfield  {journal} {\bibinfo
  {journal} {Nat. Commun.}\ }\textbf {\bibinfo {volume} {7}},\ \bibinfo {pages}
  {13880} (\bibinfo {year} {2016})}\BibitemShut {NoStop}%
\bibitem [{\citenamefont {Pan}\ \emph {et~al.}(2001)\citenamefont {Pan},
  \citenamefont {O'Neal}, \citenamefont {Badzey}, \citenamefont {Chamon},
  \citenamefont {Ding}, \citenamefont {Engelbrecht}, \citenamefont {Wang},
  \citenamefont {Eisaki}, \citenamefont {Uchida}, \citenamefont {Gupta},
  \citenamefont {Hudson}, \citenamefont {Lang},\ and\ \citenamefont
  {Davis}}]{Pan2001}%
  \BibitemOpen
  \bibfield  {author} {\bibinfo {author} {\bibfnamefont {S.~H.}\ \bibnamefont
  {Pan}}, \bibinfo {author} {\bibfnamefont {J.~P.}\ \bibnamefont {O'Neal}},
  \bibinfo {author} {\bibfnamefont {R.~L.}\ \bibnamefont {Badzey}}, \bibinfo
  {author} {\bibfnamefont {C.}~\bibnamefont {Chamon}}, \bibinfo {author}
  {\bibfnamefont {H.}~\bibnamefont {Ding}}, \bibinfo {author} {\bibfnamefont
  {J.~R.}\ \bibnamefont {Engelbrecht}}, \bibinfo {author} {\bibfnamefont
  {Z.}~\bibnamefont {Wang}}, \bibinfo {author} {\bibfnamefont {H.}~\bibnamefont
  {Eisaki}}, \bibinfo {author} {\bibfnamefont {S.-i.}\ \bibnamefont {Uchida}},
  \bibinfo {author} {\bibfnamefont {A.~K.}\ \bibnamefont {Gupta}}, \bibinfo
  {author} {\bibfnamefont {E.~W.}\ \bibnamefont {Hudson}}, \bibinfo {author}
  {\bibfnamefont {K.~M.}\ \bibnamefont {Lang}},\ and\ \bibinfo {author}
  {\bibfnamefont {J.~C.}\ \bibnamefont {Davis}},\ }\href
  {https://www.nature.com/articles/35095012} {\bibfield  {journal} {\bibinfo
  {journal} {Nature}\ }\textbf {\bibinfo {volume} {413}},\ \bibinfo {pages}
  {282} (\bibinfo {year} {2001})}\BibitemShut {NoStop}%
\bibitem [{\citenamefont {McElroy}\ \emph {et~al.}(2005)\citenamefont
  {McElroy}, \citenamefont {Lee}, \citenamefont {Slezak}, \citenamefont {Lee},
  \citenamefont {Eisaki}, \citenamefont {Uchida},\ and\ \citenamefont
  {Davis}}]{McElroy2005}%
  \BibitemOpen
  \bibfield  {author} {\bibinfo {author} {\bibfnamefont {K.}~\bibnamefont
  {McElroy}}, \bibinfo {author} {\bibfnamefont {J.}~\bibnamefont {Lee}},
  \bibinfo {author} {\bibfnamefont {J.}~\bibnamefont {Slezak}}, \bibinfo
  {author} {\bibfnamefont {D.-H.}\ \bibnamefont {Lee}}, \bibinfo {author}
  {\bibfnamefont {H.}~\bibnamefont {Eisaki}}, \bibinfo {author} {\bibfnamefont
  {S.}~\bibnamefont {Uchida}},\ and\ \bibinfo {author} {\bibfnamefont {J.~C.}\
  \bibnamefont {Davis}},\ }\href
  {https://www.science.org/doi/10.1126/science.1113095} {\bibfield  {journal}
  {\bibinfo  {journal} {Science}\ }\textbf {\bibinfo {volume} {309}},\ \bibinfo
  {pages} {1048} (\bibinfo {year} {2005})}\BibitemShut {NoStop}%
\bibitem [{\citenamefont {Campi}\ \emph {et~al.}(2015)\citenamefont {Campi},
  \citenamefont {Bianconi}, \citenamefont {Poccia}, \citenamefont {Bianconi},
  \citenamefont {Barba}, \citenamefont {Arrighetti}, \citenamefont {Innocenti},
  \citenamefont {Karpinski}, \citenamefont {Zhigadlo}, \citenamefont {Kazakov}
  \emph {et~al.}}]{Campi2015}%
  \BibitemOpen
  \bibfield  {author} {\bibinfo {author} {\bibfnamefont {G.}~\bibnamefont
  {Campi}}, \bibinfo {author} {\bibfnamefont {A.}~\bibnamefont {Bianconi}},
  \bibinfo {author} {\bibfnamefont {N.}~\bibnamefont {Poccia}}, \bibinfo
  {author} {\bibfnamefont {G.}~\bibnamefont {Bianconi}}, \bibinfo {author}
  {\bibfnamefont {L.}~\bibnamefont {Barba}}, \bibinfo {author} {\bibfnamefont
  {G.}~\bibnamefont {Arrighetti}}, \bibinfo {author} {\bibfnamefont
  {D.}~\bibnamefont {Innocenti}}, \bibinfo {author} {\bibfnamefont
  {J.}~\bibnamefont {Karpinski}}, \bibinfo {author} {\bibfnamefont {N.~D.}\
  \bibnamefont {Zhigadlo}}, \bibinfo {author} {\bibfnamefont {S.~M.}\
  \bibnamefont {Kazakov}}, \emph {et~al.},\ }\href
  {https://www.nature.com/articles/nature14987} {\bibfield  {journal} {\bibinfo
   {journal} {Nature}\ }\textbf {\bibinfo {volume} {525}},\ \bibinfo {pages}
  {359} (\bibinfo {year} {2015})}\BibitemShut {NoStop}%
\bibitem [{\citenamefont {Sinha}\ and\ \citenamefont
  {Wietek}(2024)}]{Sinha2024}%
  \BibitemOpen
  \bibfield  {author} {\bibinfo {author} {\bibfnamefont {A.}~\bibnamefont
  {Sinha}}\ and\ \bibinfo {author} {\bibfnamefont {A.}~\bibnamefont {Wietek}},\
  }\href@noop {} {\bibinfo {title} {Forestalled phase separation as the
  precursor to stripe order}} (\bibinfo {year} {2024}),\ \Eprint
  {https://arxiv.org/abs/arXiv:2411.15158} {arXiv:2411.15158} \BibitemShut
  {NoStop}%
\end{thebibliography}%


\begin{thebibliography}{20}%
\makeatletter
\providecommand \@ifxundefined [1]{%
 \@ifx{#1\undefined}
}%
\providecommand \@ifnum [1]{%
 \ifnum #1\expandafter \@firstoftwo
 \else \expandafter \@secondoftwo
 \fi
}%
\providecommand \@ifx [1]{%
 \ifx #1\expandafter \@firstoftwo
 \else \expandafter \@secondoftwo
 \fi
}%
\providecommand \natexlab [1]{#1}%
\providecommand \enquote  [1]{``#1''}%
\providecommand \bibnamefont  [1]{#1}%
\providecommand \bibfnamefont [1]{#1}%
\providecommand \citenamefont [1]{#1}%
\providecommand \href@noop [0]{\@secondoftwo}%
\providecommand \href [0]{\begingroup \@sanitize@url \@href}%
\providecommand \@href[1]{\@@startlink{#1}\@@href}%
\providecommand \@@href[1]{\endgroup#1\@@endlink}%
\providecommand \@sanitize@url [0]{\catcode `\\12\catcode `\$12\catcode
  `\&12\catcode `\#12\catcode `\^12\catcode `\_12\catcode `\%12\relax}%
\providecommand \@@startlink[1]{}%
\providecommand \@@endlink[0]{}%
\providecommand \url  [0]{\begingroup\@sanitize@url \@url }%
\providecommand \@url [1]{\endgroup\@href {#1}{\urlprefix }}%
\providecommand \urlprefix  [0]{URL }%
\providecommand \Eprint [0]{\href }%
\providecommand \doibase [0]{https://doi.org/}%
\providecommand \selectlanguage [0]{\@gobble}%
\providecommand \bibinfo  [0]{\@secondoftwo}%
\providecommand \bibfield  [0]{\@secondoftwo}%
\providecommand \translation [1]{[#1]}%
\providecommand \BibitemOpen [0]{}%
\providecommand \bibitemStop [0]{}%
\providecommand \bibitemNoStop [0]{.\EOS\space}%
\providecommand \EOS [0]{\spacefactor3000\relax}%
\providecommand \BibitemShut  [1]{\csname bibitem#1\endcsname}%
\let\auto@bib@innerbib\@empty
\bibitem [{\citenamefont {Baldelli}\ \emph {et~al.}(2025)\citenamefont
  {Baldelli}, \citenamefont {Karlsson}, \citenamefont {Kloss}, \citenamefont
  {Fishman},\ and\ \citenamefont {Wietek}}]{Baldelli2025}%
  \BibitemOpen
  \bibfield  {author} {\bibinfo {author} {\bibfnamefont {N.}~\bibnamefont
  {Baldelli}}, \bibinfo {author} {\bibfnamefont {H.}~\bibnamefont {Karlsson}},
  \bibinfo {author} {\bibfnamefont {B.}~\bibnamefont {Kloss}}, \bibinfo
  {author} {\bibfnamefont {M.}~\bibnamefont {Fishman}},\ and\ \bibinfo {author}
  {\bibfnamefont {A.}~\bibnamefont {Wietek}},\ }\href
  {https://doi.org/10.1038/s41535-024-00718-3} {\bibfield  {journal} {\bibinfo
  {journal} {npj Quantum Mater.}\ }\textbf {\bibinfo {volume} {10}},\ \bibinfo
  {pages} {22} (\bibinfo {year} {2025})}\BibitemShut {NoStop}%
\bibitem [{\citenamefont {Yu}\ \emph {et~al.}(2014)\citenamefont {Yu},
  \citenamefont {Xie}, \citenamefont {Meurice}, \citenamefont {Liu},
  \citenamefont {Denbleyker}, \citenamefont {Zou}, \citenamefont {Qin},
  \citenamefont {Chen},\ and\ \citenamefont {Xiang}}]{Yu2014}%
  \BibitemOpen
  \bibfield  {author} {\bibinfo {author} {\bibfnamefont {J.~F.}\ \bibnamefont
  {Yu}}, \bibinfo {author} {\bibfnamefont {Z.~Y.}\ \bibnamefont {Xie}},
  \bibinfo {author} {\bibfnamefont {Y.}~\bibnamefont {Meurice}}, \bibinfo
  {author} {\bibfnamefont {Y.}~\bibnamefont {Liu}}, \bibinfo {author}
  {\bibfnamefont {A.}~\bibnamefont {Denbleyker}}, \bibinfo {author}
  {\bibfnamefont {H.}~\bibnamefont {Zou}}, \bibinfo {author} {\bibfnamefont
  {M.~P.}\ \bibnamefont {Qin}}, \bibinfo {author} {\bibfnamefont
  {J.}~\bibnamefont {Chen}},\ and\ \bibinfo {author} {\bibfnamefont
  {T.}~\bibnamefont {Xiang}},\ }\href
  {https://doi.org/10.1103/PhysRevE.89.013308} {\bibfield  {journal} {\bibinfo
  {journal} {Phys. Rev. E}\ }\textbf {\bibinfo {volume} {89}},\ \bibinfo
  {pages} {013308} (\bibinfo {year} {2014})}\BibitemShut {NoStop}%
\bibitem [{\citenamefont {Vanderstraeten}\ \emph
  {et~al.}(2019{\natexlab{a}})\citenamefont {Vanderstraeten}, \citenamefont
  {Vanhecke}, \citenamefont {L\"auchli},\ and\ \citenamefont
  {Verstraete}}]{Vanderstraeten2019}%
  \BibitemOpen
  \bibfield  {author} {\bibinfo {author} {\bibfnamefont {L.}~\bibnamefont
  {Vanderstraeten}}, \bibinfo {author} {\bibfnamefont {B.}~\bibnamefont
  {Vanhecke}}, \bibinfo {author} {\bibfnamefont {A.~M.}\ \bibnamefont
  {L\"auchli}},\ and\ \bibinfo {author} {\bibfnamefont {F.}~\bibnamefont
  {Verstraete}},\ }\href {https://doi.org/10.1103/PhysRevE.100.062136}
  {\bibfield  {journal} {\bibinfo  {journal} {Phys. Rev. E}\ }\textbf {\bibinfo
  {volume} {100}},\ \bibinfo {pages} {062136} (\bibinfo {year}
  {2019}{\natexlab{a}})}\BibitemShut {NoStop}%
\bibitem [{\citenamefont {Song}\ and\ \citenamefont {Zhang}(2021)}]{Song2021}%
  \BibitemOpen
  \bibfield  {author} {\bibinfo {author} {\bibfnamefont {F.-F.}\ \bibnamefont
  {Song}}\ and\ \bibinfo {author} {\bibfnamefont {G.-M.}\ \bibnamefont
  {Zhang}},\ }\href {https://doi.org/10.1103/PhysRevB.103.024518} {\bibfield
  {journal} {\bibinfo  {journal} {Phys. Rev. B}\ }\textbf {\bibinfo {volume}
  {103}},\ \bibinfo {pages} {024518} (\bibinfo {year} {2021})}\BibitemShut
  {NoStop}%
\bibitem [{\citenamefont {Haegeman}\ and\ \citenamefont
  {Verstraete}(2017)}]{Haegeman2017}%
  \BibitemOpen
  \bibfield  {author} {\bibinfo {author} {\bibfnamefont {J.}~\bibnamefont
  {Haegeman}}\ and\ \bibinfo {author} {\bibfnamefont {F.}~\bibnamefont
  {Verstraete}},\ }\href
  {https://doi.org/https://doi.org/10.1146/annurev-conmatphys-031016-025507}
  {\bibfield  {journal} {\bibinfo  {journal} {Annu. Rev. Condens. Matter
  Phys.}\ }\textbf {\bibinfo {volume} {8}},\ \bibinfo {pages} {355} (\bibinfo
  {year} {2017})}\BibitemShut {NoStop}%
\bibitem [{\citenamefont {Zauner-Stauber}\ \emph {et~al.}(2018)\citenamefont
  {Zauner-Stauber}, \citenamefont {Vanderstraeten}, \citenamefont {Fishman},
  \citenamefont {Verstraete},\ and\ \citenamefont
  {Haegeman}}]{Zauner-Stauber2018}%
  \BibitemOpen
  \bibfield  {author} {\bibinfo {author} {\bibfnamefont {V.}~\bibnamefont
  {Zauner-Stauber}}, \bibinfo {author} {\bibfnamefont {L.}~\bibnamefont
  {Vanderstraeten}}, \bibinfo {author} {\bibfnamefont {M.~T.}\ \bibnamefont
  {Fishman}}, \bibinfo {author} {\bibfnamefont {F.}~\bibnamefont
  {Verstraete}},\ and\ \bibinfo {author} {\bibfnamefont {J.}~\bibnamefont
  {Haegeman}},\ }\href {https://doi.org/10.1103/PhysRevB.97.045145} {\bibfield
  {journal} {\bibinfo  {journal} {Phys. Rev. B}\ }\textbf {\bibinfo {volume}
  {97}},\ \bibinfo {pages} {045145} (\bibinfo {year} {2018})}\BibitemShut
  {NoStop}%
\bibitem [{\citenamefont {Vanderstraeten}\ \emph
  {et~al.}(2019{\natexlab{b}})\citenamefont {Vanderstraeten}, \citenamefont
  {Haegeman},\ and\ \citenamefont {Verstraete}}]{Vanderstraeten2019tan}%
  \BibitemOpen
  \bibfield  {author} {\bibinfo {author} {\bibfnamefont {L.}~\bibnamefont
  {Vanderstraeten}}, \bibinfo {author} {\bibfnamefont {J.}~\bibnamefont
  {Haegeman}},\ and\ \bibinfo {author} {\bibfnamefont {F.}~\bibnamefont
  {Verstraete}},\ }\href {https://doi.org/10.21468/SciPostPhysLectNotes.7}
  {\bibfield  {journal} {\bibinfo  {journal} {SciPost Phys. Lect. Notes}\ ,\
  \bibinfo {pages} {7}} (\bibinfo {year} {2019}{\natexlab{b}})}\BibitemShut
  {NoStop}%
\bibitem [{\citenamefont {Nietner}\ \emph {et~al.}(2020)\citenamefont
  {Nietner}, \citenamefont {Vanhecke}, \citenamefont {Verstraete},
  \citenamefont {Eisert},\ and\ \citenamefont {Vanderstraeten}}]{Nietner2020}%
  \BibitemOpen
  \bibfield  {author} {\bibinfo {author} {\bibfnamefont {A.}~\bibnamefont
  {Nietner}}, \bibinfo {author} {\bibfnamefont {B.}~\bibnamefont {Vanhecke}},
  \bibinfo {author} {\bibfnamefont {F.}~\bibnamefont {Verstraete}}, \bibinfo
  {author} {\bibfnamefont {J.}~\bibnamefont {Eisert}},\ and\ \bibinfo {author}
  {\bibfnamefont {L.}~\bibnamefont {Vanderstraeten}},\ }\href
  {https://doi.org/10.22331/q-2020-09-21-328} {\bibfield  {journal} {\bibinfo
  {journal} {{Quantum}}\ }\textbf {\bibinfo {volume} {4}},\ \bibinfo {pages}
  {328} (\bibinfo {year} {2020})}\BibitemShut {NoStop}%
\bibitem [{\citenamefont {Vidal}\ \emph {et~al.}(2003)\citenamefont {Vidal},
  \citenamefont {Latorre}, \citenamefont {Rico},\ and\ \citenamefont
  {Kitaev}}]{Vidal2003}%
  \BibitemOpen
  \bibfield  {author} {\bibinfo {author} {\bibfnamefont {G.}~\bibnamefont
  {Vidal}}, \bibinfo {author} {\bibfnamefont {J.~I.}\ \bibnamefont {Latorre}},
  \bibinfo {author} {\bibfnamefont {E.}~\bibnamefont {Rico}},\ and\ \bibinfo
  {author} {\bibfnamefont {A.}~\bibnamefont {Kitaev}},\ }\href
  {https://doi.org/10.1103/PhysRevLett.90.227902} {\bibfield  {journal}
  {\bibinfo  {journal} {Phys. Rev. Lett.}\ }\textbf {\bibinfo {volume} {90}},\
  \bibinfo {pages} {227902} (\bibinfo {year} {2003})}\BibitemShut {NoStop}%
\bibitem [{\citenamefont {Pollmann}\ \emph {et~al.}(2009)\citenamefont
  {Pollmann}, \citenamefont {Mukerjee}, \citenamefont {Turner},\ and\
  \citenamefont {Moore}}]{Pollmann2009}%
  \BibitemOpen
  \bibfield  {author} {\bibinfo {author} {\bibfnamefont {F.}~\bibnamefont
  {Pollmann}}, \bibinfo {author} {\bibfnamefont {S.}~\bibnamefont {Mukerjee}},
  \bibinfo {author} {\bibfnamefont {A.~M.}\ \bibnamefont {Turner}},\ and\
  \bibinfo {author} {\bibfnamefont {J.~E.}\ \bibnamefont {Moore}},\ }\href
  {https://doi.org/10.1103/PhysRevLett.102.255701} {\bibfield  {journal}
  {\bibinfo  {journal} {Phys. Rev. Lett.}\ }\textbf {\bibinfo {volume} {102}},\
  \bibinfo {pages} {255701} (\bibinfo {year} {2009})}\BibitemShut {NoStop}%
\bibitem [{\citenamefont {Song}\ and\ \citenamefont
  {Zhang}(2022{\natexlab{a}})}]{Song2022}%
  \BibitemOpen
  \bibfield  {author} {\bibinfo {author} {\bibfnamefont {F.-F.}\ \bibnamefont
  {Song}}\ and\ \bibinfo {author} {\bibfnamefont {G.-M.}\ \bibnamefont
  {Zhang}},\ }\href {https://doi.org/10.1103/PhysRevLett.128.195301} {\bibfield
   {journal} {\bibinfo  {journal} {Phys. Rev. Lett.}\ }\textbf {\bibinfo
  {volume} {128}},\ \bibinfo {pages} {195301} (\bibinfo {year}
  {2022}{\natexlab{a}})}\BibitemShut {NoStop}%
\bibitem [{\citenamefont {Song}\ and\ \citenamefont
  {Zhang}(2022{\natexlab{b}})}]{Song2022ffxy}%
  \BibitemOpen
  \bibfield  {author} {\bibinfo {author} {\bibfnamefont {F.-F.}\ \bibnamefont
  {Song}}\ and\ \bibinfo {author} {\bibfnamefont {G.-M.}\ \bibnamefont
  {Zhang}},\ }\href {https://doi.org/10.1103/PhysRevB.105.134516} {\bibfield
  {journal} {\bibinfo  {journal} {Phys. Rev. B}\ }\textbf {\bibinfo {volume}
  {105}},\ \bibinfo {pages} {134516} (\bibinfo {year}
  {2022}{\natexlab{b}})}\BibitemShut {NoStop}%
\bibitem [{\citenamefont {Song}\ and\ \citenamefont {Zhang}(2023)}]{Song2023}%
  \BibitemOpen
  \bibfield  {author} {\bibinfo {author} {\bibfnamefont {F.-F.}\ \bibnamefont
  {Song}}\ and\ \bibinfo {author} {\bibfnamefont {G.-M.}\ \bibnamefont
  {Zhang}},\ }\href {https://doi.org/10.1103/PhysRevB.108.014424} {\bibfield
  {journal} {\bibinfo  {journal} {Phys. Rev. B}\ }\textbf {\bibinfo {volume}
  {108}},\ \bibinfo {pages} {014424} (\bibinfo {year} {2023})}\BibitemShut
  {NoStop}%
\bibitem [{\citenamefont {Morita}\ and\ \citenamefont
  {Kawashima}(2019)}]{Morita2019}%
  \BibitemOpen
  \bibfield  {author} {\bibinfo {author} {\bibfnamefont {S.}~\bibnamefont
  {Morita}}\ and\ \bibinfo {author} {\bibfnamefont {N.}~\bibnamefont
  {Kawashima}},\ }\href
  {https://doi.org/https://doi.org/10.1016/j.cpc.2018.10.014} {\bibfield
  {journal} {\bibinfo  {journal} {Comput. Phys. Commun.}\ }\textbf {\bibinfo
  {volume} {236}},\ \bibinfo {pages} {65} (\bibinfo {year} {2019})}\BibitemShut
  {NoStop}%
\bibitem [{\citenamefont {Morita}\ and\ \citenamefont
  {Kawashima}(2025)}]{Morita2025}%
  \BibitemOpen
  \bibfield  {author} {\bibinfo {author} {\bibfnamefont {S.}~\bibnamefont
  {Morita}}\ and\ \bibinfo {author} {\bibfnamefont {N.}~\bibnamefont
  {Kawashima}},\ }\href {https://doi.org/10.1103/PhysRevB.111.054433}
  {\bibfield  {journal} {\bibinfo  {journal} {Phys. Rev. B}\ }\textbf {\bibinfo
  {volume} {111}},\ \bibinfo {pages} {054433} (\bibinfo {year}
  {2025})}\BibitemShut {NoStop}%
\bibitem [{\citenamefont {Kramers}\ and\ \citenamefont
  {Wannier}(1941)}]{Kramers1941}%
  \BibitemOpen
  \bibfield  {author} {\bibinfo {author} {\bibfnamefont {H.~A.}\ \bibnamefont
  {Kramers}}\ and\ \bibinfo {author} {\bibfnamefont {G.~H.}\ \bibnamefont
  {Wannier}},\ }\href {https://doi.org/10.1103/PhysRev.60.252} {\bibfield
  {journal} {\bibinfo  {journal} {Phys. Rev.}\ }\textbf {\bibinfo {volume}
  {60}},\ \bibinfo {pages} {252} (\bibinfo {year} {1941})}\BibitemShut
  {NoStop}%
\bibitem [{\citenamefont {Li}\ \emph {et~al.}(2020)\citenamefont {Li},
  \citenamefont {Yang}, \citenamefont {Xie}, \citenamefont {Tu}, \citenamefont
  {Liao},\ and\ \citenamefont {Xiang}}]{Li2020}%
  \BibitemOpen
  \bibfield  {author} {\bibinfo {author} {\bibfnamefont {Z.-Q.}\ \bibnamefont
  {Li}}, \bibinfo {author} {\bibfnamefont {L.-P.}\ \bibnamefont {Yang}},
  \bibinfo {author} {\bibfnamefont {Z.~Y.}\ \bibnamefont {Xie}}, \bibinfo
  {author} {\bibfnamefont {H.-H.}\ \bibnamefont {Tu}}, \bibinfo {author}
  {\bibfnamefont {H.-J.}\ \bibnamefont {Liao}},\ and\ \bibinfo {author}
  {\bibfnamefont {T.}~\bibnamefont {Xiang}},\ }\href
  {https://doi.org/10.1103/PhysRevE.101.060105} {\bibfield  {journal} {\bibinfo
   {journal} {Phys. Rev. E}\ }\textbf {\bibinfo {volume} {101}},\ \bibinfo
  {pages} {060105} (\bibinfo {year} {2020})}\BibitemShut {NoStop}%
\bibitem [{\citenamefont {Migdal}(1975)}]{Migdal1975}%
  \BibitemOpen
  \bibfield  {author} {\bibinfo {author} {\bibfnamefont {A.}~\bibnamefont
  {Migdal}},\ }\href {http://jetp.ras.ru/cgi-bin/dn/e_042_04_0743.pdf}
  {\bibfield  {journal} {\bibinfo  {journal} {Zh. Eksp. Teor. Fiz}\ }\textbf
  {\bibinfo {volume} {69}},\ \bibinfo {pages} {1457} (\bibinfo {year}
  {1975})}\BibitemShut {NoStop}%
\bibitem [{\citenamefont {Kadanoff}(1976)}]{Kadanoff1976}%
  \BibitemOpen
  \bibfield  {author} {\bibinfo {author} {\bibfnamefont {L.~P.}\ \bibnamefont
  {Kadanoff}},\ }\href
  {https://www.sciencedirect.com/science/article/pii/000349167690066X}
  {\bibfield  {journal} {\bibinfo  {journal} {Ann. Phys.}\ }\textbf {\bibinfo
  {volume} {100}},\ \bibinfo {pages} {359} (\bibinfo {year}
  {1976})}\BibitemShut {NoStop}%
\bibitem [{\citenamefont {Ortiz}\ \emph {et~al.}(2012)\citenamefont {Ortiz},
  \citenamefont {Cobanera},\ and\ \citenamefont {Nussinov}}]{Ortiz2012}%
  \BibitemOpen
  \bibfield  {author} {\bibinfo {author} {\bibfnamefont {G.}~\bibnamefont
  {Ortiz}}, \bibinfo {author} {\bibfnamefont {E.}~\bibnamefont {Cobanera}},\
  and\ \bibinfo {author} {\bibfnamefont {Z.}~\bibnamefont {Nussinov}},\ }\href
  {https://doi.org/https://doi.org/10.1016/j.nuclphysb.2011.09.012} {\bibfield
  {journal} {\bibinfo  {journal} {Nucl. Phys. B}\ }\textbf {\bibinfo {volume}
  {854}},\ \bibinfo {pages} {780} (\bibinfo {year} {2012})}\BibitemShut
  {NoStop}%
\end{thebibliography}%
\end{document}


\title{Supplemental Material for \\``Superfluid dome in the spatially modulated two-dimensional XY model''}

\author{Feng-Feng Song} 
\affiliation{Institute for Solid State Physics, The University of Tokyo, Kashiwa, Chiba 277-8581, Japan}

\author{Aditya Chugh}
\affiliation{Max Planck Institute for the Physics of Complex Systems, N\"{o}thnitzer Strasse 38, Dresden 01187, Germany}

\author{Hanggai Nuomin}
\affiliation{Department of Chemistry, Duke University, Durham, North Carolina 27708, United States}

\author{Naoki Kawashima}
\affiliation{Institute for Solid State Physics, The University of Tokyo, Kashiwa, Chiba 277-8581, Japan}
\affiliation{Trans-scale Quantum Science Institute, The University of Tokyo, Bunkyo, Tokyo 113-0033, Japan}

\author{Alexander Wietek} 
\affiliation{Max Planck Institute for the Physics of Complex Systems, N\"{o}thnitzer Strasse 38, Dresden 01187, Germany}

\date{\today}

\maketitle

\tableofcontents

\section{Intertwined Ginzburg-Landau Theory and Modulated XY Model}

The interplay between superconductivity (SC) and charge density wave (CDW) order can be effectively described by an intertwined Ginzburg-Landau (GL) theory~\cite{Baldelli2025}
\be
\label{eq:intertwinedgl}
\mathcal{F}[\psi] = \alpha(\bm{r})|\psi|^2 + \frac{\beta}{2} |\psi|^4 + \frac{1}{2m^*} \left| \left(-i \hbar \bm{\nabla} + 2e \bm{A}\right) \psi \right|^2,
\ee
where $\alpha(\bm{r})$ and $\beta$ are phenomenological expansion coefficients, $m^*$ is the effective mass of the Cooper pairs, and $\bm{A}$ is the electromagnetic vector potential. Here, the coefficient $\alpha(\bm{r})$ is assumed to be spatially modulated, obeying $\alpha(\bm{r}+\bm{\lambda}) = \alpha(\bm{r})$ for some modulation vector $\bm{\lambda}$.

To proceed, we separate the amplitude and phase of the SC order parameter via $\psi(\bm{r}) = \rho(\bm{r}) e^{i\theta(\bm{r})}$, where $\rho(\bm{r}) = |\psi(\bm{r})|$ and $\theta(\bm{r})$ is the local phase. Substituting this form into Eq.~\eqref{eq:intertwinedgl}, the free energy functional becomes
\be
\label{eq:GL-amp-phase}
\mathcal{F}[\rho, \theta] = \alpha(\bm{r}) \rho^2 + \frac{\beta}{2} \rho^4 + \frac{1}{2m^*} \left[ \hbar^2 (\nabla \rho)^2 + \rho^2 \left(\hbar \nabla \theta - 2e \bm{A}\right)^2 \right].
\ee

We now discretize the system on a 2D square lattice, labeling each site by $i$, with lattice constant $a$. The total free energy reads
\be
F = \int d^2 r\, \mathcal{F}[\rho, \theta] \approx \sum_i \alpha_i \rho_i^2 + \frac{\beta}{2} \rho_i^4 + \frac{\hbar^2}{2m^*} \left[ (\nabla \rho)^2 + \rho^2 \left(\nabla \theta - \frac{2e}{\hbar} \bm{A}\right)^2 \right],
\ee
where $\alpha_i = \alpha(\bm{r}_i)$ and $\rho_i = \rho(\bm{r}_i)$. The discrete gradient is implemented as $d\rho/dr \approx (\rho_{i+a} - \rho_i)/a$.

Explicitly evaluating the discrete derivatives and expanding the terms, we find
\begin{align}
F ={} & \sum_{\langle i, j\rangle} \alpha_i \rho_i^2 + \frac{\beta}{2} \rho_i^4 + \frac{\hbar^2}{2m^* a^2} 
\Big[ \rho_i^2 + \rho_j^2 - 2 \rho_i \rho_j \cos\left(\theta_i - \theta_j + \frac{2e}{\hbar} A a\right) \Big] \\
={} & \sum_{\langle i, j\rangle} \alpha_i' \rho_i^2 + \frac{\beta}{2} \rho_i^4 - J \rho_i \rho_j \cos\left(\theta_i - \theta_j + A_{ij}\right),
\end{align}
where $J = \hbar^2/(m^* a^2)$ and $A_{ij} = 2e A a / \hbar$.

In the regime where amplitude fluctuations are small compared to phase fluctuations (near the SC transition temperature), the GL theory can be mapped onto a modulated XY model on the lattice. The spatial modulation of the SC order parameter, induced by the modulated $\alpha(\bm{r})$, leads to a spatially varying coupling $J_{ij}=J\rho_i\rho_j$,
\be
\label{eq:cdwxy}
H = -\sum_{\langle i, j\rangle} J_{ij} \cos\left(\theta_i - \theta_j\right),
\ee
where $J_{ij} = J_0 [1 + \alpha \sin(k x_i)]$ encodes the modulation along the $x$-direction, with amplitude $\alpha$ and wavelength $\lambda = 2\pi / k$. Here, $x_i$ denotes the $x$-coordinate of the nearest-neighbor (NN) bond (the midpoint of the pair $\langle i, j\rangle$). The sum runs over all NN pairs $\langle i, j\rangle$, and $\theta_i$ is the $U(1)$ spin (phase) variable at site $i$. Note that the symbol $\alpha$ in $J_{ij}$ denotes a constant for modulation amplitude, distinct from the spatially varying GL coefficient $\alpha(\bm{r})$. Although both arise from the presence of the CDW, they correspond to different quantities in the theory. Since the influence of the CDW is predominantly along the $x$ axis, we set $J_{ij} = J_0$ for bonds along the $y$-direction in the main text. For completeness, including modulation along the $y$-direction does not qualitatively alter the main results, as will be shown in Sec.~\ref{sec:mJy}.

\section{Tensor network methods}
\begin{figure}[tbp]
    \centering
    \includegraphics[width=0.58\linewidth]{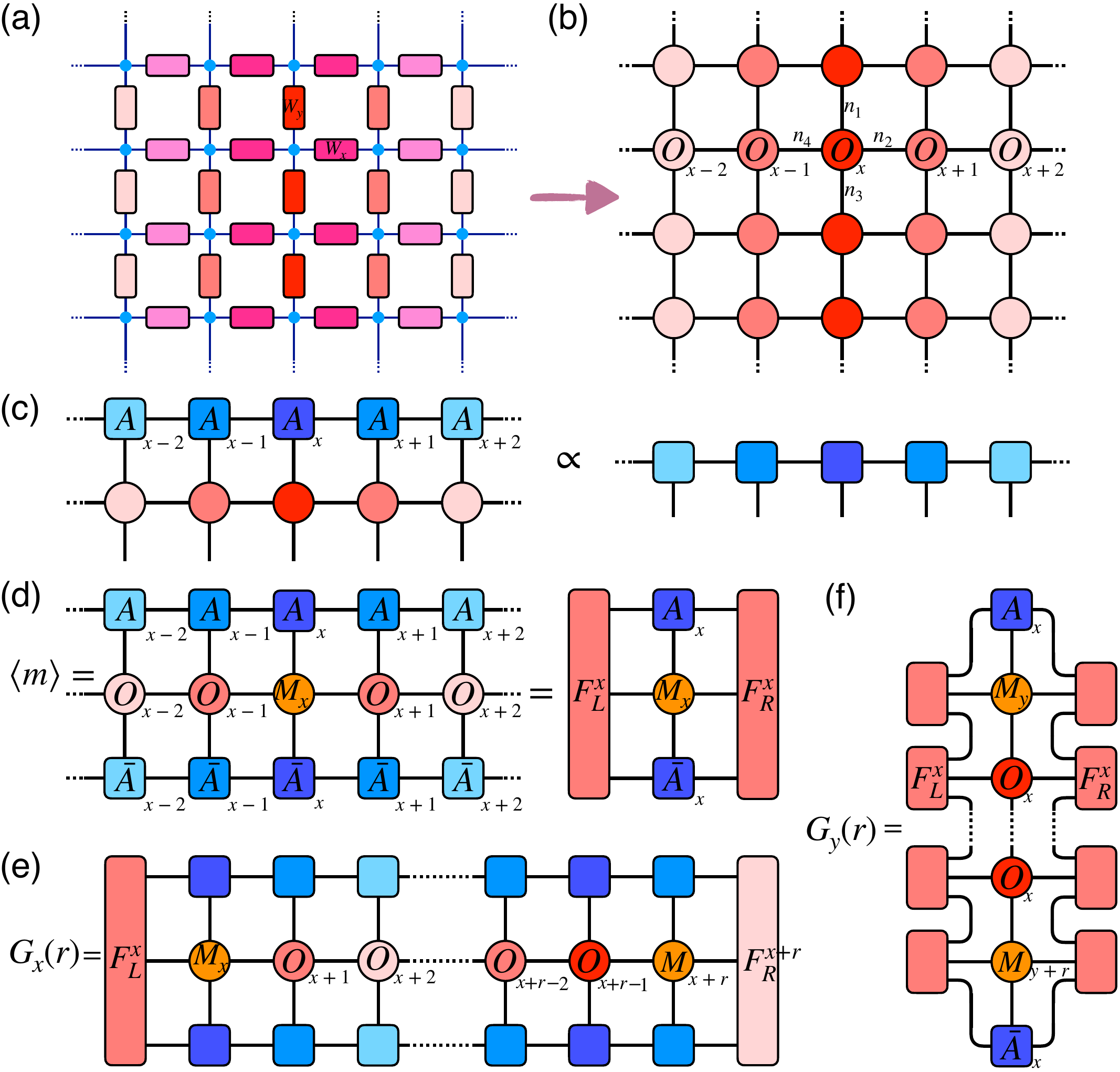}
    \caption{
    (a) The tensor network representation of the partition function with interaction matrices on the links accounting for the Boltzmann weight. The varying colors denote the modulation of $J_{ij}$ along $x$-direction. 
    (b) An infinite tensor network composed of local $O$ tensors from dual transformation. 
    (c) Eigenequation for the fixed-point MPS $\ket{\Psi(A)}$ of the 1D quantum transfer operator $\mathcal{T}$ with a unit cell made up of $\lambda$ local $A$ tensors.
    (d) Expectation of a local operator calculated by contracting the leading eigenvectors of the channel operators.
    (e) Two-point correlation function in $x$-direction represented by contracting a row of channel operators.
    (f) Two-point correlation function in $y$-direction obtained by contracting a column of tensors satisfying the fixed-point condition.}
    \label{fig:tensor}
\end{figure}

In the framework of tensor network method, we construct the partition function of the modulated XY model as a contraction of an infinite tensor network as shown in Fig.~\ref{fig:tensor} (a) and (b)
\be
Z=\prod_{i} \int \frac{\mathrm{d} \theta_{i}}{2 \pi}\prod_{\langle i j\rangle} W(\theta_i,\theta_j)=\mathrm{tTr}\prod_i O_{n_1,n_2}^{n_3,n_4}(i),
\ee
where ``tTr'' denotes the tensor contraction over all auxiliary links, $W(\theta_i,\theta_j)=e^{J_{ij}/T \cos(\theta_i-\theta_j)}$ are interaction matrices with continuous indices, and the temperature $T$ is in units of $J_0/k_B$. To transform continuous phase variables on a discrete basis, we apply a duality transformation
\be
e^{x\cos\theta}=\sum_{n=-\infty}^{\infty} I_n(x) e^{in\theta}
\label{eq:bessel}
\ee
where $I_n(x)$ are modified Bessel functions of the first kind~\cite{Yu2014,Vanderstraeten2019,Song2021}. By integrating out the $U(1)$ variables $\theta$, the local tensor of the partition function is obtained as
\be
O_{n_1, n_2}^{n_3, n_4}(i)=\sqrt{I_{n_1}(1/T)I_{n_2}((1+\alpha\cos(k(x_i+1))/T)I_{n_3}(1/T)I_{n_4}((1+\alpha\cos(kx_i)/T)}\delta_{n_1+n_2}^{n_3+n_4},
\ee
where $n_1, n_2, n_3$ and $n_4$ are the bond indices which can be truncated to a finite range since $I_n(x)$ decreases exponentially with increasing $n$. The conservation law of $U(1)$ charges is encoded in the $\delta_{n_1+n_2}^{n_3+n_4}$ tensors resulting from the continuous symmetry. Due to the periodic modulation of $J_{ij}$ in $x$-direction, $O$ tensors form a stripe structure with horizontal period $\lambda$ as displayed in Fig.~\ref{fig:tensor} (b).

In the thermodynamic limit, the partition function is determined by the dominant eigenvalues of the row-to-row transfer operator consisting of an infinite train of periodic $O$ tensors 
\be
\mathcal{T}=\mathrm{tTr}[\cdots O(x-2)O(x-1)O(x)O(x+1)O(x+2)\cdots].
\ee
As shown in Fig.~\ref{fig:tensor} (c), the fixed-point equation
\be
\mathcal{T}|\Psi(A)\rangle=\Lambda_{\max}\ket{\Psi(A)}
\ee
can be accurately solved by the multiple lattice-site VUMPS algorithm, where $\ket{\Psi(A)}$ is the leading eigenvector represented by matrix product states (MPS) made up of periodic local $A$ tensors with auxilary bond dimension $D$~\cite{Haegeman2017,Zauner-Stauber2018,Vanderstraeten2019tan,Nietner2020}.

\begin{figure}[tbp]
\centering
\includegraphics[width=.99\linewidth]{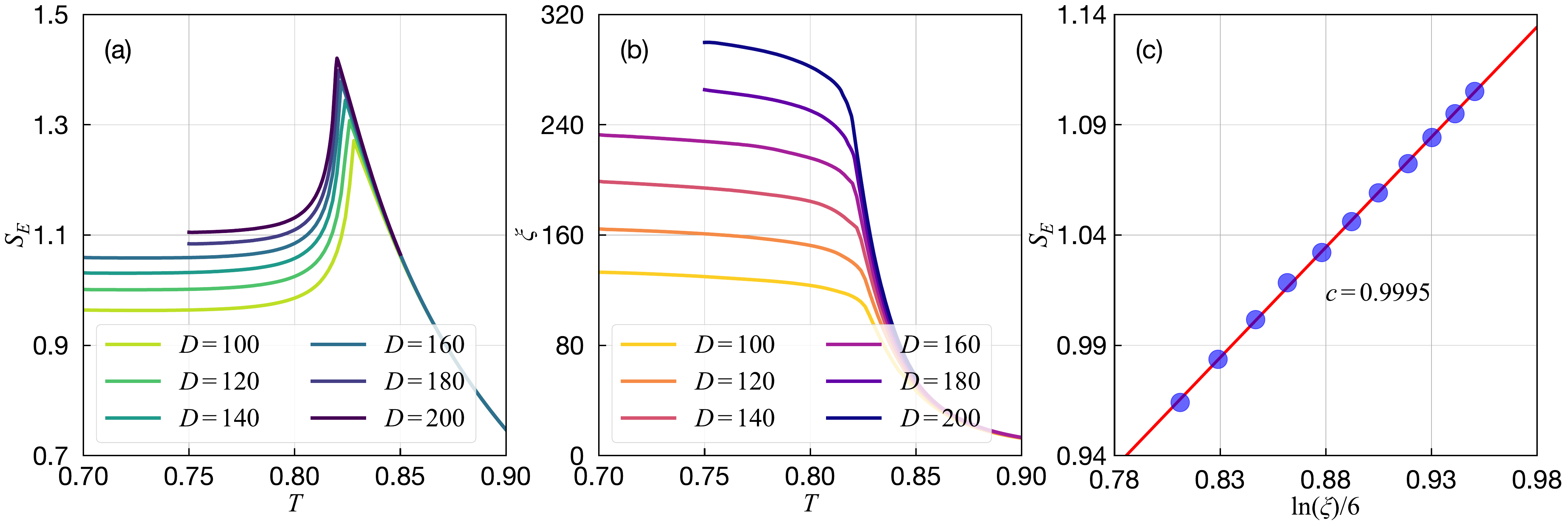}
\caption{ 
    (a) The entanglement entropy as a function of temperature for $\alpha=0.5$ and $\lambda=2$, with singularities indicating the BKT transition.
    (b) The corresponding correlation length as a function of temperature.
    (c) The central charge is fitted to be $c\simeq 1$ at $T=0.75$ in the critical phase.}
\label{fig:ccharge}
\end{figure}

Based on fixed-point MPS, various physical quantities can be accurately calculated. The quantum entanglement entropy can be directly obtained from the singular values of the Schmidt decomposition of $\ket{\Psi(A)}$ as 
\be
S_E = -\sum_{\alpha=1}^D s_\alpha^2 \ln s_\alpha^2,
\ee
whose singularity provides an efficient measure to determine phase transitions~\cite{Vidal2003,Pollmann2009}. Figure~\ref{fig:ccharge} (a) shows the entanglement entropy for $\alpha = 0.5$ and $\lambda = 2$ computed with bond dimensions ranging from $D = 100$ to $200$. Sharp peaks emerge at the critical points, and their positions converge systematically with increasing $D$. Finite bond dimension effects do not qualitatively alter the phase structure, with deviations of at most 3\% compared to extrapolations to infinite bond dimension. Therefore, for consistency and computational efficiency, we extract the transition temperature $T_c$ reported in the main text using $D = 100$.

We can obtain the central charge of the corresponding conformal field theory within the critical phases. It is known that there exists a linear scaling behavior between the entanglement entropy $S_E$ and logarithmic MPS correlation lengths $\xi_D$ under different bond dimensions $D$ of the fixed-point MPS~\cite{Pollmann2009}
\begin{equation}
S_E\propto\frac{1}{6}\ln\xi_D,
\end{equation}
where $\xi_D$ is the the MPS correlation length calculated from the largest and the second largest eigenvalues of the transfer matrix
\begin{equation}
\xi_D=-1/\ln|\lambda_2/\lambda_1|.
\end{equation}
The correlation lengths for $\alpha=0.5$ and $\lambda=2$ under different bond dimensions are displayed in Fig.~\ref{fig:ccharge} (b). As the temperature approaches $T_c$ from the high temperature side, the correlation length diverges exponentially as $\xi(T) \propto \exp\left( b / \sqrt{T - T_c} \right)$, where $b$ is a non-universal positive constant. In the low temperature phase, the correlation length is limited by the finite bond dimension and becomes saturated.

In Fig.~\ref{fig:ccharge} (c), we show clear evidence that the central charge is equal to $1$ in the low temperature phase, which is consistent with the prediction of conformal field theory for superfluid fields.

Moreover, the expectation value of local observables can be evaluated by inserting impurity tensors into the original tensor network for the partition function\cite{Song2021,Song2022,Song2022ffxy,Song2023,Morita2019,Morita2025}. As shown in Fig.~\ref{fig:tensor} (d), the contraction of the network containing the impurity tensor is reduced to a trace of an infinite sequence of channel operators
\be
\expval{m(\theta_x)}=\mathrm{Tr}(\cdots \mathbb{T}_{O_{x-2}}\mathbb{T}_{O_{x-1}}\mathbb{T}_{M_x}\mathbb{T}_{O_{x+1}}\mathbb{T}_{O_{x+2}}\cdots),
\ee
where the channel operators have a sandwich structure $\mathbb{T}_X=\sum_{i,j}A^i\otimes X^{i,j}\otimes A^j$. Using the fixed points of channel operators, we can further squeeze the long train to a small network as $\expval{m(\theta_x)}=\bra{F_L^x}M_x\ket{F_R^x}$, where $\bra{F_L^x}$ and $\ket{F_R^x}$ represent the left and right environments. 

In the same way, the two-point correlation function $G_x(r)=\expval{m(\theta_x)m(\theta_{x+r})}$ in the $x$-direction can be represented as a contraction of a sequence of channel operators with two local impurity tensors $M_x$ and $M_{x+r}$, 
\be
G_x(r)=\bra{F_L^x}\mathbb{T}_{M_x}\cdots \mathbb{T}_{O_{x+r-2}}\mathbb{T}_{O_{x+r-1}}\mathbb{T}_{M_{x+r}}\ket{F_R^{x+r}},
\ee
as shown in Fig.~\ref{fig:tensor} (e). To obtain the correlation function in the $y$-direction, we use the fix-point condition that $A_x$ is the upper fixed point of the local environment composed by $\bra{F_L^x}$, $O_x$ and $\ket{F_R^x}$~\cite{Zauner-Stauber2018,Vanderstraeten2019tan}. Then, the correlation function $G_y(r)$ is reduced to a trace of a column tensor network containing two impurity tensors as shown in Fig.~\ref{fig:tensor} (f).

\begin{figure}[tbp]
\centering
\includegraphics[width=.75\linewidth]{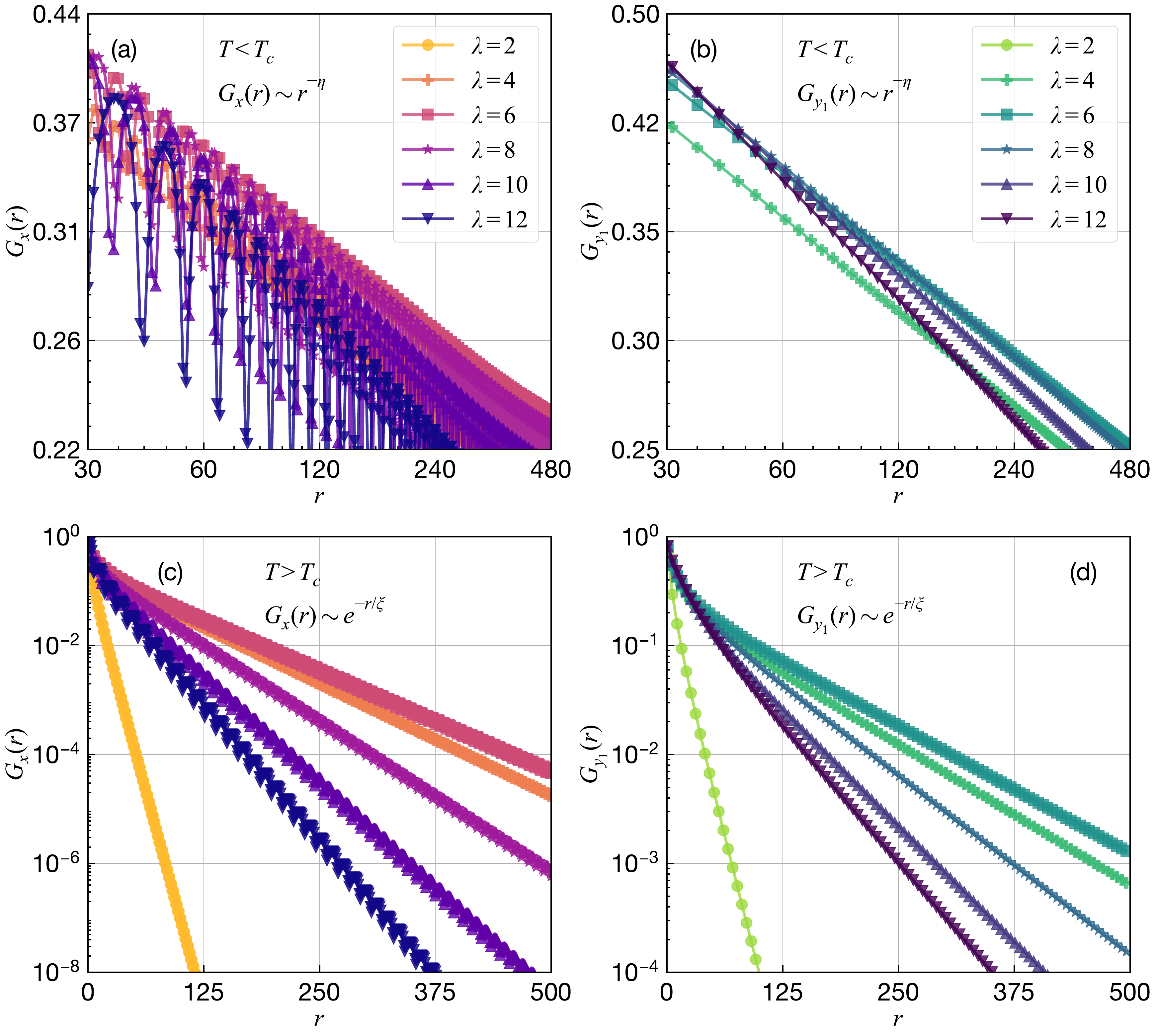}
\caption{ 
    (a) and (b) Correlation functions $G_x(r)$ along the $x$-direction and $G_{y_1}(r)$ along the first vertical stripe exhibit power-law decay, $G(r) \sim r^{-\eta}$, at temperature $T = 0.65 < T_c$, with modulation amplitude $\alpha = 0.8$ and varying modulation period $\lambda$. The decay exponent $\eta$ reaches its minimum at $\lambda = 6$, and increases for both smaller and larger values of $\lambda$. Note that $G(r)$ for $\lambda = 2$ is not shown because it decays exponentially at this temperature.
    (c) and (d) The correlation functions exhibit exponential decay, $G(r) \sim e^{-r/\xi}$, at temperature $T = 0.75 > T_c$, using the same modulation parameters with varying $\lambda$. Correlations decay more rapidly along the $x$-direction than along the $y$-direction. The correlation length $\xi$ is maximized around $\lambda = 6$ and becomes shorter for both smaller and larger $\lambda$.
    }
\label{fig:clfn}
\end{figure}

In the main text, we demonstrate that for the fixed modulation amplitude $\alpha$ and the period $\lambda$, both the two-point correlation functions $G_x(r)$ and $G_{y(i)}(r)$ exhibit power law decay in the low temperature phase. Specifically, we consider
\be
G_{x}(r)=\expval{\cos\left[\theta(x,y) - \theta(x+r,y)\right]},
\ee
which measures correlations along the $x$-direction (the direction of modulation), and
\be
G_{y(i)}(r) = \expval{\cos\left[\theta(x_i,y) - \theta(x_i,y+r)\right]},
\ee
which measures correlations along the $y$-direction (vertical), taken at fixed horizontal positions $x_i$ such that $G_{y(i)}(r)=G_{y({i+\lambda})}(r)$ for $i = 1, \ldots, \lambda$. Both $G_x(r)$ and $G_{y(i)}(r)$ decay as power laws $G(r) \sim r^{-\eta}$ with a common exponent $\eta$ in the low-temperature quasi-long-range ordered phase. Above the critical temperature $T_c$, all correlations decay exponentially, with $G_x(r)$ (along the modulated direction) decaying faster than $G_{y(i)}(r)$.

Here, we explore how the correlation functions depend on the modulation wavelength $\lambda$. Since all $G_{y(i)}(r)$ show the same decay behavior, it is enough to just focus on $G_{x}(r)$ and $G_{y_1}(r)$. As shown in Fig.~\ref{fig:clfn}(a) and (b), for $\alpha=0.8$, both $G_{x}(r)$ and $G_{y}(r)$ display algebraic behavior in a double-logarithmic plot at $T=0.65$ (below $T_c$). A closer inspection of the slopes reveals that the decay exponent varies with $\lambda$. As shown in Figs.~\ref{fig:clfn} (c) and (d), this dependence on the modulation period becomes more pronounced at $T=0.75$ (above $T_c$), where the slope in the semi-logarithmic graph decreases with increasing $\lambda$ for $\lambda \le 6$, and then increases for $\lambda \ge 6$. This trend aligns well with the observed dependence of $T_c$ on $\lambda$. For $\alpha=0.8$, the $T_c$ dome reaches its maximum at $\lambda=6$.

\section{Modulated clock model}
\begin{figure}[tbp]
    \centering
    \includegraphics[width=0.66\linewidth]{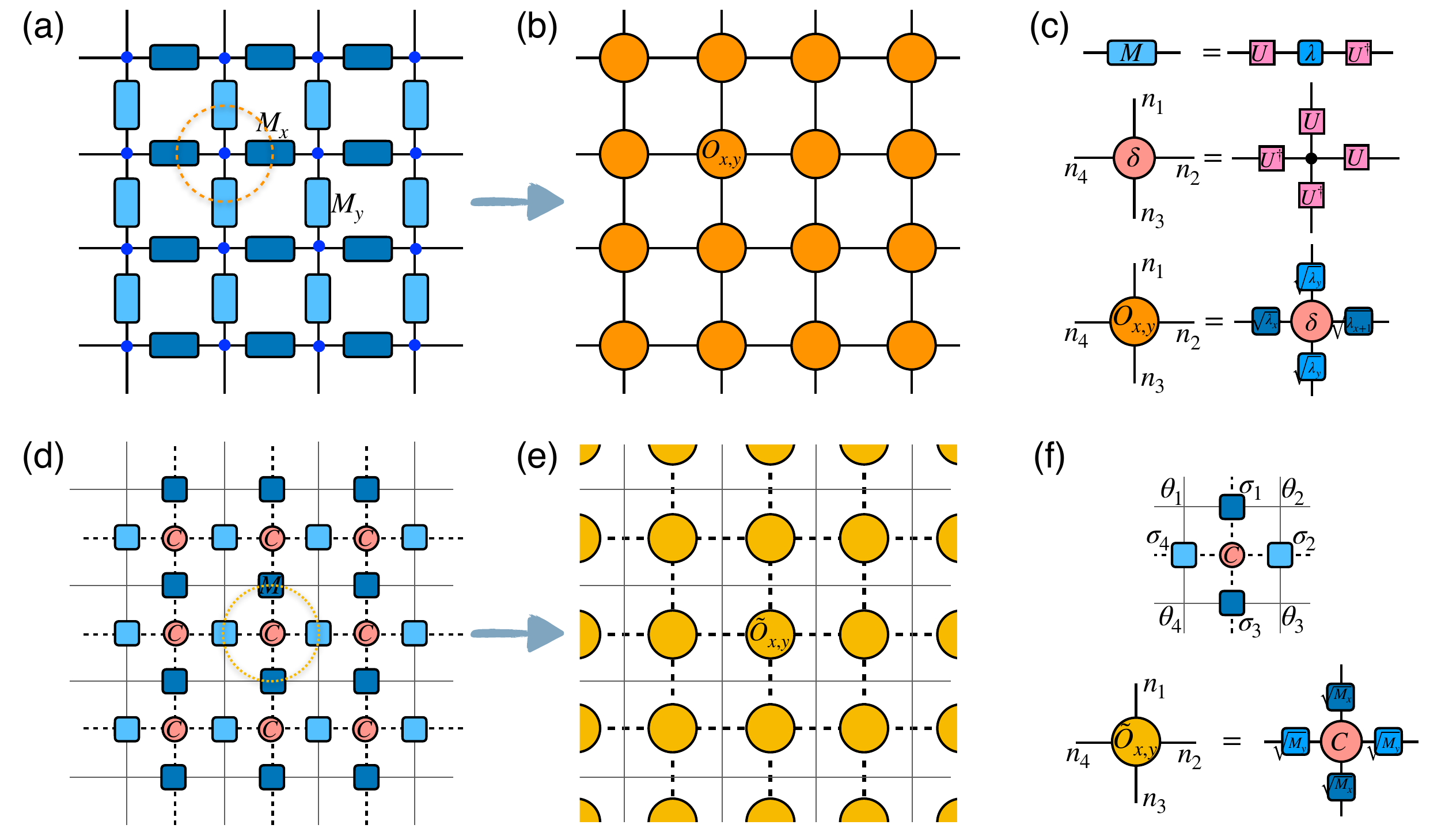}
    \caption{
    (a) and (d) Tensor network representation of the partition function for the modulated $q$-state clock model as a product of interaction matrices on the original lattice and the dual lattice, respectively. 
    (b) and (e) The uniform tensor network representation of the partition function for the modulated $q$-state clock model with local tensor $O_{n_1,n_2}^{n_3,n_4}$ on the original lattice and $\tilde{O}_{n_1,n_2}^{n_3,n_4}$ on the dual lattice, respectively. 
    (c) and (f) The construction of the local tensors for the partition function on the original lattice and the dual lattice, respectively.
    Dotted circles indicate the correspondence between the interaction matrices and the uniform local tensors.}
    \label{fig:clock}
\end{figure}
The modulated $q$-state clock model on the 2D square lattice is defined by the Hamiltonian
\be
H = -\sum_{\langle ij\rangle}J_{ij}\cos(\theta_i-\theta_j),
\ee
where $J_{ij}$ is defined in the same way as in modulated XY models \eqref{eq:cdwxy}, $\theta_i=2\pi k_i/q$ are discrete spin variables with $k=0,1,\ldots,q-1$, and the sum is over all NN pairs $\expval{i,j}$.

The tensor network representation of the partition function on the original lattice in Fig.~\ref{fig:clock} (a) can be derived by eigenvalue decomposition of the interaction matrices
\be
    M_{\theta_i,\theta_j}=\exp(K\cos(\theta_i-\theta_j))=\sqrt{q}\sum_{n=0}^{q-1}U_{n,\theta_i}\lambda_n(K) U_{n,\theta_j}^*,
\ee
where $K_x=J_0(1+\sin(kx))/T$ and $K_y=J_0/T$. The unitary matrix is $U_{n,\theta}=e^{in\theta}/\sqrt{q}$ and the diagonal matrix is given by
\be
\lambda_n(K)=\sum_{\theta}e^{K\cos\theta}U_{n,\theta}^*=\sum_{\theta}\cos(n\theta)e^{K\cos\theta}/\sqrt{q}.
\label{eq:lambda}
\ee
As shown in Fig.~\ref{fig:clock} (c), using the orthogonal relation
\be
\sum_\theta U_{n_1,\theta}U_{n_2,\theta}U_{n_3,\theta}^*U_{n_4,\theta}^*=\frac{1}{q}\delta_{n_1+n_2-n_3-n_4\bmod q},
\ee
we arrive at the local tensors on the original lattice
\be
O_{x,y}=\sqrt{\lambda_x\lambda_y\lambda_{x+1}\lambda_y}\delta_{n_1+n_2-n_3-n_4\bmod q}.
\ee
The partition function is then transformed into a uniform tensor network $Z=\prod_i O_{n_1,n_2}^{n_3,n_4}(i)$ as shown in Fig.~\ref{fig:clock} (b).

The tensor network representation of the partition function on the dual lattice can be defined with dual variables
\be
\sigma_1=\theta_2-\theta_1,\quad
\sigma_2=\theta_3-\theta_2,\quad
\sigma_3=\theta_3-\theta_4,\quad
\sigma_4=\theta_4-\theta_1,
\ee
as displayed in Fig.~\ref{fig:clock} (d) and (f). The interaction matrices $M_{\theta_i,\theta_j}$ changes to diagonal matrices $M_{\sigma_i}=\exp(K\cos\sigma_i)$ connected by a constraint tensor $C_{\sigma_1,\sigma_2,\sigma_3,\sigma_4}=\delta_{\sigma_1+\sigma_2-\sigma_3-\sigma_4\mod 2\pi}$. By splitting the interaction matrices we obtain the local tensors for the dual representation
\be
\tilde{O}_{x,y}=\sqrt{M_yM_xM_yM_x}\delta_{\sigma_1+\sigma_2-\sigma_3-\sigma_4\bmod{ 2\pi}},
\ee
and the partition function is also a contraction of a uniform network $Z=\prod_i\tilde{O}_{n_1,n_2}^{n_3,n_4}(i)$ as shown in Fig.~\ref{fig:clock} (e).

Without modulation, i.e., $\alpha=0$, the $q$-state clock model has self-dual points, which can be observed from the local tensor structures. Since two representations are related by the Kramer-Wannier dual transformation~\cite{Kramers1941} exchanging the low- and high-temperature phases, the critical temperatures can arise only in pairs (or degenerate at the same temperature). For $q=2,3,4$, there is a single phase transition and the critical temperatures can be derived directly from $\lambda_n(K_c)/\lambda_0(K_c)=\tilde{\lambda}_n(K_c)/\tilde{\lambda}_0(K_c)$ with $\tilde{\lambda}_n\equiv M_{2\pi n/q}$ as $T_c=2/\ln(\sqrt{2}+1), 3/2\ln(\sqrt{3}+1), 1/\ln(\sqrt{2}+1)$, respectively.

\begin{figure}[tbp]
    \centering
    \includegraphics[width=0.99\linewidth]{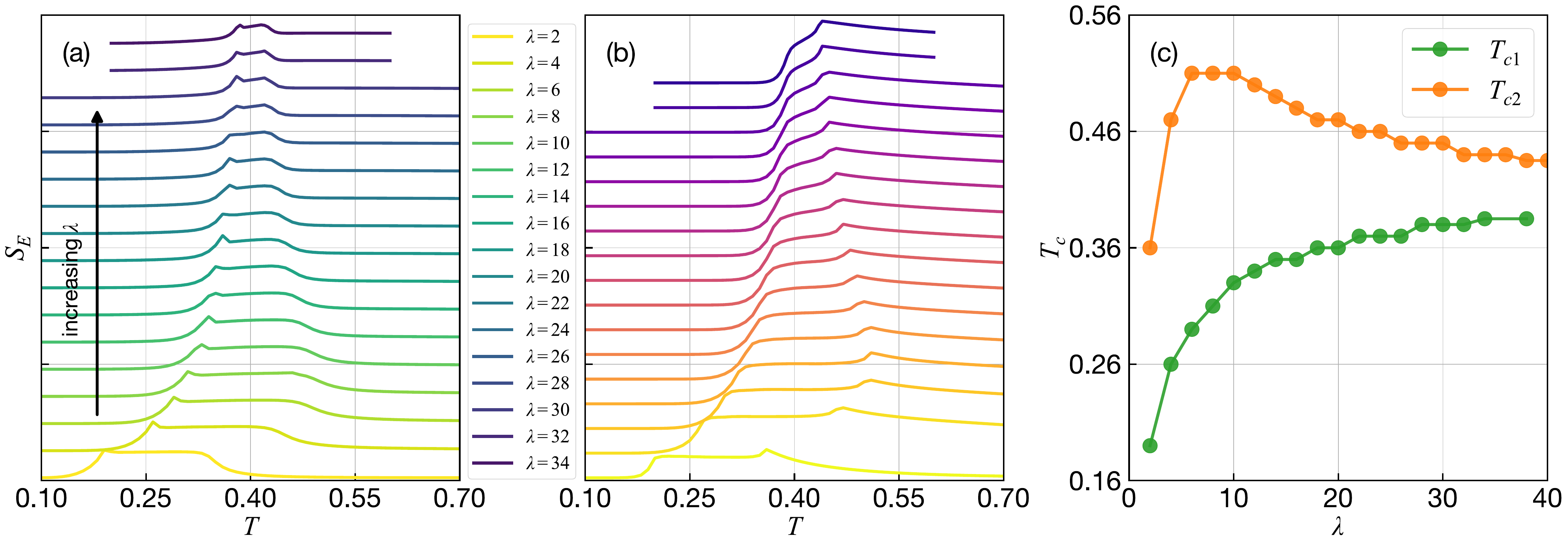}
    \caption{
    (a) and (b) The entanglement entropy as a function of temperature for the modulated clock models and their corresponding dual models with $\alpha=0.95$ and $q=8$, where peak positions indicates the transition temperatures. The modulation periods $\lambda$ increase from bottom to top. For clarity, the curves are vertically shifted by a fixed offset $\Delta S_E$ to better illustrate the overall trend.
    (c) The critical temperature $T_{c1}$ increases monotonically while $T_{c2}$ displays a dome structures dependent on the modulation wave lengths $\lambda$. As $\lambda$ increases further, both transition temperatures approach fixed values while remaining distinct. }
    \label{fig:clockTc}
\end{figure}

For $\alpha\ne0$, the local tensors $O_{x,y}$ and $\tilde{O}_{x,y}$ have different structures without a clear dual relationship. We determine the critical temperatures numerically from the singularities of the entanglement entropy~\cite{Li2020}. As shown in Fig.~\ref{fig:clockTc} (a) and (b), the entanglement entropy displays singular behavior at temperatures $T_{c1}$ and $T_{c2}$ for the modulated $q = 8$ clock model with $\alpha = 0.95$, on the dual lattice and the original lattice, respectively. The dependence of $T_{c1}$ and $T_{c2}$ on $\lambda$ is summarized in Fig.~\ref{fig:clockTc} (c), where $T_{c1}$ increases monotonically with $\lambda$, while $T_{c2}$ exhibits a non-monotonic, dome-like profile. Both $T_{c1}$ and $T_{c2}$ tend to stabilize at fixed values as $\lambda$ becomes large.

\section{Renormaization group and duality equation}
\begin{figure}[tbp]
    \centering
    \includegraphics[width=0.75\linewidth]{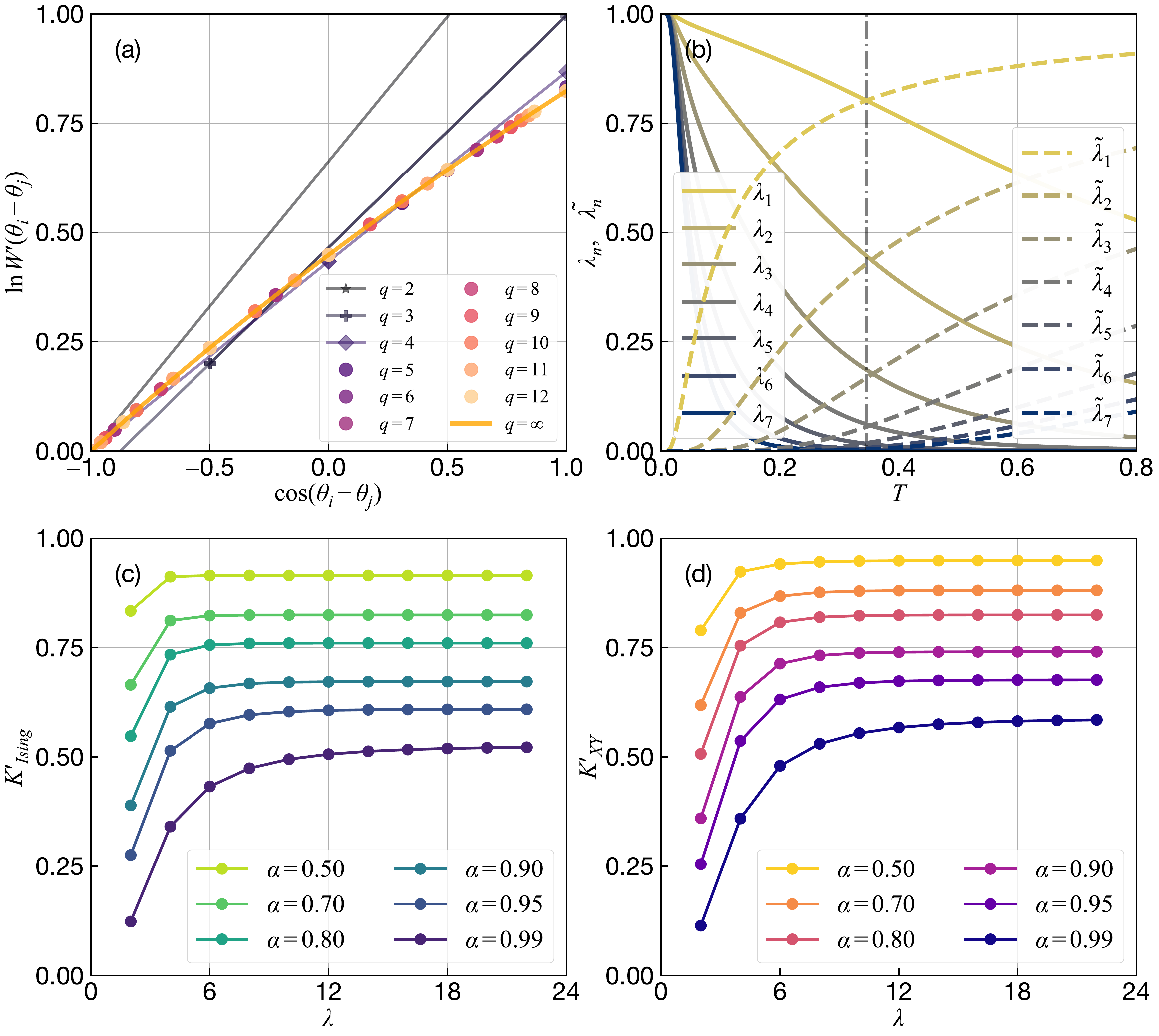}
    \caption{
    (a) Verification of whether the renormalized interaction matrix $W'(\theta_i, \theta_j)$, obtained after decimating two spins, retains the form $\exp(K' \cos(\theta_i - \theta_j))$. For $q = 2, 3, 4$, the renormalized matrices follow this exponential cosine form, as evidenced by the linear relationship between $\ln W'(\theta_i, \theta_j)$ and $\cos(\theta_i - \theta_j)$. In contrast, for $q \ge 5$, the matrix structure deviates from the ideal form and instead aligns with the nonlinear curve characteristic of the XY model ($q = \infty$), indicating convergence in behavior for large $q$.
    (b) Solution of the duality equation~\eqref{eq:dual} for the $q = 16$ clock model, illustrated through the crossover of the corresponding functions. No unique solution exists, as the curves for different $n$ intersect the duality condition at different temperatures, indicating inconsistency across $n$. The vertical black line serves as a visual guide, indicating the solution for $n = 1$.
    (c) and (d) Dependence of the renormalized interaction strength $K_x$ on the modulation parameters for $q = 2$ and $q = \infty$ (XY model) with $T=2$ and $T=1/3$, respectively. In both cases, $K_x$ increases with the modulation period $\lambda$ and decreases with the modulation amplitude $\alpha$. The saturation behavior becomes more apparent at larger $\lambda$, and the saturation point shifts to higher $\lambda$ values as $\alpha$ increases.}
    \label{fig:rg}
\end{figure}

In order to study the dependence of the phase transition temperatures on the modulation, we apply a a hybrid approach that combines a single Migdal-Kadanoff (MK) renormalization group (RG)~\cite{Migdal1975,Kadanoff1976} transformation with the Kramers-Wannier duality~\cite{Kramers1941,Ortiz2012}.

We first use MK-RG to map the modulated system onto an anisotropic model by the 1D decimation tracing over $\lambda$ different horizontal interactions $K_n$, with $K_n \equiv J_{ij}/T$ associated with each NN pair $\expval{i,j}$. For the clock model, this corresponds to a matrix product
\be
W' = \prod_{n=1}^{\lambda} W_n,
\ee
where $W_n(\theta_i, \theta_j) = \exp(K_n \cos(\theta_i - \theta_j))$.

We aim to determine an effective coupling constant $K'$ such that the renormalized interaction takes the form $W'(\theta_i, \theta_j) \propto \exp(K' \cos(\theta_i - \theta_j))$. For $q = 2, 3, 4$, the effective coupling $K'$ resulting from the 1D decimation can be calculated exactly. This can be verified by evaluating the product of two transfer matrices $W(K') = W_1(K_1) W_2(K_2)$, where $W_1$ and $W_2$ correspond to the original couplings $K_1$ and $K_2$, respectively. Instead of evaluating $W(K')$ directly, we can consider the spectrum of the renormalized matrix. Specifically, we compute the new eigenvalues via $\lambda_n(K') = \lambda_n(K_1)\lambda_n(K_2)$. For the case $q = 2$, this takes the form
\be
\begin{pmatrix}
    \lambda_0(K') & \\
    & \lambda_1(K')
\end{pmatrix}
=
\begin{pmatrix}
    2\cosh(K') & \\
    & 2\sinh(K')
\end{pmatrix}
\propto
\begin{pmatrix}
    \cosh(K_1) & \\
    & \sinh(K_1)
\end{pmatrix}
\begin{pmatrix}
    \cosh(K_2) & \\
    & \sinh(K_2)
\end{pmatrix}.
\ee
Since the interaction $W'(\theta_i, \theta_j) \propto \exp(K' \cos(\theta_i - \theta_j))$ implies that the normalized eigenvalues satisfy 
\be
\frac{\lambda_n(K')}{\lambda_0(K')} = \frac{\lambda_n(K_1)\lambda_n(K_2)}{\lambda_0(K_1)\lambda_0(K_2)},
\label{eq:K1K2}
\ee
we can solve for the renormalized coupling $K'$ and obtain
\be
\tanh(K') = \tanh(K_1)\tanh(K_2).
\ee

For $q = 3$, the equation for $K'$ can be derived from \eqref{eq:lambda} and \eqref{eq:K1K2} as
\be
\exp\left(\frac{3K'}{2}\right) = \frac{\exp(K_1 + K_2) + 2\exp\left(-\frac{1}{2}(K_1 + K_2)\right)}{\exp\left(-\frac{1}{2}(K_1 + K_2)\right) + \exp\left(K_1 - \frac{K_2}{2}\right) + \exp\left(K_2 - \frac{K_1}{2}\right)}.
\ee
For $q = 4$, the model can be mapped to two independent Ising models via the identity
\be
\cos(\theta_i - \theta_j) = \frac{1}{2}\left(\sigma_i^a \sigma_j^a + \sigma_i^b \sigma_j^b\right),
\ee
where $\sigma_i^a, \sigma_i^b = \pm 1$ are the Ising variables. As a result, the analysis for $q = 4$ becomes equivalent to the $q = 2$ case, but with the effective coupling halved $K_{q=4} = K_{q=2}/2$. This relationship is also evident from the spectrum identity $\lambda_n^{q=4} \propto \lambda_n^{q=2} \otimes \lambda_n^{q=2}$.

However, for $q \ge 5$, the relation $W'(\theta_i, \theta_j) \propto \exp(K' \cos(\theta_i - \theta_j))$ no longer holds exactly. In Fig.~\ref{fig:rg} (a), we compare $\ln W'(\theta_i, \theta_j)$, obtained from the product of two $W(\theta_i, \theta_j)$ matrices, with $\cos(\theta_i - \theta_j)$. For $q = 2, 3, 4$, all matrix elements lie along a straight line, indicating that the effective interaction retains the same functional form. In contrast, for $q \ge 5$, the points trace out a curve, revealing deviations from the ideal exponential cosine form. The case $q = \infty$ corresponds to the XY model, whose spectrum is obtained from Eq.~\eqref{eq:bessel}. Notably, all clock models with $q \ge 5$ converge toward the same curve as $q$ increases, suggesting that their behavior increasingly resembles that of the XY model.

For the modulated model with modulation period $\lambda$, applying 1D decimation along the $x$-direction yields an effective coupling $K_x'$. Along the $y$-direction, a simple bunching results in $K_y' = \lambda K_0$. This leads to an anisotropic model with uniform horizontal and vertical interactions. However, this approach introduces significant errors, since the interaction matrices generally have bond dimensions greater than one. To reduce such errors, we can revert to the unrenormalized description by effectively setting the RG scale factor $\lambda \to 1$. In other words, our goal is to find a uniform coupling $K_x$ that produces the same renormalized $K'$ as the original modulated sequence. This requirement can be expressed as
\be
\prod_{n=1}^{\lambda} W(K_n) = \prod_{n=1}^{\lambda} W(K_x),
\label{eq:Kx}
\ee
with $K_y$ remains the original one.

After transforming the modulated model into an anisotropic model with couplings $K_x$ and $K_y$, the self-dual temperatures can be determined directly from the duality equation
\be
\frac{\lambda_n(K_x)}{\lambda_0(K_x)} = \frac{\tilde{\lambda}_n(K_y)}{\tilde{\lambda}_0(K_y)},
\label{eq:dual}
\ee
since the horizontal bond spectra in the original tensor network correspond to the vertical bond spectra in the dual representation. For systems exhibiting a single phase transition, the self-dual point corresponds to the critical temperature. However, in systems with multiple transitions, the self-dual point does not necessarily coincide with any of the critical temperatures. Eq.~\eqref{eq:dual} typically admits a unique solution for $q \le 5$. For $q > 5$, however, no exact solution exists, as the duality condition does not hold simultaneously for all $n$. This is illustrated in Fig.~\ref{fig:rg} (b), where for $q = 16$, different values of $n$ yield crossing points at different temperatures.

The critical temperatures for $q = 2$ and $q = 3$ presented in the main text are determined by combining the MK-RG equation~\eqref{eq:Kx} with the duality condition~\eqref{eq:dual}, yielding results in good agreement with tensor network calculations. For instance, in the $q = 2$ clock model with modulation period $\lambda = 2$, the critical temperature $T_c$ satisfies the following self-consistent equations
\begin{align}
    \tanh(J_x/T_c)&=\sqrt{\tanh((1+\alpha)J_0/T_c)\tanh((1-\alpha)J_0/T_c)},\\
    &\sinh(2J_x/T_c)\sinh(2J_0/T_c)=1.
\end{align}
These equations can be solved self-consistently through several iterations to give $J_x\approx0.3595$ and $T_c\approx1.4282$.

Furthermore, we can show that the effective coupling $K_x$ after renormalization always increases with the modulation period $\lambda$. For the case of $q = 2$, the renormalized coupling satisfies
\be
\ln \tanh(K_x) = \frac{1}{\lambda} \sum_{n=1}^{\lambda} \ln \tanh(K_n).
\ee
Here, we define the function $f(x) = \ln \tanh(x)$, which is strictly increasing and concave for $x > 0$, as indicated by its derivatives
\be
f'(x) = \frac{2}{\sinh(2x)} > 0, \quad f''(x) = -\frac{4 \cosh(2x)}{\sinh^2(2x)} < 0.
\ee
Because $f(x)$ is concave, $\ln \tanh(K_x)$ decreases as the distribution of $\{K_n\}$ becomes more spread out, i.e., as the modulation amplitude $\alpha$ increases. Meanwhile, as $\lambda$ increases, more of the $K_n$ values cluster near the center of the interval $[J_0(1 - \alpha)/T,\; J_0(1 + \alpha)/T]$, reducing the spread with an increase in $\ln \tanh(K_x)$, and hence in $K_x$ itself. As $\lambda$ increases, $\ln \tanh(K_x)$ quickly converges to the integration of $1/2\pi\int_0^{2\pi} \ln\tanh((1+\alpha\cos\theta)J_0/T)d\theta$, which implies that $K_x$ saturates rapidly to a constant value.

More rigorously, we show that for modulation periods $\lambda = 2m$ and $\lambda = 2m + 2$, the following inequality holds
\be
\frac{1}{2m} \left[ f(x_0) + 2f(x_1) + \cdots + 2f(x_{m-1}) + f(x_m) \right]
<
\frac{1}{2m+2} \left[ f(y_0) + 2f(y_1) + \cdots + 2f(y_m) + f(y_{m+1}) \right],
\ee
where $x_i = \left(1 + \alpha \cos\left(\frac{2\pi i}{2m}\right)\right) J_0 / T$ and $y_j = \left(1 + \alpha \cos\left(\frac{2\pi j}{2m + 2}\right)\right) J_0 / T$.
The inequality above can be rewritten as
\be
\frac{1}{2m(m + 1)} f(x_0) + \frac{1}{m} \sum_{i = 1}^{m - 1} f(x_i) + \frac{1}{2m(m + 1)} f(x_m)
<
\frac{1}{m + 1} \sum_{j = 1}^{m} f(y_j),
\label{eq:ineq}
\ee
where we have used the fact that $f(x_0) = f(y_0)$ and $f(x_m) = f(y_{m+1})$. Due to $x_0<y_1<x_1<y_2<\cdots<y_n<x_n$, we can properly split the coefficients before $f(x_1),\cdots,f(x_{n-1})$ to satisfy the Jensen's inequality for each local pair to obtain the desired inequality~\eqref{eq:ineq}.

For the modulated XY model, the renormalized spectrum is given by
\be
\ln \lambda_n' = \frac{1}{m} \sum_{i=1}^{m} \ln I_n(K_i),
\ee
where $I_n(x)$ is the modified Bessel function of the first kind. The effective coupling $K_x$ can then be obtained from the ratio of matrix elements $W'(0, 0)/W'(0, \pi)$ via the expression
\be
\exp(2K_x) = \frac{\sum_n \lambda_n'}{\sum_n (-1)^n \lambda_n'}.
\ee
One can also show that $K_x$ increases with the modulation period $\lambda$, using the fact that $\ln I_n(x)$ is concave for $n > 1$.

The dependence of the renormalized interaction strength $K_x$ on the modulation parameters is shown in Figs.~\ref{fig:rg} (c) and (d) for $q = 2$ and $q = \infty$, respectively. In both cases, the effective coupling $K_x$ increases with larger $\lambda$, but decreases as the modulation amplitude $\alpha$ increases. As a result, the $T_c$ obtained from the duality equations always grows with larger $K_x$. In order to explain the non-monotonic behavior of $T_c(\lambda)$ observed in the modulated clock model for $q \ge 5$ and in the modulated XY model, a more refined RG analysis that incorporates vortex excitations is required. Notably, the saturation point of $K_x$ shifts to larger $\lambda$ values as $\alpha$ increases. This shift is closely related to the movement of the peak positions of $T_c(\lambda)$ dome, which also shift toward larger $\lambda$ with increasing $\alpha$.

\section{Monte Carlo results}
\begin{figure}[tbp]
    \centering
    \includegraphics[width=0.8\linewidth]{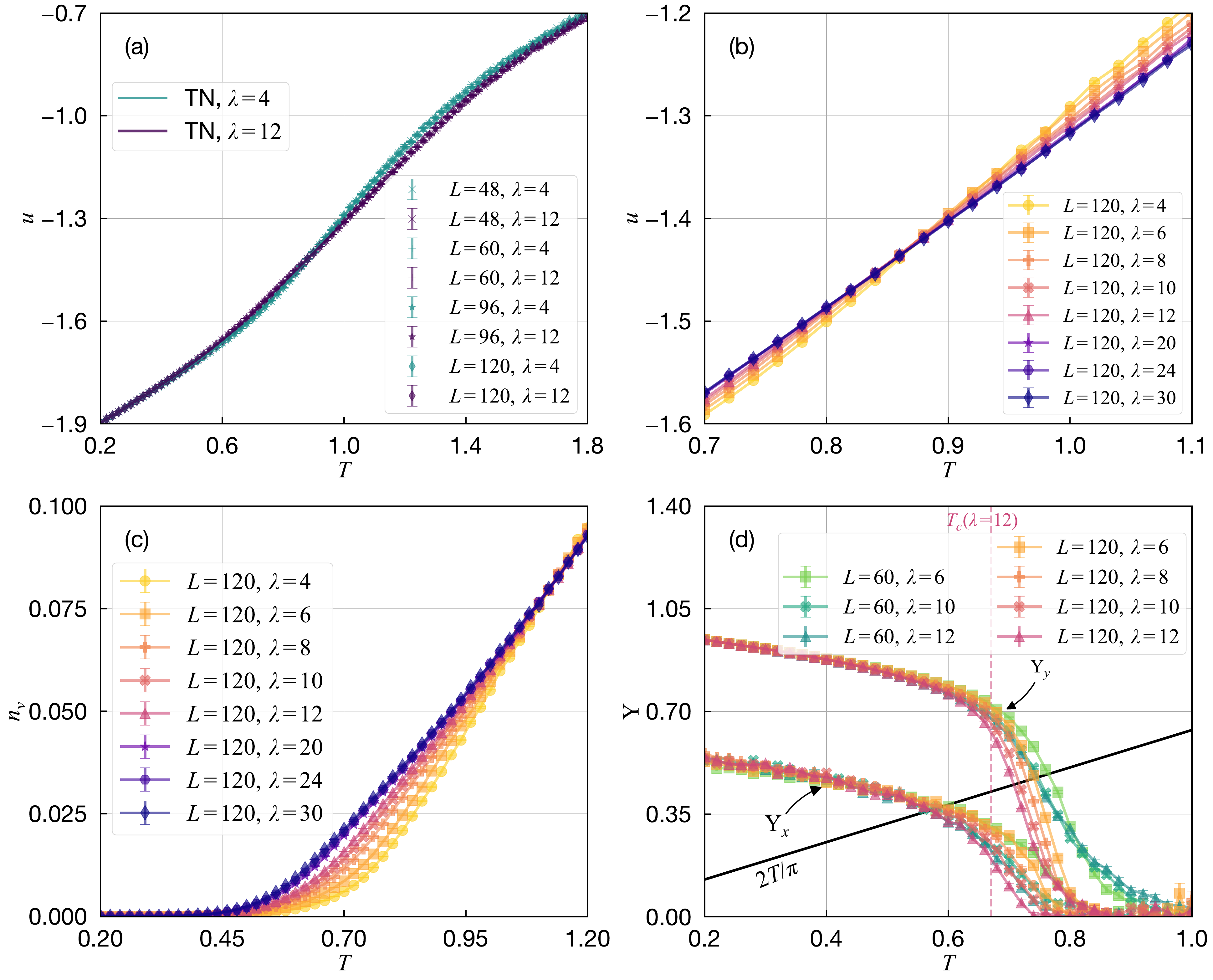}
    \caption{
    (a) The internal energy of the modulated XY model with a modulation amplitude of $\alpha = 0.8$, obtained from Monte Carlo (MC) simulations for various system sizes $L$. Solid lines represent results from tensor network (TN) calculations, showing good agreement with the MC data.
    (b) A zoomed-in view of the internal energy for different modulation periods $\lambda$ reveals that the rate of increase (slope) of the internal energy slows down with increasing $\lambda$, with a crossover observed around $T = 0.9$.
    (c) The vortex density as a function of temperature increases with larger modulation wavelengths $\lambda$, eventually converging at high temperatures.
    (d) The helicity modulus $\Upsilon_x$ in the $x$-direction is smaller than $\Upsilon_y$ in the $y$-direction, reflecting anisotropic stiffness in the modulated XY model. The straight line $2T/\pi$ is plotted as a reference, known to intersect the helicity modulus at the critical temperature in the standard XY model. The critical temperature $T_c$ from TN calculations for $\lambda = 12$ is indicated by the red vertical dashed line, lying between the intersection points of $\Upsilon_x$ and $\Upsilon_y$ with $2T/\pi$.}
    \label{fig:mc}
\end{figure}

The distribution of vortex excitations in the modulated XY model is investigated through Monte Carlo (MC) simulations using the Metropolis algorithm at high temperatures and the Wolff algorithm at low temperatures.

The temperature dependence of the internal energy density $u = \langle H \rangle / N$ for modulation amplitude $\alpha = 0.8$ is shown in Fig.~\ref{fig:mc} (a), where the total number of spins is $N = L \times L$. Finite-size effects are negligible, as results for different system sizes $L$ exhibit good overlap. The data agree well with tensor network results. At both low and high temperatures, the internal energy for different modulation periods $\lambda$ converges, while in the intermediate temperature range, slightly above the critical temperature $T_c$, noticeable differences emerge. The zoomed-in view in Fig.~\ref{fig:mc} (b) shows that the rate of increase of the internal energy decreases with increasing $\lambda$, with a crossover near $T = 0.9$. This behavior may be attributed to the unbinding of vortex-antivortex pairs, which occurs more abruptly for smaller $\lambda$, indicating a more rapid transition. 

The average vortex density is shown in Fig.~\ref{fig:mc} (c). It increases with modulation period $\lambda$, indicating that vortices are more easily excited due to the larger regions of weaker interactions. Moreover, the rate of increase becomes more pronounced in the temperature range slightly above the critical temperature $T_c$ for smaller $\lambda$, consistent with the behavior observed in the internal energy.

The anisotropic effects induced by modulation can be characterized by the helicity modulus
\be
\Upsilon = \left. \frac{1}{N} \frac{\partial^2 F(\phi)}{\partial \phi^2} \right|_{\phi=0},
\ee
which quantifies the response of the system's free energy $F$ to a small uniform twist $\phi$ applied to the XY spins along a given direction. Since the modulation is applied along the $x$-direction and the system remains uniform along $y$, the response to a twist becomes direction-dependent.
The modified Hamiltonians under a uniform twist in the $x$- or $y$-direction are given by
\begin{align}
H(\phi_x) &= -\sum_{\langle i,j \rangle_x} J_{ij} \cos(\theta_i - \theta_j + \phi_x) - \sum_{\langle i,j \rangle_y} J_{ij} \cos(\theta_i - \theta_j), \\
H(\phi_y) &= -\sum_{\langle i,j \rangle_x} J_{ij} \cos(\theta_i - \theta_j) - \sum_{\langle i,j \rangle_y} J_{ij} \cos(\theta_i - \theta_j + \phi_y),
\end{align}
where $\phi_x$ is applied between NN $\langle i,j \rangle_x$ along the $x$-direction, and $\phi_y$ is applied similarly in the $y$-direction.
Using the $U(1)$ symmetry of the spins,
\be
\sum_{\langle i,j \rangle} J_{ij} \left\langle \sin(\theta_i - \theta_j) \right\rangle = 0,
\ee
we obtain the helicity moduli in the $x$- and $y$-directions
\begin{align}
\Upsilon_x &= -\frac{1}{N} \left\langle \sum_{\langle i,j \rangle_x} J_{ij} \cos(\theta_i - \theta_j) \right\rangle - \frac{1}{N T} \left\langle \left( \sum_{\langle i,j \rangle_x} J_{ij} \sin(\theta_i - \theta_j) \right)^2 \right\rangle, \\
\Upsilon_y &= -\frac{1}{N} \left\langle \sum_{\langle i,j \rangle_y} J_{ij} \cos(\theta_i - \theta_j) \right\rangle - \frac{1}{N T} \left\langle \left( \sum_{\langle i,j \rangle_y} J_{ij} \sin(\theta_i - \theta_j) \right)^2 \right\rangle.
\end{align}

The helicity moduli are shown in Fig.~\ref{fig:mc} (d), whose jump at $T_c$ due to unbinding of vortex-antivortex pairs is a characteristic feature for the BKT transition. The helicity modulus $\Upsilon_x$ in the $x$-direction is smaller than $\Upsilon_y$ in the $y$-direction, indicating an anisotropic response induced by the spatial modulation. This anisotropy is also consistent with the direction-dependent behavior observed in the correlation functions. The difference arises because strong modulation reduces the effective coupling between vertical stripes, allowing them to behave more independently and thereby weakening the system's stiffness in the $x$-direction. Both $\Upsilon_x$ and $\Upsilon_y$ reach their maximum at $\lambda=6$, mirroring the same dependence of $T_c(\lambda)$ for $\alpha=0.8$. The drop in the moduli becomes sharper as the system size increases, indicating the finite-size effect. For comparison, we also plot the line $2T/\pi$, which is predicted to intersect the helicity modulus at the BKT transition temperature in the standard XY model. The intersection points differ between $\Upsilon_x$ and $\Upsilon_y$, with the critical temperature $T_c$ obtained from tensor network methods lying between them, and closer to the intersection with $\Upsilon_y$.

\begin{figure}[tbp]
    \centering
    \includegraphics[width=0.99\linewidth]{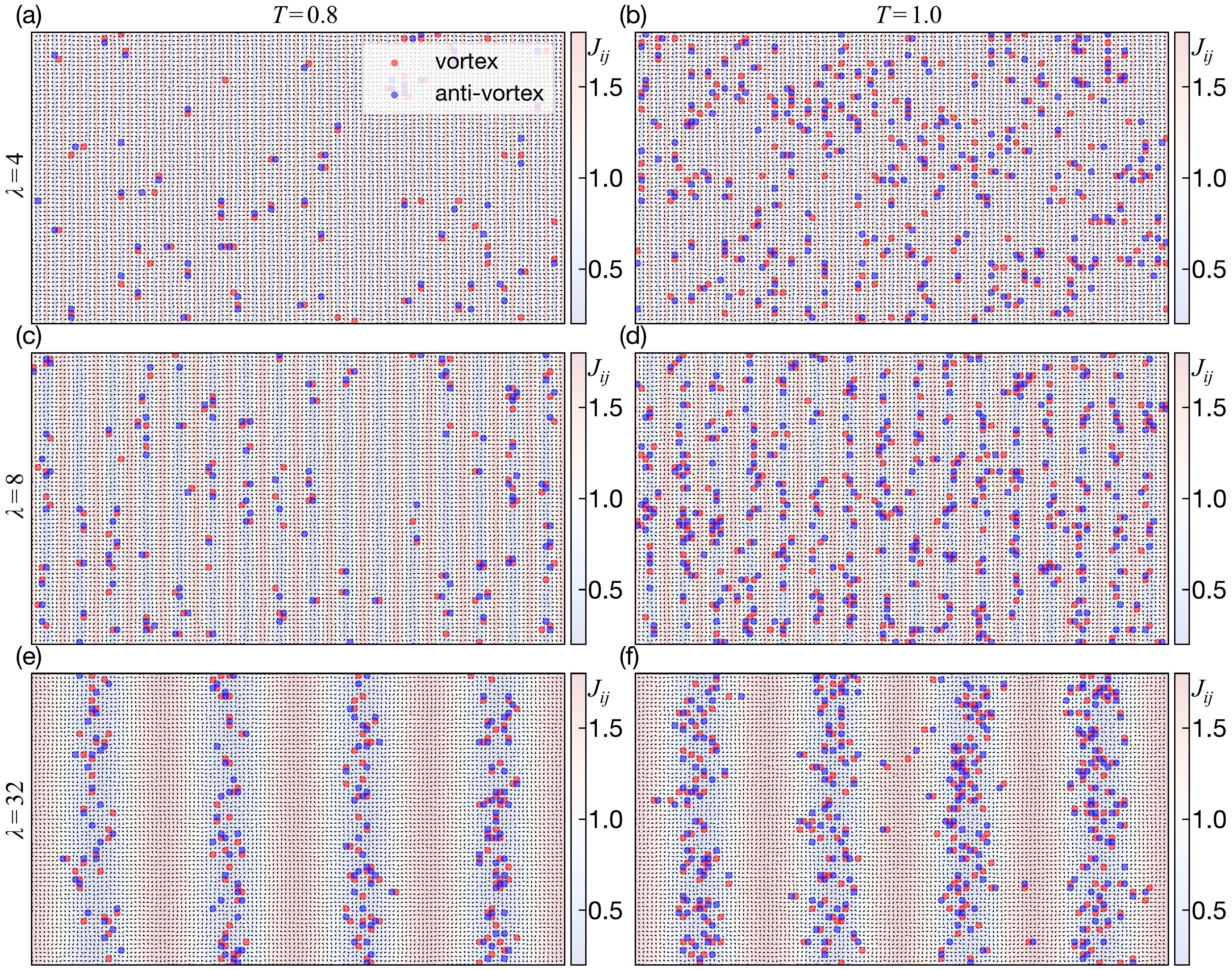}
    \caption{
    Snapshots of spin configurations and vortex distributions from Monte Carlo simulations on a system of size $L = 128$ with modulation amplitude $\alpha = 0.8$. The first and second columns correspond to temperatures $T = 0.8$ and $T = 1.0$, both above the critical temperatures $T_c$. Different rows are for different modulation periods $\lambda=4,\ 8,\ 32$. Background color indicates the spatial modulation of interaction strength $J_{ij}$. The configuration for $\lambda = 16$ is shown in the main text.}
    \label{fig:vortex}
\end{figure}

Moreover, typical vortex configurations from Monte Carlo simulations on a system of size $L = 128$ with modulation amplitude $\alpha = 0.8$ are shown in Fig.~\ref{fig:mc}, where the spatial distribution of the NN interaction strength $J_{ij}$ is indicated by the background color. A vortex is a topological defect characterized by spins rotating around a central point, resulting in a nonzero winding number. The local vorticity is quantified by 
\be
v = \frac{1}{2\pi} \sum_{\langle i,j \rangle\in\square} \Delta_{ij},
\ee
where the sum is taken counterclockwise around a plaquette, and $\Delta_{ij} = (\theta_i - \theta_j)$ is the angle difference between neighboring spins, wrapped to the interval $[-\pi, \pi]$. An integer value of $v$ identifies a vortex core, i.e., $v = +1$ for a vortex and $v = -1$ for an antivortex.

The anisotropic effects become more pronounced as the modulation period $\lambda$ increases. This is evident from the spatial distribution of vortices, which tend to localize within the valleys of the modulation (blue regions where $J_{ij} < J_0$). The total vortex number also increases with $\lambda$ since valleys provide more space to accommodate vortices. 

A detailed comparison of vortex configurations at different temperatures reveals distinct behaviors depending on $\lambda$. For smaller $\lambda$, vortices more readily penetrate the peak regions (red areas where $J_{ij} > J_0$), leading to a more rapid onset of disorder throughout the system. In contrast, for larger $\lambda$, vortices predominantly remain confined within the valleys even at higher temperatures, and the transition to a disordered phase occurs more gradually. In this case, vortex proliferation proceeds by progressively filling the wider valleys as temperature increases, delaying their spread into the peak regions.

This difference in vortex dynamics provides insight into the crossover behavior observed in the internal energy shown in Fig.~\ref{fig:mc} (b). At low temperatures, systems with smaller $\lambda$ exhibit a lower vortex density, as the narrow valleys restrict vortex formation and inhibit their entry into the stronger coupling regions. As the temperature rises, the tendency to increase entropy drives vortices to enter the peak regions, where more configurational degrees of freedom are available. This leads to a rapid increase in internal energy, even though the total number of vortices remains lower than in systems with larger $\lambda$.

Overall, the modulation period $\lambda$ plays a crucial role in controlling the spatial distribution, excitation dynamics, and thermodynamic impact of vortices. It significantly influences the phase transition and give rise the emergent dome-shaped $T_c$ dependence.

\section{Modulation in vertical interactions}\label{sec:mJy}
\begin{figure}[tbp]
    \centering
    \includegraphics[width=0.66666\linewidth]{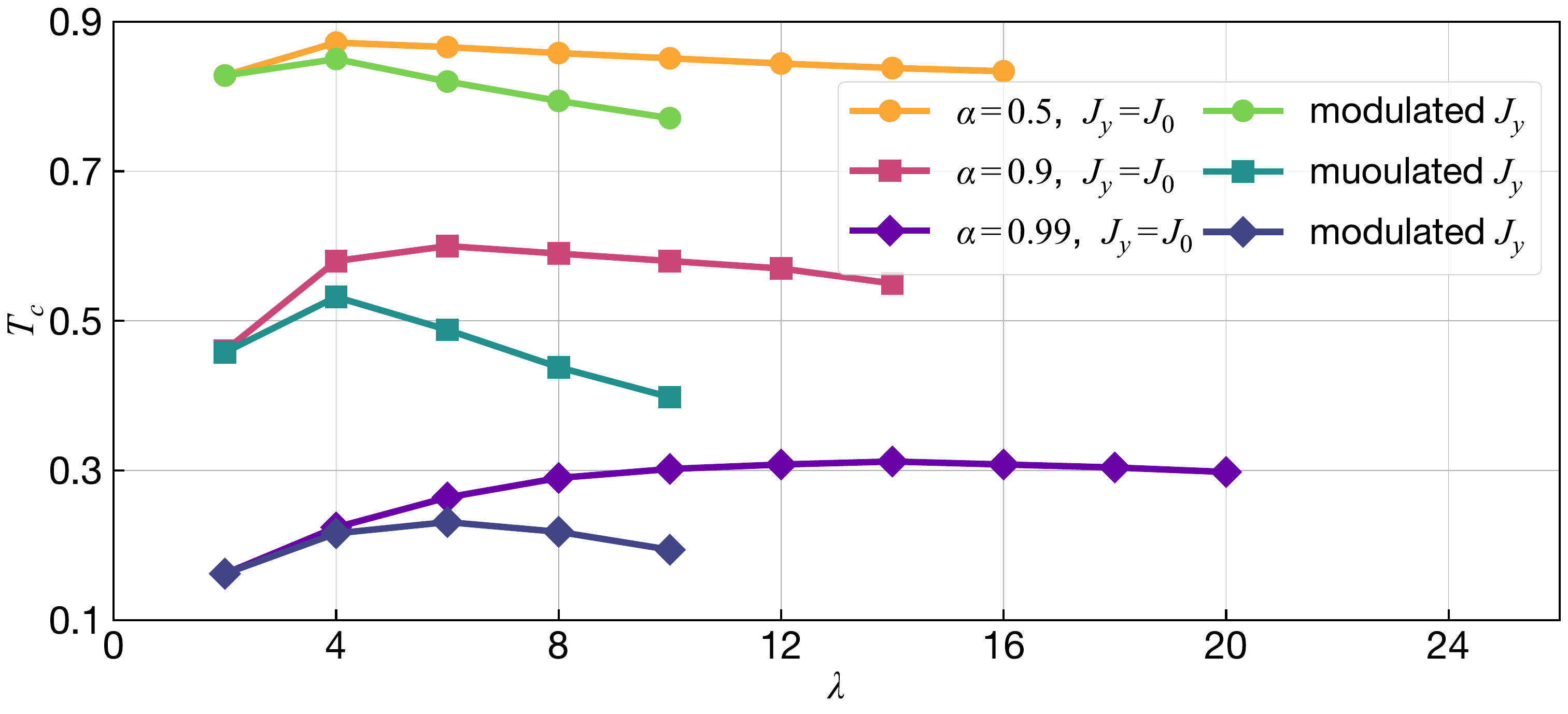}
    \caption{
    Dome structures of $T_c$ as a function of modulation amplitude $\alpha$ and wavelength $\lambda$. Compared to the uniform case ($J_y = J_0$), modulating $J_y$ enhances the modulation effect, shifting the dome toward smaller $\lambda$ and reducing its maximum.}
    \label{fig:mJy}
\end{figure}

In the main text, we set the interaction along the $x$-direction as $J_x = J_0 [1 + \alpha \sin(k x_i)]$, while keeping $J_y = J_0$ along the $y$-direction, since the influence of the CDW is predominantly along the $x$-axis. This choice preserves the essential physics but makes the modulation dependence of the dome structure more apparent.

In Fig.~\ref{fig:mJy}, we present results for the modulated XY model with a modulated vertical interaction defined as $J_y = 0.5(J_x + J_{x+1})$, i.e., the average of two neighboring horizontal $J_x$ values. We observe that the peak height of $T_c$ decreases, and its position shifts to larger $\lambda$ as $\alpha$ increases. Compared to the constant $J_y = J_0$ case, introducing modulation in $J_y$ shifts the peak to smaller $\lambda$, indicating a stronger modulation effect. As a result, a much larger $\alpha$ is required to produce a noticeable shift in the peak position. Notably, for $\lambda = 2$, the modulated and unmodulated models are equivalent, since $J_y = J_0$ corresponds to the average of $J_0(1+\alpha)$ and $J_0(1-\alpha)$. For all other $\lambda$, $T_c$ is generally reduced.

The effect of modulating $J_y$ can also be understood via the MK-RG analysis, which accounts for the spatial variation and bunching of vertical ($y$-direction) bonds, in addition to the uniform case with $J_y = J_0$. Specifically, the renormalization condition becomes
\be
\bigotimes_{m=1}^{\lambda} W(K_m) = \bigotimes_{n=1}^{\lambda} W(K_y),
\label{eq:Ky}
\ee
with $K_m=J_y/T$, different from Eq.~\eqref{eq:Kx} for the 1D decimation. This leads to a system of $\lambda d$ equations with $d$ unknowns, which generally has no exact solution (where $d \times d$ is the size of the interaction matrix $W$). In the special case of bond dimension $d=1$, the renormalized coupling simplifies to $K_y = (\prod_{n=1}^{\lambda} K_m )^{1/\lambda}$. By symmetry considerations, $K_y$ increases with $\lambda$, similar to the behavior of $K_x$. Consequently, when both $J_x$ and $J_y$ are modulated, the overall modulation effect becomes more sensitive to $\lambda$. Since modulation typically reduces the effective coupling strength compared to the uniform case, the presence of vertical modulation further suppresses $T_c$.

\bibliographystyle{apsrev4-2}
\bibliography{ref}